\documentclass{article}
\usepackage{graphicx}%
\usepackage{multirow}%
\usepackage{amsmath,amssymb,amsfonts}%
\usepackage{amsthm}%
\usepackage{mathrsfs}%
\usepackage[title]{appendix}%
\usepackage{xcolor}%
\usepackage{textcomp}%
\usepackage{manyfoot}%
\usepackage{booktabs}%
\usepackage{algorithm}%
\usepackage{algorithmicx}%
\usepackage{algpseudocode}%
\usepackage{adjustbox}
\usepackage{listings}%
\usepackage{mathtools}
\usepackage{bbm}
\usepackage{bm}
\usepackage[labelsep=space]{caption}
\usepackage{subcaption}
\usepackage{url}
\usepackage{multirow}
\usepackage{geometry}
\usepackage[numbers]{natbib}

\newtheorem{Theorem}{Theorem}

\newtheorem{Example}{Example}
\newtheorem{Lemma}{Lemma}

\title{A Diffusion-Based Approach for Simulating Forward-in-Time State-Dependent Speciation and Extinction Dynamics}
\author{
	Albert C. Soewongsono
	\thanks{Department of Biology, Washington University in St. Louis, Rebstock Hall, St. Louis, Missouri, 63130, USA, email: soewongsono@wustl.edu.}
	\and
	Michael J. Landis
        \thanks{Department of Biology, Washington University in St. Louis, Rebstock Hall, St. Louis, Missouri, 63130, USA}
}
\date{}
\begin{document}

\maketitle

\begin{abstract}
We establish a general framework using a diffusion approximation to simulate forward-in-time state counts or frequencies for cladogenetic state-dependent speciation-extinction (ClaSSE) models. We apply the framework to various two- and three-region geographic-state speciation-extinction (GeoSSE) models. We show that the species range state dynamics simulated under tree-based and diffusion-based processes are comparable. We derive a method to infer rate parameters that are compatible with given observed stationary state frequencies and obtain an analytical result to compute stationary state frequencies for a given set of rate parameters. We also describe a procedure to find the time to reach the stationary frequencies of a ClaSSE model using our diffusion-based approach, which we demonstrate using a worked example for a two-region GeoSSE model. Finally, we discuss how the diffusion framework can be applied to formalize relationships between evolutionary patterns and processes under state-dependent diversification scenarios. 
\end{abstract}



\section{Introduction}\label{sec:intro}

The branching events of a phylogenetic tree exhibit a pattern that stores information about the underlying speciation and extinction processes~\citep{nee1994reconstructed}. In~\cite{nee1994reconstructed}, they first considered a model where both speciation and extinction are treated as a constant-rate birth-death process by which lineages give birth to new lineages (speciation) at a rate $\lambda$ and lineages die (extinction) at a rate $\mu$. Speciation and extinction rates, however, are expected to vary idiosyncratically among phylogenetic lineages and over geological timescales. For example, \cite{nee1994reconstructed} also considered another model in which speciation and extinction rates vary over time. Workers have designed birth-death models to study a variety of intrinsic and extrinsic factors that might shape diversification rates. Species age~\citep{Hagen2015,alexander2016quantifying,soewongsono2022shape} and inherited traits~\citep{Kontoleon,maddison2007estimating,fitzjohn2010quantitative,fitzjohn2012diversitree,soewongsono2023matrix} are two types of intrinsic factors thought to drive diversification rates, whereas environment ~\citep{condamine2013,quintero2023} and geography~\citep{goldberg2011phylogenetic,landis2022phylogenetic,swiston2023testing} are common extrinsic factors of interest. In the end, a common goal of these models is to infer the underlying event rates given an observed phylogenetic pattern either through likelihood-based~\citep{morlon2010inferring,stadler2013can,louca2020} or likelihood-free approaches~\citep{nee1994reconstructed,voznica2022deep,he2023approximate,lambert2023,thompson2023deep}. \\

Fundamentally, birth-death processes model the random arrival times of discrete events that generate or ``build'' a phylogenetic tree over time \citep{nee1994reconstructed, maddison2007estimating}. As an alternative to this tree-based representation of the process, recent work~\citep{chevin2016species} introduced an equivalent diffusion-based representation for a class of birth-death models with state-dependent rates, known as state-dependent speciation-extinction (SSE) models~\citep{maddison2007estimating}. As noted by~\cite{chevin2016species}, population genetics theory has benefited immensely from diffusion-based approximations to population-based models of allele frequency change, yet diffusion-based approximations of birth-death models remain underexplored in the phylogenetics literature. Despite the widespread popularity of birth-death models among evolutionary biologists, these models recently entered a phase of intense but overdue scrutiny to better understand what the models can and cannot estimate reliably when fitted to real biological datasets~\citep{louca2020extant,morlon2022,vasconcelos2022flexible,dragomir2023parameter,kopperud2023,legried2023, truman2024fossilised, celentano2024exact,Tarasov2022}. This has created demand for new frameworks to understand the mathematical properties of these complex stochastic processes to guide biological research programs. \\
 
As mentioned above, applying diffusion processes in the macroevolutionary context is not new, and was recently applied by~\cite{chevin2016species} to study the properties of the BiSSE~\citep{maddison2007estimating} and QuaSSE~\citep{fitzjohn2010quantitative} models.
Our work begins by extending the diffusion-based BiSSE representation of~\cite{chevin2016species} to a general multi-state SSE model that allows for both cladogenetic and anagenetic state changes, known as the ClaSSE model~\citep{goldberg2012tempo}. 
We then show how our formulation may be used to determine the relationship between a set of SSE rates and their implied stationary state frequencies.
Inverting this perspective, we show that our framework correctly delimits classes of SSE rate values that yield a given set of stationary frequencies.
This establishes a many-to-one mapping of SSE rates on to stationary frequencies.
After introducing our general framework for ClaSSE models, we apply it to a special geographical case of the ClaSSE model, known as the GeoSSE model~\citep{goldberg2011phylogenetic}. We choose the GeoSSE model because it possesses a complex but structured relationship among its parameters and its constituent events -- i.e. dispersal, within-region speciation, between-region speciation, and local extinction -- that impact lineages over evolutionary time. We then validate our theoretical results by simulating state frequency trajectories using both tree-based and diffusion-based simulators. \\

The rest of the paper is organized as follows. Firstly, in Section~\ref{sec:overview}, we give a brief overview of SSE models in general. In Section~\ref{sec:transform} we visit relevant results in the theory of stochastic process, then in Section~\ref{subsec::classe} we apply our framework to analyze the ClaSSE model, and later for the GeoSSE model with arbitrary number of regions in Section~\ref{subsec::diffusiongeosse}. Following these, in Sections~\ref{sec::Pro1} and~\ref{subsec::derivestationary} we present a method for simulating state dynamics under our framework and deriving rate parameters given stationary state frequencies. In Section~\ref{subsec::derivestationary-inverse}, we derive a result to compute theoretical stationary state frequencies given rate parameters. Moreover, in Section~\ref{subsec::derivestationary-time}, we describe a procedure to compute time to reach stationary frequencies in a 2-region GeoSSE system using results derived in Section~\ref{subsec::derivestationary-inverse}. Furthermore, in Section~\ref{subsec::resultdiffusion}, we show, through simulation examples, that our diffusion-based framework offers a good approximation for simulating range state dynamics when comparing to tree-based approach. In Section~\ref{subsec::resultstationary}, using an example, we show the existence of alternative rate scenarios that lead to the same stationary state frequencies. Additionally, we apply results derived in Section~\ref{subsec::derivestationary-inverse} and Section~\ref{subsec::derivestationary-time} to that example in Section~\ref{subsec::resultstationary}. Lastly, in Section~\ref{sec:conclusion}, we summarize our results and discuss promising ways to study pattern-process relationships for data generated by SSE models, and ideas for future work using our framework. 

\section{Methods}\label{sec:methods}
This section describes the framework for how construct our diffusion approximation for a ClaSSE model to analyze the dynamics of states through time. Key results include derivations of the transition probabilities and the infinitesimal mean and variance parameters of the diffusion equation. We describe and implement the methods for simulating the evolution of state frequencies, and derive relevant results for the stationary conditions, focusing on two- and three-region GeoSSE models, which are special cases of the ClaSSE model. 



\subsection{Overview of state-dependent speciation and extinction models}\label{sec:overview}

In this section, we give a brief overview of SSE models by highlighting the key assumptions and different events occurring along lineages. Then, we briefly re-visit a particular SSE model type, the GeoSSE model~\citep{goldberg2011phylogenetic}. Then, we guide towards how to shift from tree-based perspective to non-tree-based perspective to derive our object of interest. \\

\begin{figure}[h!]
	\centering
        \includegraphics[width=1\linewidth]{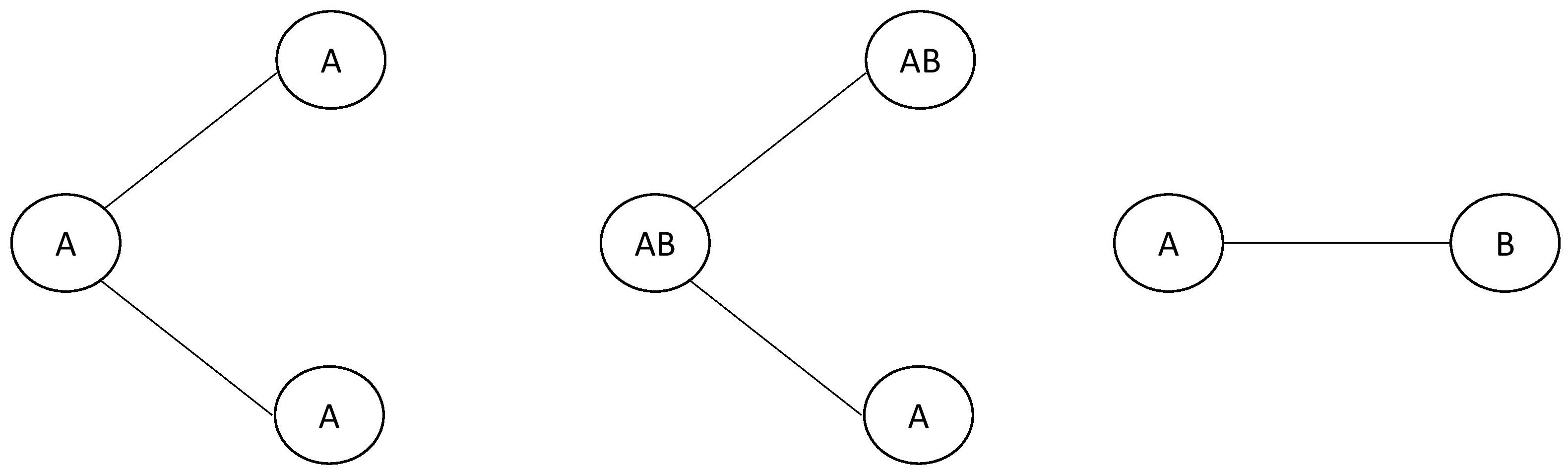}
	\nobreak
	\caption{From left to right: a speciation event without cladogenetic state changes, a speciation event with cladogenetic state changes, an anagenetic state change}
	\label{anagenetic_cladogenetic}
\end{figure}

In general, SSE models are stochastic branching processes with state-dependent birth (speciation) and death (extinction) rates. The states can either be discrete or continuous~\citep{maddison2007estimating,fitzjohn2010quantitative,fitzjohn2012diversitree} and can represent various things, ranging from phenotypic traits to geographical ranges~\citep{goldberg2011phylogenetic}. Some SSE models have processes that are only defined by anagenetic process and state-dependent diversification process~\citep{maddison2007estimating}, while others have processes that are defined by both anagenetic and cladogenetic processes~\citep{goldberg2011phylogenetic,goldberg2012tempo} shown in Fig.~\ref{anagenetic_cladogenetic}. An anagenetic process is defined as a process of trait evolution within lineages, between branching events. In the BiSSE model~\citep{maddison2007estimating}, this corresponds to trait transition events of going from a discrete trait $A$ to another discrete trait $B$ or vice versa. These trait-dependent transition rates are encoded in the infinitesimal rate matrix $\bm Q$, for which the off-diagonal entry $q_{ij}$ defines the rate of transitioning from state $i$ to $j$. A cladogenetic process is defined as a process in which state transition occurs in conjunction with a branching event (with speciation) of a lineage. SSE models with anagenetic and cladogenetic events are referred to as ClaSSE models. \\

Part of this paper will consider a special case of the ClaSSE model, the GeoSSE model~\citep{goldberg2011phylogenetic}. A GeoSSE model describes how species move and evolve among a sets of discrete geographical regions, called species ranges. Species that occur in just one region are said to be endemic to that region. Species occurring in two or more regions are said to be widespread. \\

GeoSSE events can be classified as anagenetic or cladogenetic events. Anagenetic events in GeoSSE include dispersal events and local extinction (sometimes called extirpation) events. Dispersal events add one region to a species range. Local extinction remove one region from a species range. A species experiences complete extinction (i.e. it is removed from the species pool) when it goes locally extinct in the last region in its range. Note that widespread species cannot experience complete extinction through a single event under a GeoSSE model; their widespread ranges must first be reduced to a single region before complete extinction is a possibility. \\

Cladogenetic events under GeoSSE include within-region speciation and between-region speciation events. Each within-region speciation event creates a new species within any single region of the parental species range. Each between-region speciation event causes a widespread parental species and its range to split, such that all regions in the parental range are distributed among the two new daughter lineages. Section~\ref{subsec::diffusiongeosse} defines how GeoSSE assigns rates to different events. \\

\begin{figure}[h!]
	\centering
        \includegraphics[scale=.1]
        {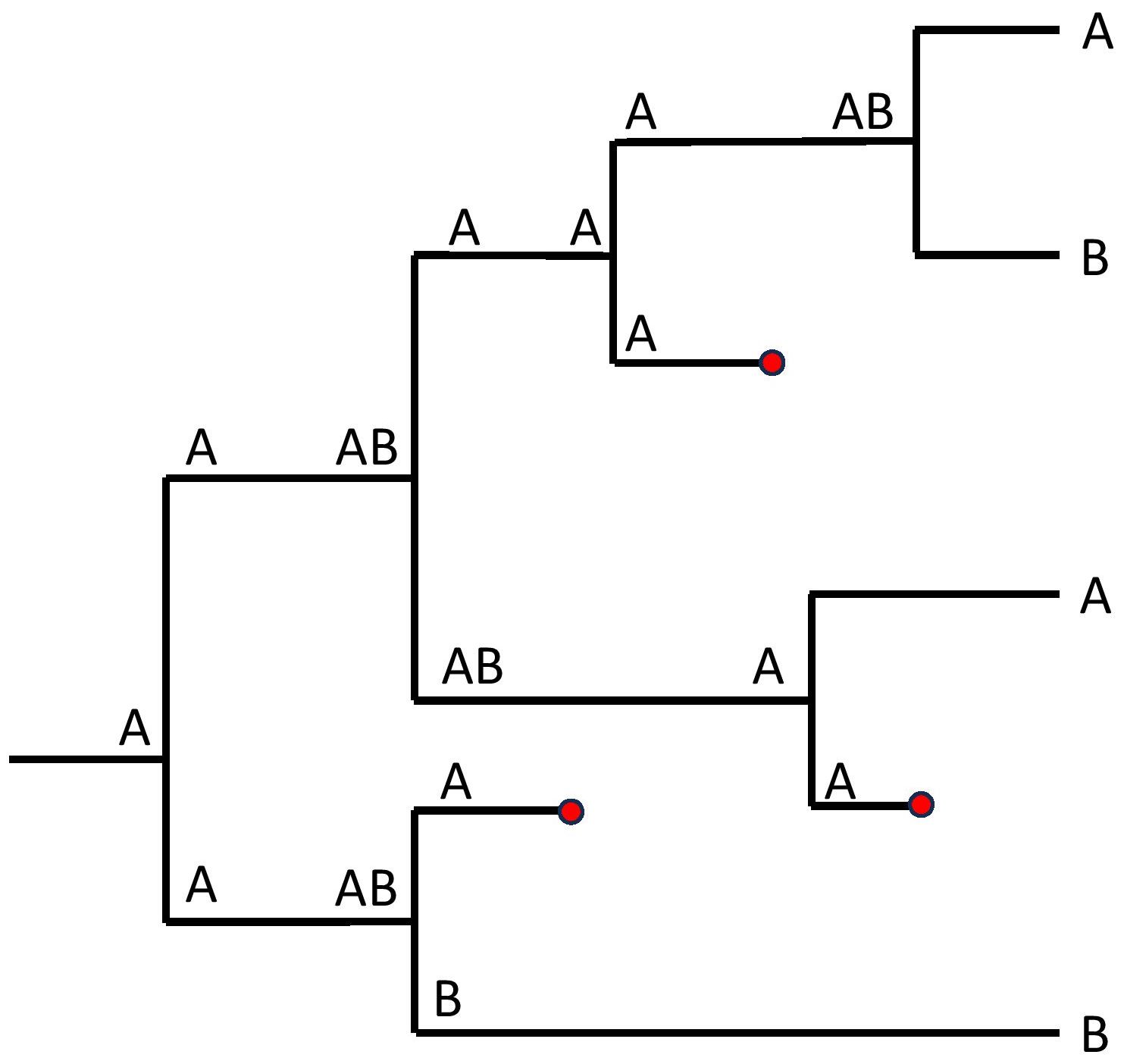}
	\nobreak
	\caption{An illustration of GeoSSE events on a phylogeny with range state information}
	\label{geosse_anagenetic_cladogenetic}
\end{figure}

Given a phylogeny with range state information as seen in Fig.~\ref{geosse_anagenetic_cladogenetic}, one can observe the dynamics of range states accumulated by species though time. In Section~\ref{sec:transform}, we present the necessary theory that will later be used to allow us transitioning from a tree-based process to an alternative, diffusion-based process to simulate the dynamics. 
\subsection{Transforming a stochastic process}\label{sec:transform}

In this section, we briefly describe the relevant results in the theory of stochastic processes that enable us to transform one stochastic process into another stochastic process. In the context of the ClaSSE model described in Section~\ref{sec:overview}, we want to define a process that simulates the (discrete) count of species with state $i$ through time. This process can then be used to define a second process that simulates the (continuous) frequency of species with state $i$ over time. \\

\begin{Theorem}{It\^o's transformation formula}{\text{}}
\label{thm:1}

Consider a stochastic process $\{Z(t)\}$ with infinitesimal parameters $\mu(z)$ and $\sigma^{2}(z)$. Define a new stochastic process $\{Y(t)\}$ with $Y(t)=g(Z(t))$ where $g$ is a strictly monotone continuous and twice-differentiable function. Then, the new process $\{Y(t)\}$ has infinitesimal parameters given by,
\begin{itemize}
    \item $\mu_{Y}(y)=\mu(z)g'(z)+\frac{1}{2}\sigma^{2}(z)g''(z)$,
    \item $\sigma^{2}_{Y}(y)= \sigma^{2}(z)\left[g'(z)\right]^{2}.$
\end{itemize}
\end{Theorem}
{\bf Proof:}
This theorem is also known as It\^o's formula or It\^o's lemma. The proof is given in~\cite{ito1951stochastic,karlin1981second} .\qed\\
\begin{Lemma}{\text{}}

\label{lemma::stochtrans}
    Given a stochastic process  $\{N_{i}(t):=n_{i}(t)\}$ with infinitesimal mean and variance parameters $\mu_{i}=\mathbb{E}(dn_{i}/dt)$ and $\sigma^{2}_{i}=var(dn_{i}/dt)$, respectively.
    Define a stochastic process $\{X(t)\}$ derived using the following transformation.
    \begin{eqnarray}
        X(t)&=&g(N)\nonumber\\
            &=&g\left(\sum_{i}n_{i}\right)\nonumber\\
            &=& \sum_{i}h(n_{i}),
    \label{eq1}
    \end{eqnarray}
    where $\{N(t):=\sum_{i}n_{i}(t)\}$ is a stochastic process with infinitesimal parameters defined as follows, 
    \begin{eqnarray*}
        \mu(N) &=& \mu\left({\sum_{i}n_{i}(t)}\right)\\
               &=& \sum_{i}\mu\left(n_{i}(t)\right)\\
               &=& \sum_{i}\mu_{i}.\\
        \sigma^{2}(N) &=& \sigma^{2}\left({\sum_{i}n_{i}(t)}\right)\\
                      &=& \sum_{i}\sigma^{2}(n_{i}(t)) + \sum_{\substack{i,j \\ i\neq j}}\sigma_{ij}\\
                      &=& \sum_{i}\sigma^{2}(n_{i}(t))\\
                      &=& \sum_{i}\sigma^{2}_{i}.
    \end{eqnarray*}
    Note here we have used the fact that $\sigma_{ij}=0$ for $i \neq j$ to account for independent birth-death processes. The infinitesimal mean and variance parameters for $\{X(t)\}$ are given by,
    \begin{eqnarray}
    \mu_{X} &=& \sum_{i}\frac{\partial X}{\partial n_{i}}\mu_{i} + \frac{1}{2}\sum_{i}\frac{\partial^{2}X}{\partial n_{i}^{2}}\sigma^{2}_{i}\label{eq::itomean} \\
    \sigma^{2}_{X} &=& \sum_{i}\left(\frac{\partial X}{\partial n_{i}}\right)^{2}\sigma^{2}_{i}     \label{eq::itovariance}.
\end{eqnarray}
\end{Lemma}
{\bf Proof:}
Proof of Lemma~\ref{lemma::stochtrans} is given in Appendix~\ref{subsec::stochtrans}.\qed


\subsection{Diffusion-based framework for state-dependent diversification model}
\label{subsec::classe}

In this section, we establish the framework for simulating state dynamics for state-dependent speciation and extinction models using diffusion processes. We show how to implement the framework in the ClaSSE model introduced in~\cite{goldberg2012tempo}. Then, we relate our framework to earlier research~\citep{chevin2016species} using a diffusion process for the BiSSE model~\citep{maddison2007estimating} and, later on, for the GeoSSE model~\citep{goldberg2011phylogenetic}. 

Our first goal is to define the stochastic process $\{N_{i}(t)\}$, which describes the number of species with state $i \in S$ at time $t$, where $S$ is the state space of the model. Then, using the method presented in Section~\ref{sec:transform}, we can obtain the stochastic process $\{\Pi_{i}(t)\}$, which describes the frequency of species with state $i$ at time $t$. Using these two processes, we then derive results that directly link model parameters with stationary state frequency patterns that the model generates. \\

To proceed, we define the following probabilities: 
\begin{eqnarray}
    Prob(\{N_{i} \rightarrow N_{i}+1 \text{ in } \Delta t\}) = Prob(N_{i}(t+\Delta t)=n_{i}+1 \mid N_{i}(t)=n_i) &:=& \mathbb{P}_{i}^{+} \Delta t, \nonumber \\
    Prob(\{N_{i} \rightarrow N_{i}-1 \text{ in } \Delta t\}) &:=& \mathbb{P}_{i}^{-} \Delta t, \nonumber \\
    Prob(\{N_{i} \rightarrow N_{i} \text{ in } \Delta t\}) &:=& \mathbb{P}_{i}\Delta t.\nonumber\\
    \label{eq:transition}
\end{eqnarray}
These probabilities correspond to gaining a new species in state $i \left(\mathbb{P}_{i}^{+}\right)$, losing a species in state $i\left(\mathbb{P}_{i}^{-}\right)$, and neither losing nor gaining a new species in state $i\left( \mathbb{P}_{i}\right)$ within an infinitesimal time step $\Delta t$.\\

For the ClaSSE model, we can write those probabilities as follows, 
\begin{eqnarray}
    \mathbb{P}_{i}^{+}\Delta t &=& S_{i}^{+} + E_{i}^{+} + Q_{i}^{+}, \nonumber \\
    \mathbb{P}_{i}^{-}\Delta t &=& S_{i}^{-} + E_{i}^{-} + Q_{i}^{-}, \nonumber \\
    \mathbb{P}_{i}\Delta t &=& 1 - \left(\mathbb{P}_{i}^{+} + \mathbb{P}_{i}^{-}\right) \Delta t, \nonumber \\
\end{eqnarray}
where
\begin{eqnarray}
    S_{i}^{+} &=& \text{Probability of events that lead to an increase in the number of species in state $i$ through }\nonumber\\
    &&\text{state-dependent speciation and speciation in conjunction with \textit{cladogenetic} state change.} \nonumber \\
    E_{i}^{+} &=& \text{Probability of events that lead to an increase in the number of species in state $i$ through} \nonumber \\
    &&\text{extinction.}\nonumber\\
    Q_{i}^{+} &=& \text{Probability of events that lead to an increase in the number of species in state $i$ through} \nonumber \\
    && \text{\textit{anagenetic} state change.}\nonumber\\
    S_{i}^{-} &=& \text{Probability of events that lead to a decrease in the number of species in state $i$ through }\nonumber\\
    &&\text{state-dependent speciation and speciation in conjunction with \textit{cladogenetic} state change.} \nonumber \\
    E_{i}^{-} &=& \text{Probability of events that lead to a decrease in the number of species in state $i$ through } \nonumber \\
    && \text{extinction.}\nonumber\\
    Q_{i}^{-} &=& \text{Probability of events that lead to a decrease in the number of species in state $i$ through} \nonumber \\
    &&\text{\textit{anagenetic} state change.}\nonumber\\
    \label{eq::generalclasse}
\end{eqnarray}
Next, we define the infinitesimal mean $\mu_{i} = \mathbb{E}\left(dN_{i}/dt\right)$ and variance $\sigma_{i}^{2} = var\left(dN_{i}/dt\right)$ for the stochastic process $\{N_{i}(t):t>0\}$.\\ 

\begin{Lemma}{\text{}}

\label{lemma::geosseNpar}
    The infinitesimal mean $\mu_i$ and variance $\sigma_{i}^{2}$ for the stochastic process $\{N_{i}(t):t>0\}$ is given by 
    \begin{eqnarray}
        \mu_{i}         &=& \mathbb{P}_{i}^{+} - \mathbb{P}_{i}^{-}, \label{eq::mu_geosse}\\
    \sigma_{i}^2    &=& \mathbb{P}_{i}^{+} + \mathbb{P}_{i}^{-}. \label{eq::var_geosse}
    \end{eqnarray}
\end{Lemma}
{\bf Proof:}
Proof of Lemma is given in Appendix~\ref{subsec::geosseNpar}.\qed\\

Next, we define a stochastic process $\{\Pi_{i}(t): t>0\}$ where
\begin{eqnarray*}
    \Pi_{i} = \frac{N_{i}}{\sum_{j \in S} N_{j}} = \frac{N_{i}}{N}.
\end{eqnarray*}
$\Pi_{i}(t)$ denotes the frequency of species being in state $i$ at time $t$. We define the infinitesimal mean and variance for the process in Lemma~\ref{Lemma::geossepipars}.\\
\begin{Lemma}{\text{}}
\label{Lemma::geossepipars}

    The infinitesimal mean $\mu_{\Pi_i}$ and variance $\sigma_{\Pi_i}^{2}$ for the stochastic process $\{\Pi_{i}(t): t>0\}$ is given by 
\begin{eqnarray}
\mu_{\Pi_{i}}       
                    &=& \frac{1}{N}\left(\mu_i - \frac{\sigma^{2}_{i}}{N}\right) + \frac{\Pi_i}{N}\sum_{j \in S}\left(-\mu_j + \frac{\sigma_{j}^2}{N}\right),
                    \label{eq::mu_geossefreq}\\
\sigma_{\Pi_{i}}^{2} 
                     &=& \left(\frac{\sigma_i}{N}\right)^{2}(1-2\Pi_i) + \left(\frac{\Pi_i}{N}\right)^{2}\sum_{j\in S}\sigma_{j}^2.\label{eq::var_geossefreq}
\end{eqnarray}
\end{Lemma}
{\bf Proof:} Proof of Lemma~\ref{Lemma::geossepipars} is given in Appendix~\ref{subsec::geossepipars}.\qed\\

From Eqs.~\eqref{eq::mu_geossefreq}-\eqref{eq::var_geossefreq}, it is clear that the diffusion parameters $\left(\text{i.e. }\mu_{\Pi_i}, \sigma^{2}_{\Pi_i}\right)$ are undefined under a total extinction scenario of a tree (i.e. where $N=0$ appears in multiple denominators).\\

To demonstrate the generality of the framework, we show the BiSSE model~\citep{maddison2007estimating} (and similarly for the MuSSE model~\citep{fitzjohn2012diversitree}) can be represented as a diffusion process as follows. Under the BiSSE model, species possess binary traits with values in the state space $S = \{ 1, 2 \}$. BiSSE is a special case of the ClaSSE model that, while it allows anagenetic trait transition and extinction events, its speciation events do not cause cladogenetic trait changes. That is, daughter lineages identically inherit the parent lineage state following speciation. Readers can refer to the supplementary material from~\cite{goldberg2012tempo} for its derivation. For the BiSSE model, we have  
\begin{eqnarray}
    S_{1}^{+} = \lambda_{1}N_{1}\Delta t, \:
    E_{1}^{+} = 0, \: Q_{1}^{+} = q_{21}N_{2}\Delta t, \nonumber\\
    S_{1}^{-} = 0, \: E_{1}^{-} = \mu_{1}N_{1}\Delta t, \: Q_{1}^{-} = q_{12}N_{1}\Delta t, \nonumber
\end{eqnarray}
where $\lambda_1$ and $\mu_1$ ($b_1$ and $d_1$ in~\cite{chevin2016species}) are speciation and extinction rates for trait $1$, respectively. $q_{12}$ and $q_{21}$ ($\tau_{12}$ and $\tau_{21}$ in~\cite{chevin2016species}) are anagenetic trait transition from $1$ to $2$ and from $2$ to $1$, respectively. Similarly, following the definitions in Eq.~\eqref{eq::generalclasse}, we also have\\
\begin{eqnarray}
    S_{2}^{+} = \lambda_{2}N_{2}\Delta t, \:
    E_{2}^{+} = 0, \: Q_{2}^{+} = q_{12}N_{1}\Delta t, \nonumber\\
    S_{2}^{-} = 0, \: E_{2}^{-} = \mu_{2}N_{2}\Delta t, \: Q_{2}^{-} = q_{21}N_{2}\Delta t, \nonumber
\end{eqnarray}
Using Eq.~\eqref{eq::mu_geosse} and Eq.~\eqref{eq::var_geosse} we have the infinitesimal mean and variance of $N_1$,
\begin{eqnarray}
    \mu_1 &=& \left(\lambda_{1} - \mu_{1} - q_{12}\right)N_{1} + q_{21}N_{2},\\
    \sigma_{1}^{2} &=& \left(\lambda_{1} + \mu_{1} + q_{12}\right)N_{1} + q_{21}N_{2},
\end{eqnarray}
and similarly for $N_2$ with indices changed accordingly. These are the same $\mu_{1}$ and $\sigma^{2}_{1}$ as described in Eq.~(2) in~\cite{chevin2016species}.
\qed


\subsection{Diffusion-based framework for the GeoSSE model}
\label{subsec::diffusiongeosse}

In this section, we use the framework established in Section~\ref{subsec::classe} for general ClaSSE models to the GeoSSE model.
The procedure we apply here is also compatible with any model from the ClaSSE family. For the GeoSSE model, unlike the BiSSE model described in Section~\ref{subsec::classe}, some speciation events also cause \textit{cladogenetic} state changes. Thus, following the notation used in the previous section we have,
\begin{eqnarray*}
    S_{i}^{+} &=& W_{i}^{+} + B_{i}^{+} \\
    E_{i}^{+} &=& E_{i}^{+}\\
    Q_{i}^{+} &=& D_{i}^{+} + E_{i}^{+} \\[10pt]
    S_{i}^{-} &=& W_{i}^{-} + B_{i}^{-} \\
    E_{i}^{-} &=& E_{i}^{-}\\
    Q_{i}^{-} &=& D_{i}^{-} + E_{i}^{-},
\end{eqnarray*}
where
\begin{eqnarray}
    W_{i}^{+} &=& \text{Probability of events that lead to an increase in the number of species in range state $i$}\nonumber\\
    &&\text{through within-region speciation for either widespread or endemic species.} \nonumber \\
    B_{i}^{+} &=& \text{Probability of events that lead to an increase in the number of species in range state $i$}\nonumber\\
    &&\text{through between-region speciation for widespread species.} \nonumber \\
    E_{i}^{+} &=& \text{Probability of events that lead to an increase in the number of species in range state $i$} \nonumber \\
    &&\text{through extinction for either widespread species (local extinction) or endemic species}\nonumber \\
    &&\text{(species extinction).}\nonumber\\
    D_{i}^{+} &=& \text{Probability of events that lead to an increase in the number of species in range state $i$} \nonumber \\
    && \text{through range dispersal event for endemic species.}\nonumber\\[10pt]
    W_{i}^{-} &=& \text{Probability of events that lead to a decrease in the number of species in range state $i$}\nonumber\\
    &&\text{through within-region speciation for either widespread or endemic species.} \nonumber \\
    B_{i}^{-} &=& \text{Probability of events that lead to a decrease in the number of species in range state $i$}\nonumber\\
    &&\text{through between-region speciation for widespread species.} \nonumber \\
    E_{i}^{-} &=& \text{Probability of events that lead to a decrease in the number of species in range state $i$} \nonumber \\
    && \text{through extinction for either widespread species (local extinction) or endemic species}\nonumber\\
    &&\text{(species extinction).}\nonumber\\
    D_{i}^{-} &=& \text{Probability of events that lead to a decrease in the number of species in range state $i$} \nonumber \\
    && \text{through range dispersal event for endemic species.}\nonumber
\end{eqnarray}
Next, consider an $n$-region GeoSSE model where $n\in \mathbb{Z}^{+}$, we define the following state space and variable,
\begin{eqnarray*}
    R &=& \text{state space for regions e.g., } R = \{A,B\}.\\
    S &=& \text{state space for species ranges e.g., } S = \{ \{A\}, \{B\}, \{A,B\}\}  \\
    N_{i} &=& \text{number of species with range state $i$ where $i \in S$}.
\end{eqnarray*}
Then, we define the following rate parameters,
\begin{eqnarray*}
    d_{k\ell} &=& \text{per lineage dispersal rate of any species in region $k$ to colonize region $\ell$}.\\
    w_{\ell} &=& \text{per lineage within-region speciation rate of any species in region $\ell$}.\\
    b^{i}_{j} &=& \text{per lineage between-region speciation rate of a widespread species into}. \\
    &&\text{two daughter species with ranges $i$ and $j$, respectively. Note that }b^{i}_{j} \equiv b^{j}_{i}.\\
    e_{\ell} &=& \text{local extinction rate of any species in region $\ell$}.
\end{eqnarray*}
Thus, both $ w_{\ell}$ and $ b^{i}_{j}$ determine state-dependent speciation rate, $e_{\ell}$ determines state-dependent extinction rate, and $d_{k\ell}$ and (among widespread species) $e_\ell$ determine the anagenetic state transition rate.\\ 

We define a stochastic process $\{N_{i}(t)\}$ with infinitesimal mean $\mu_{i}=\mathbb{E}(dN_{i}/dt)$ and variance $\sigma^{2}_{i}=var(dN_{i}/dt)$. Here, $N_{i}(t)$ represents the number of species with range state $i$ at time $t$. The infinitesimal mean $\mu_{i}$ and variance $\sigma^{2}_{i}$ follow directly from Lemma~\ref{lemma::geosseNpar}. We derive the transition probabilities described in Eq.~\eqref{eq:transition} in the context of the GeoSSE model, as shown in Eqs.~\eqref{eq::piup}-\eqref{eq::pistay}. \\

Each of these probabilities describe possible events in a GeoSSE model occurring within an infinitesimal time step that result in gaining a new species with range state $i$ ($\mathbb{P}_{i}^{+}$), losing a species with range state $i$ ($\mathbb{P}_{i}^{-}$), and neither losing nor gaining a species with range state $i$ ($\mathbb{P}_{i}$).
\begin{eqnarray}
     \mathbb{P}_{i}^{+}\Delta t &=& W_i^{+} + D_{i}^{+} + B_{i}^{+} +       
                            E_{i}^{+} \nonumber \\
                        &=& \underbrace{\sum_{\mathclap{\substack{
                          j \in S
                        }}}
                        \sum_{\mathclap{\substack{
                          \ell \in j \\
                          \{ \ell \} = i
                        }}}N_j w_\ell \Delta t}_{W_{i}^{+}}
                        + 
                        \underbrace{\sum_{\mathclap{\substack{
                          k \in i}}}
                        \sum_{\mathclap{\substack{
                          \ell \in i \\
                          \ell \neq k}}}
                        N_{i \backslash \{ \ell \}} d_{k\ell} \Delta t}_{D_{i}^{+}}  
                        + 
                        \underbrace{\sum_{\mathclap{\substack{
                          j \in S \\
                          i \subset j
                        }}} N_j b^{i}_{j \backslash i}\Delta t}_{B_{i}^{+}} 
                        + \underbrace{\sum_{\substack{
                          j \in S \\
                          |j \backslash i| = 1
                        }}
                        \sum_{\substack{
                          \ell \in j \backslash i
                        }} N_j e_\ell\Delta t}_{E_{i}^{+}}
                        \label{eq::piup} \nonumber\\
                        \\
                        [10pt]
    \mathbb{P}_{i}^{-}\Delta t  &=& W_{i}^{-} + D_{i}^{-} + B_{i}^{-} + E_{i}^{-} \nonumber \\
                        &=& \underbrace{0 \vphantom{\sum_{\substack{\\
                          k \in i \\
                          |i| < R}} 
                        \sum_{\ell \in R \backslash \{k\}}
                        N_i d_{k\ell}}}_{W_{i}^{-}} 
                        + 
                        \underbrace{\sum_{\substack{\\
                          k \in i \\
                          |i| < R}} 
                        \sum_{\ell \in R \backslash \{k\}}
                        N_i d_{k\ell}\Delta t}_{D_{i}^{-}} 
                        +
                        \underbrace{\sum_{\substack{
                          j \in S \\
                          j \subset i
                        }}\frac{1}{2} N_i b^j_{i \backslash j}\Delta t}_{B_{i}^{-}}
                        + 
                        \underbrace{\sum_{\substack{
                          \ell \in R \\
                          \ell \in i
                        }} N_i e_\ell\Delta t}_{E_{i}^{-}}
                        \label{eq::pidown}
                        \\[10pt]
    \mathbb{P}_{i}\Delta t    &=& 1-\left(\mathbb{P}_{i}^{+} + \mathbb{P}_{i}^{-}\right)\Delta t.
    \label{eq::pistay}
\end{eqnarray}

\begin{figure}[h!]
	\centering
        \includegraphics[scale=.3]{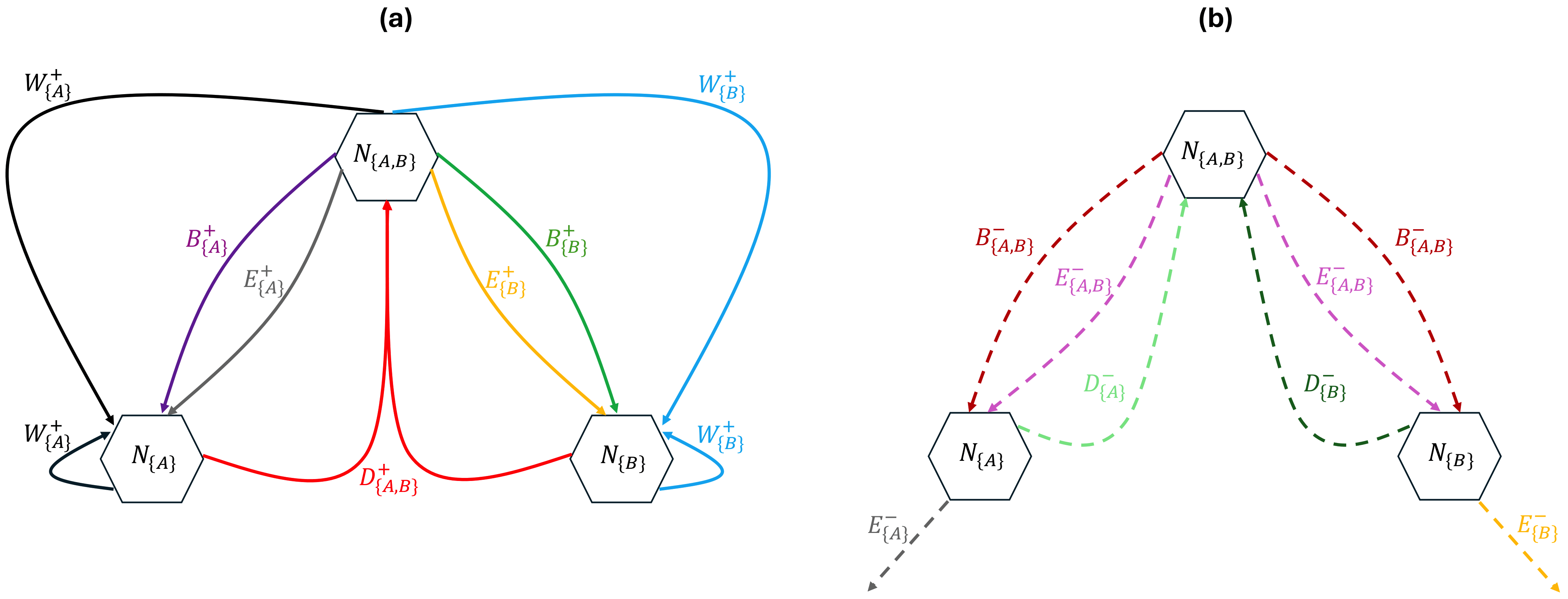}
	\nobreak
	\caption{Graphical illustrations of probabilities of events following Eq.~\eqref{eq::piup} shown in (a), and Eq.~\eqref{eq::pidown} shown in (b) for a 2-region GeoSSE system with state space $S = \{\{A\},\{B\},\{A,B\}\}$. $N_{i}$ represents the number of species with range state $i \in S$. An incoming arrow into $N_{i}$ compartment means there is an increase in species count with range state $i$ and an outgoing arrow from $N_{i}$ means there is a decrease in species count with range state $i$. All the events and arrows are color-coded accordingly.}
	\label{events_geosse}
\end{figure}

For clarity, we provide the biogeographic interpretation on how each term in Eqs.~\ref{eq::piup}-\ref{eq::pistay} is derived and a graphical illustration of the events in Fig.~\ref{events_geosse}.
\begin{enumerate}
    \item $\bm{W_{i}^{+}}$. To gain a new species with range state $i$ through a within-region speciation event, the new species range $i$ must contain only region $\ell$ ($\ell \in i$ and $|i| = 1$). This endemic species can undergo a speciation event with probability $w_\ell N_{i}$. Any species with range state $j$ that also occupies region $\ell$ can undergo a within-region speciation event with probability $w_\ell \sum_{j \in S} \mathbbm{1}_{i \subseteq j}N_{j}$. The total probability of this event occurring within $\Delta t$ is, 
    \begin{eqnarray*}
        \sum_{\mathclap{\substack{
                          j \in S}}}
                        \sum_{\mathclap{\substack{
                          \ell \in j \\
                          \{ \ell \} = i
                        }}}N_j w_\ell \Delta t.
    \end{eqnarray*}
    As an example, in a $2-$region GeoSSE system with state space~$S = \{\{A\},\{B\},\{A,B\}\}$ we have,
    \begin{eqnarray*}
        W^{+}_{\{A\}} &=& \left(N_{\{A,B\}} + N_{\{A\}} \right)w_{A}\Delta t \\
        W^{+}_{\{B\}} &=& \left(N_{\{A,B\}} + N_{\{B\}} \right)w_{B}\Delta t \\
        W^{+}_{\{A,B\}} &=& 0.
    \end{eqnarray*}
    \item $\bm{D_{i}^{+}}$. To gain a new species with range state $i$ through a dispersal event, the species adds the new region $\ell$ to its ancestral range. Species are always widespread immediately following dispersal. The total probability of this event occurring within $\Delta t$ is, 
    \begin{eqnarray*}
        \sum_{\mathclap{\substack{
                          k \in i}}}
                        \sum_{\mathclap{\substack{
                          \ell \in i \\
                          \ell \neq k}}}
                        N_{i \backslash \{ \ell \}} d_{k\ell}\Delta t.
    \end{eqnarray*}
    As an example, in a $2-$region GeoSSE system with state space~$S = \{\{A\},\{B\},\{A,B\}\}$ we have,
    \begin{eqnarray*}
        D^{+}_{\{A\}} &=& 0\\
        D^{+}_{\{B\}} &=& 0\\
        D^{+}_{\{A,B\}} &=& N_{\{B\}}d_{BA}\Delta t + N_{\{A\}}d_{AB}\Delta t.
    \end{eqnarray*}
    \item $\bm{B_{i}^{+}}$. To gain a new species with range state $i$ through a between-region speciation event, the new species can be either endemic or widespread $|i| > 0$ that originated from a widespread ancestral species with larger range state $j$ ($i \subset j$). In general, we have no information of whether the new species occurs in left or right lineage following a speciation event, so we do not consider the orientation. The total probability of this event occurring within $\Delta t$ is, 
    \begin{eqnarray*}
        \sum_{\mathclap{\substack{
                          j \in S \\
                          i \subset j
                        }}} N_j b^{i}_{j \backslash i}\Delta t.
    \end{eqnarray*}
     As an example, in a $2-$region GeoSSE system with state space~$S = \{\{A\},\{B\},\{A,B\}\}$ we have,
     \begin{eqnarray*}
         B^{+}_{\{A\}} &=& N_{\{A,B\}}b^{A}_{B}\Delta t\\
         B^{+}_{\{B\}} &=& N_{\{A,B\}}b^{B}_{A}\Delta t\\
         B^{+}_{\{A,B\}} &=& 0.
     \end{eqnarray*}
    \item $\bm{E_{i}^{+}}$. To gain a new species with range state $i$ through a  local extinction event, the ancestral species must have a larger range state $j$ with size that differs by 1 from the new species' range state $i$ such that $|j \backslash i| = 1$. The total probability of this event occurring within $\Delta t$ is, 
    \begin{eqnarray*}
        \sum_{\substack{
                          j \in S \\
                          |j \backslash i| = 1
                        }}
                        \sum_{\substack{
                          \ell \in j \backslash i
                        }} N_j e_\ell \Delta t.
    \end{eqnarray*}
    As an example, in a $2-$region GeoSSE system with state space~$S = \{\{A\},\{B\},\{A,B\}\}$ we have,
    \begin{eqnarray*}
        E^{+}_{\{A\}} &=& N_{\{A,B\}}e_B \Delta t\\
        E^{+}_{\{B\}} &=& N_{\{A,B\}}e_A \Delta t\\
        E^{+}_{\{A,B\}} &=& 0.
    \end{eqnarray*}
    \item $\bm{W_{i}^{-}}$. The probability of losing a either endemic or widespread species with range state $i$ through a within-region speciation event is $0$. This is because the event will only increase the local abundance in a region and causes the widespread abundance to remain unchanged. \\
    \item $\bm{D_{i}^{-}}$. To lose a species with range state $i$ through a dispersal event, the species must disperse to a new region. The species count remains unchanged if the species already occupies all regions ($|i| = |R|$). The total probability of this event occurring within $\Delta t$ is, 
    \begin{eqnarray*}
        \sum_{\substack{\\
                          k \in i \\
                          |i| < R}} 
                        \sum_{\ell \in R \backslash \{k\}}
                        N_i d_{k\ell}\Delta t.
    \end{eqnarray*}
    As an example, in a $2-$region GeoSSE system with state space~$S = \{\{A\},\{B\},\{A,B\}\}$ we have,
    \begin{eqnarray*}
        D^{-}_{\{A\}} &=& N_{\{A\}}d_{AB}\Delta t \\
        D^{-}_{\{B\}} &=& N_{\{B\}}d_{BA}\Delta t \\
        D^{-}_{\{A,B\}} &=& 0.
    \end{eqnarray*}
    \item $\bm{B_{i}^{-}}$. To lose a species with range state $i$ through a between-region speciation event, the species must be widespread and undergo a speciation event that gives rise to a new species in state $j$ with smaller range state size ($|j| < |i|$). The factor of 1/2 corrects for double-counting the new species with range $j$ being either the left daughter or right daughter lineage. The total probability of this event occurring within $\Delta t$ is, 
    \begin{eqnarray*}
        \sum_{\substack{
                          j \in S \\
                          |j| < |i|
                        }} \frac{1}{2}N_i b^j_{i \backslash j}\Delta t.
    \end{eqnarray*}
     As an example, in a $2-$region GeoSSE system with state space~$S = \{\{A\},\{B\},\{A,B\}\}$ we have,
     \begin{eqnarray*}
         B^{-}_{\{A\}} &=& 0 \\
         B^{-}_{\{B\}} &=& 0 \\
          B^{-}_{\{A,B\}} &=& \frac{1}{2}N_{\{A,B\}}\left(b^{A}_{B} + b^{B}_{A}\right)\Delta t.
     \end{eqnarray*}
    \item $\bm{E_{i}^{-}}$. To lose a species with range state $i$ through a local extinction event, a species must undergo an extinction event in one of its regions. If the species is endemic, this event leads to total extinction of the species. The total probability of this event occurring within $\Delta t$ is, 
    \begin{eqnarray*}
        \sum_{\substack{
                          \ell \in R \\
                          \ell \in i
                        }} N_i e_\ell \Delta t.
    \end{eqnarray*}
    As an example, in a $2-$region GeoSSE system with state space~$S = \{\{A\},\{B\},\{A,B\}\}$ we have,
    \begin{eqnarray*}
        E^{-}_{\{A\}} &=& N_{\{A\}}e_{A}\Delta t \\
        E^{-}_{\{B\}} &=& N_{\{B\}}e_{B}\Delta t \\
        E^{-}_{\{A,B\}} &=& N_{\{A,B\}}\left(e_{A} + e_{B}\right)\Delta t.
    \end{eqnarray*}
\end{enumerate}
 
The next section uses Eqs.~\ref{eq::piup}-\ref{eq::pistay} to define the stochastic process $\{\Pi_{i}(t): t>0\}$ that models the frequency of species in range state $i$ at time t. The infinitesimal mean $\mu_{\Pi_i}$ and variance $\sigma^{2}_{\Pi_i}$ follow directly from Lemma~\ref{Lemma::geossepipars}.
%


\subsection{Comparison on diffusion-based and tree-based models using simulation}
\label{sec::Pro1}

In this section we show that our diffusion-based approach correctly models the temporal behaviour of range state frequencies in a GeoSSE model. To validate, we compare our results with a tree-based approach that explicitly simulates phylogenetic trees under the same GeoSSE parameter values using the MASTER package~\citep{vaughan2013stochastic} implemented in BEAST2~\citep{bouckaert2014beast}. When simulating given a large number of species initially, $N(0)>> 0$, both diffusion-based and MASTER-based simulations are conditioned only for the process to run until a specific elapsed time $T$. Later in Appendix~\ref{subsec::singlespec}, when we simulate using both approaches starting with a single species in random state, $N(0) = 1$, we condition the process under both elapsed time and survival until the present. Details for setting up reaction equations for the MASTER simulation can be found in Appendix~\ref{subsec::master}. \\

For simulations under a diffusion, we generate sample paths on $[0,T]$, where $T$ is the simulation running time. Each simulation yields a time-series of state frequencies for the provided SSE rate values. Simulations were generated as follows:

\begin{enumerate}
    \item Given the following It\^{o} stochastic differential equation (SDE) and the initial number of species in each range state, $N_{i}(0), \forall i \in S$,
    \begin{eqnarray}
        d N_i = \mu_{i}(t)dt + \sigma_{i}(t)dW_t,
    \end{eqnarray}
    where $dW_t$ is a Wiener process, we draw a sample path by using the following approximation,
    \begin{eqnarray}
    \label{eq::ori_diff}
        N_i(t+\Delta t) = N_i(t) + \mu_{i}(t)\Delta t + \sigma_i(t)\sqrt{\Delta t}U_{t},
    \end{eqnarray}
    where $\sqrt{\Delta t}U_{t}\sim \sqrt{\Delta t}N(0,1) $ is a (discretized) standard Wiener process, and $\mu_{i}(t)$ and $\sigma_{i}(t)$ are computed using Eqs~\eqref{eq::mu_geosse}-\eqref{eq::var_geosse}, respectively.\\
    \item Given $N_{i}(t+\Delta t)$ for each $i\in S$ from step~$1$, we compute the total number of species at $t + \Delta t \in [0,T]$
    \begin{eqnarray*}
        N(t + \Delta t) = \sum_{i \in S} N_i(t + \Delta t).
    \end{eqnarray*}
    \item Next, using $N_i(t)$ and $N(t)$ from steps~1-2, we compute the infinitesimal mean, $\mu_{\Pi_{i}}(t)$, and infinitesimal variance, $\sigma_{\Pi_i}(t)$ using Eqs.~\eqref{eq::mu_geossefreq}-\eqref{eq::var_geossefreq}, respectively. Given $\mu_{\Pi_{i}}(t)$, $\sigma_{\Pi_i}(t)$, and the following It\^{o} SDE with the initial frequency of species of range state $i$, $\Pi_i(0) = \frac{N_i(0)}{N(0)}$, 
    \begin{eqnarray}
        d\Pi_i = \mu_{\Pi_i}(t)dt + \sigma_{\Pi_i}(t)dW_t,
        \label{SDE_pi}
    \end{eqnarray}
    where $dW_t$ is a Wiener process, we draw a sample path by using the following approximation,
    \begin{eqnarray}
    \label{eq::diff_eq}
        \Pi_{i}(t+\Delta t) = \Pi_{i}(t) + \mu_{\Pi_i}(t)\Delta t + \sigma_{\Pi_i}(t)\sqrt{\Delta t}U_{t},
    \end{eqnarray}
    where $\sqrt{\Delta t}U_{t} \sim \sqrt{\Delta t}N(0,1)$ is a (discretized) standard Wiener process.
\end{enumerate}
In Section~\ref{subsec::resultdiffusion}, we show that the dynamic of the range state frequencies can be well-approximated using the diffusion-based framework. We provide different examples through numerical simulations under a variety of GeoSSE scenarios to visualize this result. Specifically, we apply the following procedure, 
\begin{enumerate}
    \item We consider a 3-region GeoSSE model, then we simulate range state dynamics using tree-based approach (via the MASTER package in BEAST2) and the diffusion-based approach over 1000 replicates on $[0,10]$ time interval with 1000 time steps. Note that if one simulates over a longer time interval, then one needs to choose larger time steps to reduce the chance that multiple events occur within $\Delta t$ for the diffusion-based approach. For diffusion-based approach, at each time step, we assign a zero value to any state with a count less than zero since the number of species in any range states cannot be negative. This is reasonable because if $N_{i}(t) = 0$, then some events are not permitted such as a local extinction. Note that although some $N_i$'s might be equal to $0$, it is very unlikely for the whole clade to become extinct, i.e., $N(t)=0$, given a relatively large clade size at the beginning of each process (Fig.~\ref{sim_setup}) and value of each parameter we pick for the simulations (Figs.~\ref{fig:geossewrbet}-\ref{fig:geosseall}). We consider the following scenarios for the GeoSSE model,
        \begin{Example}
            \label{ex::geossewrbe}
            GeoSSE model with only within-region speciation and between-region speciation events (Fig.~\ref{fig:geossewrbet}).
        \end{Example}
        \begin{Example}
            \label{ex::geossewrdi}
             GeoSSE model with only within-region speciation and range dispersal events (Fig.~\ref{fig:geossewrdi}).
        \end{Example}
        \begin{Example}
            \label{ex:geossewrext}
            GeoSSE model with only within-region speciation and local extinction events (Fig.~\ref{fig:geossewrext}).
        \end{Example}
        \begin{Example}
            \label{ex:geosseall}
            GeoSSE model with all the events included (Fig.~\ref{fig:geosseall}).
        \end{Example}
        \begin{figure}[h!]
    	\centering
            \includegraphics[scale=.3]{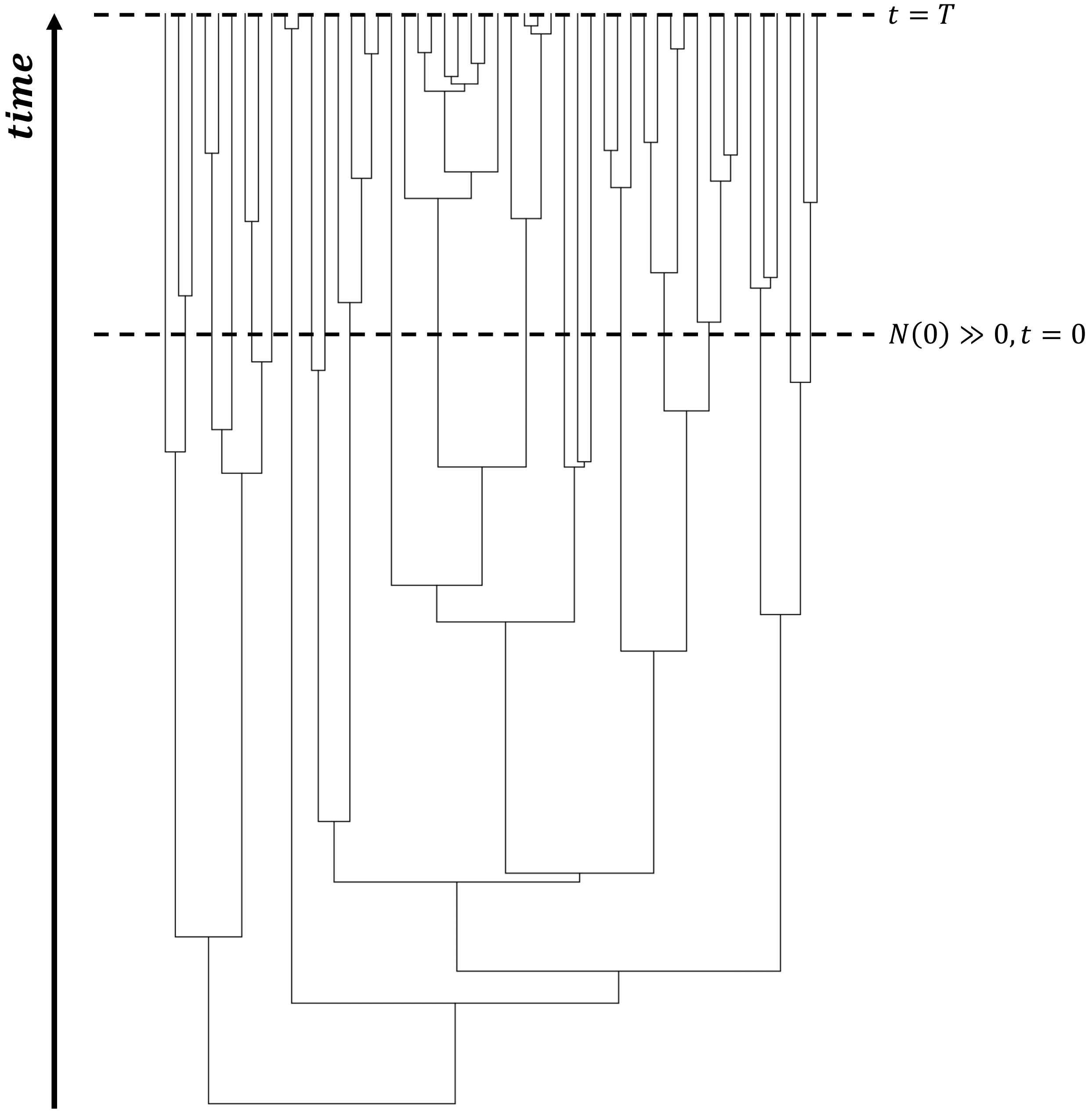}
    	\nobreak
    	\caption{For each diffusion-based and MASTER-based simulation, we assume that we start each $N_i(t)$ simulation, given a relatively large clade size at the beginning, $N(0) >> 0$}
    	\label{sim_setup}
        \end{figure}
    \item We visualize the trajectory of mean state counts for each range state from both diffusion and tree-based approaches. For each simulation, we start the forward-in-time simulation given relatively large clade size for diffusion-based approach to accurately predict the dynamics given by tree-based approach from MASTER simulations. We also visualize stacked bar charts of expected state frequencies for both approaches. To compute the state frequencies under the tree-based approach across replicates, we use the following analytical formula
    \begin{eqnarray*}
        \Pi_{i}(t) = \frac{N_{i}(t)}{\sum_{i\in S}N_{i}(t)}.
    \end{eqnarray*}
    We simulate frequency trajectories under the diffusion-based approach using Eq.~\eqref{eq::diff_eq}. Also for diffusion-based approach, we normalize $\Pi_{i}(t)$ at each time step for each $i \in S$. Thus, keeping  $\Pi_i(t) \leq 1$ at any time.\\
    \item We find the $95\%$ confidence intervals of expected state counts at the end time for both diffusion and tree-based simulations for each GeoSSE scenario described above. Then, we apply the Welch's unequal variances t-test~\citep{welch1947generalization} for testing the following hypothesis
    \begin{eqnarray*}
        H_{0} &:& \Bar{\mu}_{N_i,\text{tree}} = \Bar{\mu}_{N_i,\text{diffusion}}\\
        H_{1} &:& \Bar{\mu}_{N_i,\text{tree}} \neq \Bar{\mu}_{N_i,\text{diffusion}},
    \end{eqnarray*}
    where $\Bar{\mu}_{N_i,\text{tree}}$ and $\Bar{\mu}_{N_i,\text{diffusion}}$ are population means of state counts for range $i$ at the end time from tree and diffusion-based approaches, respectively.\\
    \item We also conduct the $F$ test for testing the following hypothesis
    \begin{eqnarray*}
        H_{0} &:& \Bar{\sigma}^{2}_{N_i,\text{tree}} = \Bar{\sigma}^{2}_{N_i,\text{diffusion}}\\
        H_{1} &:& \Bar{\sigma}^{2}_{N_i,\text{tree}} \neq \Bar{\sigma}^{2}_{N_i,\text{diffusion}},
    \end{eqnarray*}
    where $\Bar{\sigma}^{2}_{N_i,\text{tree}}$ and $\Bar{\sigma}^{2}_{N_i,\text{diffusion}}$ are population variances of state counts for range $i$ at the end time from tree and diffusion-based approaches, respectively.\\
    \item We compute ratio of two sample variances for range state $i$ as 
    \begin{eqnarray*}
        r_{i,\text{var}} = \frac{s^{2}_{i,\text{diffusion}}}{s^{2}_{i,\text{tree}}},
    \end{eqnarray*}
    where $s^{2}_{i,\text{diffusion}}$ and $s^{2}_{i,\text{tree}}$ are sample variances from diffusion- and tree-based simulations for range state $i$, respectively. Then, we construct the $95\%$ confidence interval for $r_{i,\text{var}}$.\\

\noindent If the diffusion-based and tree-based simulation methods are statistically indistinguishable, we should fail to reject all null hypotheses and that the confidence intervals of the ratios of variances include the value 1 at the appropriate significance levels.\\

While all the diffusion-based simulations presented in the main text assume that we always start with a relatively large clade size, this is not how phylogenetic trees are normally simulated. Instead, most simulations generate the entire clade, beginning with one stem or two sister lineages to represent the origin of the process. However, the diffusion approximation assumes the number of species is large. Therefore, to adapt our diffusion-based model for clade-generation scenarios where the initial number of species is small, we adapted our diffusion-based simulation method to start the process with a single species in a random state (see Appendix~\ref{subsec::singlespec}). We show that the difference between diffusion-based and tree-based simulations is reduced after applying the correction.

\end{enumerate}
   
\subsection{Deriving rate parameters that lead to stationary state frequencies when $N$ is large}
\label{subsec::derivestationary}

In this section, we derive conditions for the rate parameters such that there is no change in state frequency, $\Pi_i$, over time for a given a range state $i \in S$, assuming large $N$. That is, we derive the conditions when $\frac{d\Pi_i}{dt}=0, \forall i \in S$.\\

Knowing that $\Pi_i = \frac{N_i}{N}$, we re-write Eqs.~\eqref{eq::piup}-\eqref{eq::pidown} as follows, 
\begin{eqnarray}
         \mathbb{P}_{i}^{+} 
                        &=& N\left[\underbrace{\sum_{\mathclap{\substack{
                          j \in S}}}
                        \sum_{\mathclap{\substack{
                          \ell \in j \\
                          \{ \ell \} = i
                        }}}\Pi_j w_\ell}_{\hat{W}_{i}^{+}}
                        + 
                        \underbrace{\sum_{\mathclap{\substack{
                          k \in i}}}
                        \sum_{\mathclap{\substack{
                          \ell \in i \\
                          \ell \neq k}}}
                        \Pi_{i \backslash \{ \ell \}} d_{k\ell}}_{\hat{D}_{i}^{+}} 
                        + \underbrace{\sum_{\mathclap{\substack{
                          j \in S \\
                          i \subset j
                        }}}  \Pi_j b^{i}_{j \backslash i}}_{\hat{B}_{i}^{+}} 
                        + 
                        \underbrace{\sum_{\substack{
                          j \in S \\
                          |j \backslash i| = 1
                        }}
                        \sum_{\substack{
                          \ell \in j \backslash i
                        }} \Pi_j e_\ell}_{\hat{E}_{i}^{+}}\right]\nonumber\\
                        &=& N\hat{\mathbb{P}}_{i}^{+}   \\
        \mathbb{P}_{i}^{-}  &=& N\left[\underbrace{0 \vphantom{\sum_{\substack{\\
                          k \in i \\
                          |i| < R}} 
                        \sum_{\ell \in R \backslash \{k\}}
                        \Pi_i d_{k\ell}}}_{\hat{W}_{i}^{-}} 
        + 
        \underbrace{\sum_{\substack{\\
                          k \in i \\
                          |i| < R}} 
                        \sum_{\ell \in R \backslash \{k\}}
                        \Pi_i d_{k\ell}}_{\hat{D}_{i}^{-}}   
        +
      \underbrace{\sum_{\substack{
                          j \in S \\
                          j \subset i
                        }}\frac{1}{2} \Pi_i b^j_{i \backslash j}}_{\hat{B}_{i}^{-}} 
        + 
        \underbrace{\sum_{\substack{
                          \ell \in R \\
                          \ell \in i
                        }} \Pi_i e_\ell}_{\hat{E}_{i}^{-}} \right] \nonumber\\
                        &=& N\hat{\mathbb{P}}_{i}^{-}
\end{eqnarray}
Then, Eqs.~\eqref{eq::mu_geosse}-\eqref{eq::var_geosse} can be re-written as follows
    \begin{eqnarray}
        \mu_{i}         &=& N\left(\hat{\mathbb{P}}_{i}^{+} - \hat{\mathbb{P}}_{i}^{-} \right),\label{eq::mu_mod_geosse}\\
    \sigma_{i}^2    &=& N\left(\hat{\mathbb{P}}_{i}^{+} + \hat{\mathbb{P}}_{i}^{-} \right).\label{eq::var_mod_geosse}
    \end{eqnarray}
Given Eqs.~\eqref{eq::mu_mod_geosse}-\eqref{eq::var_mod_geosse}, as $N \rightarrow \infty$, Eqs.~\eqref{eq::mu_geossefreq}-\eqref{eq::var_geossefreq} become
\begin{eqnarray}
    \hat{\mu}_{\Pi_i} &=& \lim_{N \rightarrow \infty} \mu_{\Pi_i} \nonumber \\
                      &=& \hat{\mathbb{P}}_{i}^{+} - \hat{\mathbb{P}}_{i}^{-} ,\\
    \hat{\sigma}^{2}_{\Pi_i} &=& \lim_{N \rightarrow \infty} \sigma^{2}_{\Pi_i} \nonumber \\
                             &=& 0.
\end{eqnarray}
Moreover, we no longer have the stochastic component from the SDE given in Eq.~\eqref{SDE_pi}. Instead, we solve the following ordinary differential equation 
\begin{eqnarray}
    d\Pi_{i} &=&  \hat{\mu}_{\Pi_i}dt \nonumber \\
    \frac{d\Pi_{i}}{dt} &=&  \hat{\mu}_{\Pi_i}.
    \label{stationary_infitesize}
\end{eqnarray}

Given stationary frequency of each range state, $\hat{\Pi}_i$, where $\sum_{i}\hat{\Pi}_{i}=1$, the rate parameters must satisfy
\begin{eqnarray}
    \hat{\mu}_{\Pi_i}  =  0 \iff \hat{\mathbb{P}}_{i}^{+} = \hat{\mathbb{P}}_{i}^{-}. \nonumber 
\end{eqnarray}
Furthermore, we assume all rate parameters must be positive, as all modeled events have some non-zero probability of occurring. That is, 
\begin{eqnarray}
    w_{i} > 0, e_{i} > 0, d_{ij} > 0, \forall i,j \in R. \text{ and } b^{s}_{t} > 0, \forall s,t \in S \nonumber
\end{eqnarray}
Next,  we define total rates of all events occurring in each range state $i$, $\Phi_{total,i}$, as follows
\begin{eqnarray}
    \Phi_{total,i} = \left(r_{W^{+}_{i}} + r_{D^{+}_{i}} + r_{B^{+}_{i}} + r_{E^{+}_{i}}\right)-\left(r_{D^{-}_{i}}+r_{B^{-}_{i}}+r_{E^{-}_{i}}\right),\nonumber
\end{eqnarray}
where $r_{W^{+}_{i}},r_{D^{+}_{i}},r_{B^{+}_{i}},r_{E^{+}_{i}},r_{D^{-}_{i}},r_{B^{-}_{i}},r_{E^{-}_{i}}$ consist of sums of rates across all adjacent states that correspond to the events $W^{+}_{i},D^{+}_{i},B^{+}_{i},E^{+}_{i},D^{-}_{i},B^{-}_{i},E^{-}_{i}$, respectively. $ \Phi_{total,i}$ can also be thought as a flux for range state $i$. That is, it is a difference between total incoming rates and outgoing rates. For example, in a two-region GeoSSE model, we can define $\Phi_{total,\{A\}}$ as follows,
\begin{eqnarray*}
\Phi_{total,\{A\}} = \left(2w_{A} + 0 + b^{A}_{B} + e_{B}\right) - (d_{AB} + 0 + e_{A}),
\end{eqnarray*}
where we have $r_{W^{+}_{\{A\}}} = 2w_{A}$ because within-region speciation rate $w_{A}$ is acting on both endemic species with state $\{A\}$ and widespread species with state  $\{A,B\}$.\\
\begin{Lemma}{\text{}}
\label{lemma::system}

    Given a GeoSSE with state space $S$, set of stationary frequencies, $\{\hat{\Pi}_{i}, \forall i \in S\}$, and initial state frequencies $\Pi_{i}(0)$, the rate parameters satisfy the following system of equations
    \begin{eqnarray}
        &&\hat{\mathbb{P}}_{i}^{+} = \hat{\mathbb{P}}_{i}^{-} \nonumber\\
        &&\Phi_{total,i} 
        \begin{cases}
        = \Phi_{total,j}, \: \text{if } \: \hat{\Pi}_{i} = \hat{\Pi}_{j} \\
        > \Phi_{total,j}, \: \text{if } \: \hat{\Pi}_{i} > \hat{\Pi}_{j}\\
        < \Phi_{total,j}, \: \text{if } \: \hat{\Pi}_{i} < \hat{\Pi}_{j}
        \end{cases}\nonumber \\
        && \sum_{i \in S}\Pi_{i}(0) = 1 \nonumber \\
        &&w_{i} > 0, e_{i} > 0, d_{ij} > 0, b^{s}_{t} > 0,\Pi_{i}(0)\geq 0, \: \forall i,j \in R \text{ and } \forall s,t \in S.
        \label{eq::statio_set}
    \end{eqnarray}
\end{Lemma}

In Section~\ref{subsec::resultstationary}, we demonstrate the application of Lemma~\ref{lemma::system} for a 2-region GeoSSE model.


\subsection{Deriving stationary state frequencies given rate parameters in a GeoSSE model}
\label{subsec::derivestationary-inverse}

In this section, we use our framework to find the stationary state frequencies that result from a given set of rate parameters. This result links the configuration of a data-generating process to its expected pattern, which complements results from Section~\ref{subsec::derivestationary} that link expected patterns to data-generating processes. We present the result in Lemma~\ref{lemma::solve} for the case of a 2-region GeoSSE model for simplicity. \\

\begin{Lemma}{\text{}}
\label{lemma::solve}

    Consider a 2-region GeoSSE model with state space $S = \{\{A\},\{B\},\{A,B\}\}$. Given the rate parameters from the model and initial state frequencies, $\Pi_{\{A\}}(0)=\Pi_{A}^{0}$, $\Pi_{\{B\}}(0)=\Pi_{B}^{0}$, $\Pi_{\{A,B\}}(0)=\Pi_{AB}^{0}$, the general solution to Eq.~\eqref{stationary_infitesize} is given by, 
    \begin{eqnarray}
       \bm{\Pi} &=& \begin{bmatrix}
           \Pi_{\{A\}}(t)\\
           \Pi_{\{B\}}(t)
       \end{bmatrix} \nonumber\\
        &=& C_{1}\bm{\nu}_{1}e^{\lambda_{1}t} + C_{2}\bm{\nu}_{2}e^{\lambda_{2}t} + \bm{K},
    \end{eqnarray}
    and $\Pi_{\{A,B\}}(t) = 1-\Pi_{\{A\}}(t)- \Pi_{\{B\}}(t)$, 
 provided that $\Pi_{\{A\}}(t)+\Pi_{\{B\}}(t) \leq 1$.\\
    
   Furthermore, the stationary frequencies are given by
    \begin{eqnarray}
        \hat{\Pi}_{\{A\}} &=& \frac{num_A}{denom_A},\\[10pt]
        \hat{\Pi}_{\{B\}} &=& 1-\left(\frac{e_A + d_{AB} + b^{A}_{B} + e_B}{w_A + b^{A}_{B} + e_B}\right)\left(\frac{num_A}{denom_A}\right) ,\hspace{2em}\\[10pt]
       \hat{\Pi}_{\{A,B\}} &=& 1 - \hat{\Pi}_{\{A\}}-\hat{\Pi}_{\{B\}},
    \end{eqnarray}
    where
    \begin{eqnarray}
        num_A &=& \left(w_A + b^{A}_{B} + e_B\right)\left(e_B + d_{BA} - w_B\right), \nonumber \\[10pt]
        denom_A &=& \left(e_{A} + d_{AB} + b^{A}_{B} + e_B\right)\left(e_B + d_{BA} + b^{A}_{B} + e_A\right)-\left(w_B + b^{A}_{B} + e_A\right)\left(w_A + b^{A}_{B} + e_B\right) \nonumber,\\[10pt]
        R &=& \sqrt{R_{1} + R_{2}},\nonumber\\ [10pt]
        R_{1} &=& 4\left(b^{A}_{B}\right)^2 + 4\left(b^{A}_{B}e_{A} + b^{A}_{B}e_{B} + b^{A}_{B}w_A + b^{A}_{B}w_B\right) + 4\left(e_{A}e_{B} + e_{A}w_{A} + e_{B}w_{B} + w_{A}w_{B} \right), \nonumber \\ [10pt]
        R_{2} &=& -2 d_{AB}d_{BA} + \left(d_{AB}^{2} + d_{BA}^{2}\right), \nonumber \\ [10pt]
        \lambda_{1} &=& \frac{1}{2}\left(-2 b^{A}_{B} - d_{AB} -d_{BA} - 2e_{A} - 2e_{B} - R\right), \nonumber\\ [10pt]
        \lambda_{2} &=& \frac{1}{2}\left(-2 b^{A}_{B} - d_{AB} -d_{BA} - 2e_{A} - 2e_{B} + R\right), \nonumber \\[10pt]
        \bm{\nu}_1 &=& \begin{bmatrix}
        -\frac{1}{2\left(b^{A}_{B} + e_A + w_B\right)}\left(-d_{AB} + d_{BA} -R\right)\\[10pt]
        1
        \end{bmatrix},\nonumber \\[10pt]
        \bm{\nu}_2 &=& \begin{bmatrix}
        -\frac{1}{2\left(b^{A}_{B} + e_A + w_B\right)}\left(-d_{AB} + d_{BA} + R\right)\\[10pt]
        1
    \end{bmatrix}, \nonumber \\[10pt]
        \bm{K} &=& \begin{bmatrix}
            \hat{\Pi}_{\{A\}} \\[10pt]
             \hat{\Pi}_{\{B\}}
        \end{bmatrix}, \nonumber
    \end{eqnarray}
    \begin{eqnarray}
         C_{1} &=& \frac{\left(\Pi_{A}^{0} - K_1\right)\left(b^{A}_{B} + e_{A} + w_{B}\right)}{R} - \frac{\left(\Pi_{B}^{0}-K_{2}\right)\left(d_{AB} - d_{BA} - R\right)}{2 R}, \nonumber\\[10pt]
        C_{2} &=& \frac{\left(K_{1}-\Pi_{A}^{0}\right)\left(b^{A}_{B} + e_{A} + w_{B}\right)}{R} + \left(\Pi_{B}^{0} - K_{2}\right)\left(1 + \frac{d_{AB}-d_{BA}-R}{2 R}\right)\nonumber.
    \end{eqnarray}
\end{Lemma}
{\bf Proof:} Proof of Lemma~\ref{lemma::solve} is given in Appendix~\ref{subsec::prooflemma5}.\\

We note that this strategy can be generalized to accommodate arbitrary models within the ClaSSE family. Specifically, as seen in the proof of Lemma~\ref{lemma::solve} in Appendix~\ref{subsec::prooflemma5}, for a ClaSSE model with $|S|$ states, one only needs to find eigenvalues (either numerically or analytically) and eigenvectors that correspond to a $(|S| - 1) \times (|S| - 1)$ matrix to obtain a general solution. The resulting solution for the stationary frequencies would then reflect the parameterization of the particular ClaSSE model variant being studied. Note that this approach of  solving a matrix with one dimension lower than the state space only holds providing that the sum of the remaining frequencies is less than or equal to 1. This assumption, however, can be ignored if one is to solve the full system by finding eigenvalues and eigenvectors that correspond to a $|S| \times |S|$ matrix, and normalize the resulting stationary frequencies.\\

In Section~\ref{subsec::resultstationary}, we use Lemma~\ref{lemma::solve} using rates obtained from Lemma~\ref{lemma::system} to verify that the system, indeed, converges to the true stationary frequencies that we observe through simulations. 


\subsection{Deriving time to reach stationary state frequencies in a GeoSSE model}
\label{subsec::derivestationary-time}

In this section, we describe a method for deriving time to reach stationary state frequencies in a 2-region GeoSSE model. Note that we have assumed a relatively large clade size at the start of the process for simulating $\Pi_{i}(t)$. Thus, the following is time to stationary frequencies since some relatively large clade size (Fig.~\ref{sim_setup}).\\

From Lemma~\ref{lemma::solve} in Section~\ref{subsec::derivestationary-inverse}, we have derived an analytical expression to compute state frequencies over time, given large $N$. In order to find the time to stationarity for each range state, we define the following procedure, as follows
\begin{enumerate}
    \item Given the initial state frequencies, $\Pi_{A}^{0},\Pi_{B}^{0},\Pi_{AB}^{0}$, and that the system runs from $[0,T]$, we find the mixing time $t^{*}_{i}$ for all $i \in S$ such that, 
    \begin{eqnarray}
        \left|\Pi_{i}\left(t_{i}^{*}\right)-\Pi_{i}\left(t_{i}^{*}-\Delta t\right)\right| < \epsilon,
    \end{eqnarray}
    for some $\Delta t >0$ and $\epsilon > 0$. $t_{i}^{*}$ is the stationary time for the range state $i$, given the $\epsilon$ value.\\
    \item We visually check that $t^{*}_{i}$ derived from the theory reconciles with what we observe from simulations.
\end{enumerate}

We apply this procedure to an example in Section~\ref{subsec::resultstationary}.

\section{Results}\label{sec::results}

\subsection{Diffusion-based approach is a good approximation to tree-based approach for describing state dynamics}\label{subsec::resultdiffusion}

\begin{figure}[htbp]
	\centering
    \begin{subfigure}{.6\textwidth}
        \includegraphics[scale=0.14]{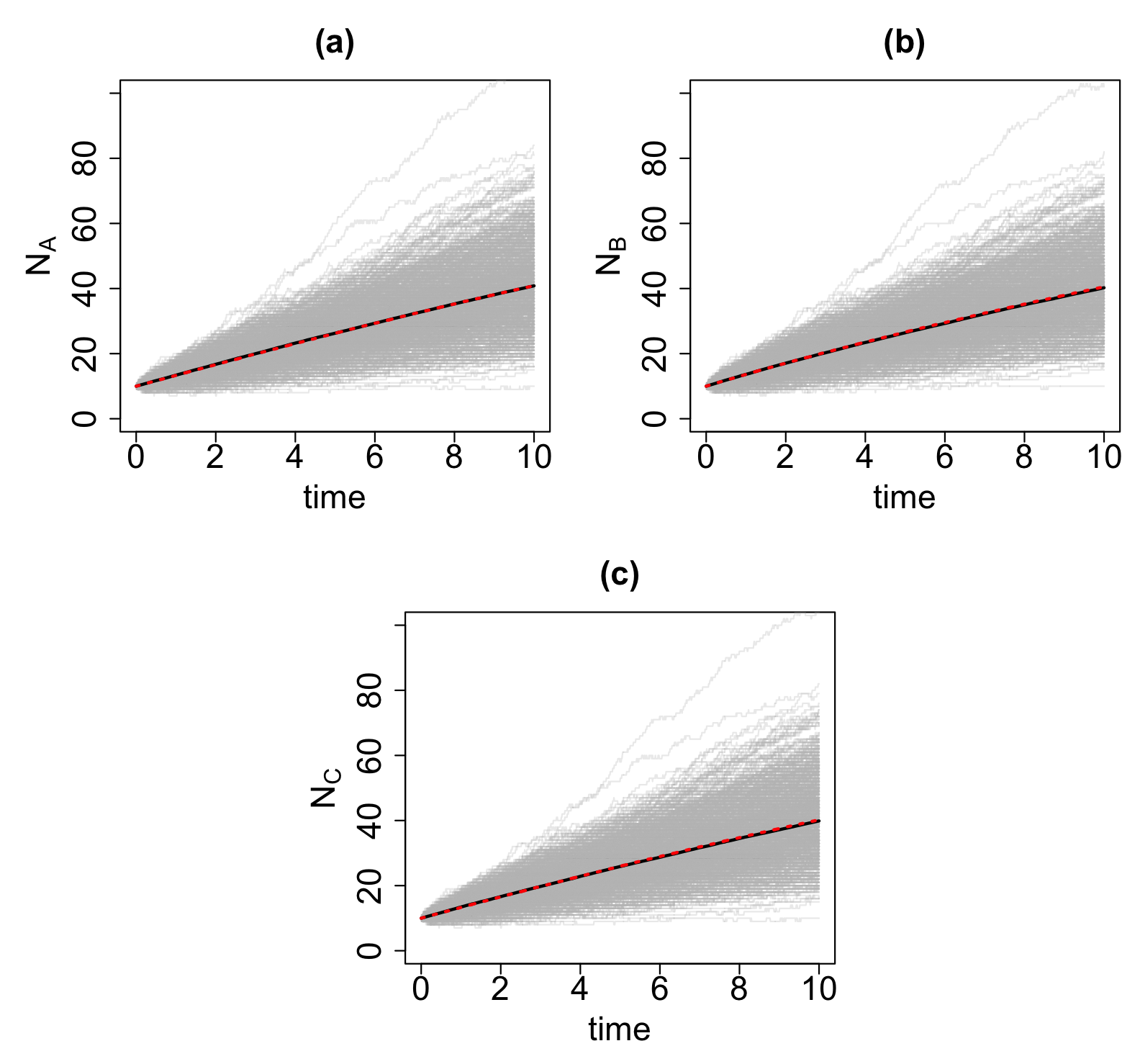}
    \end{subfigure}%
    \begin{subfigure}{.5\textwidth}
        \includegraphics[scale=0.14]{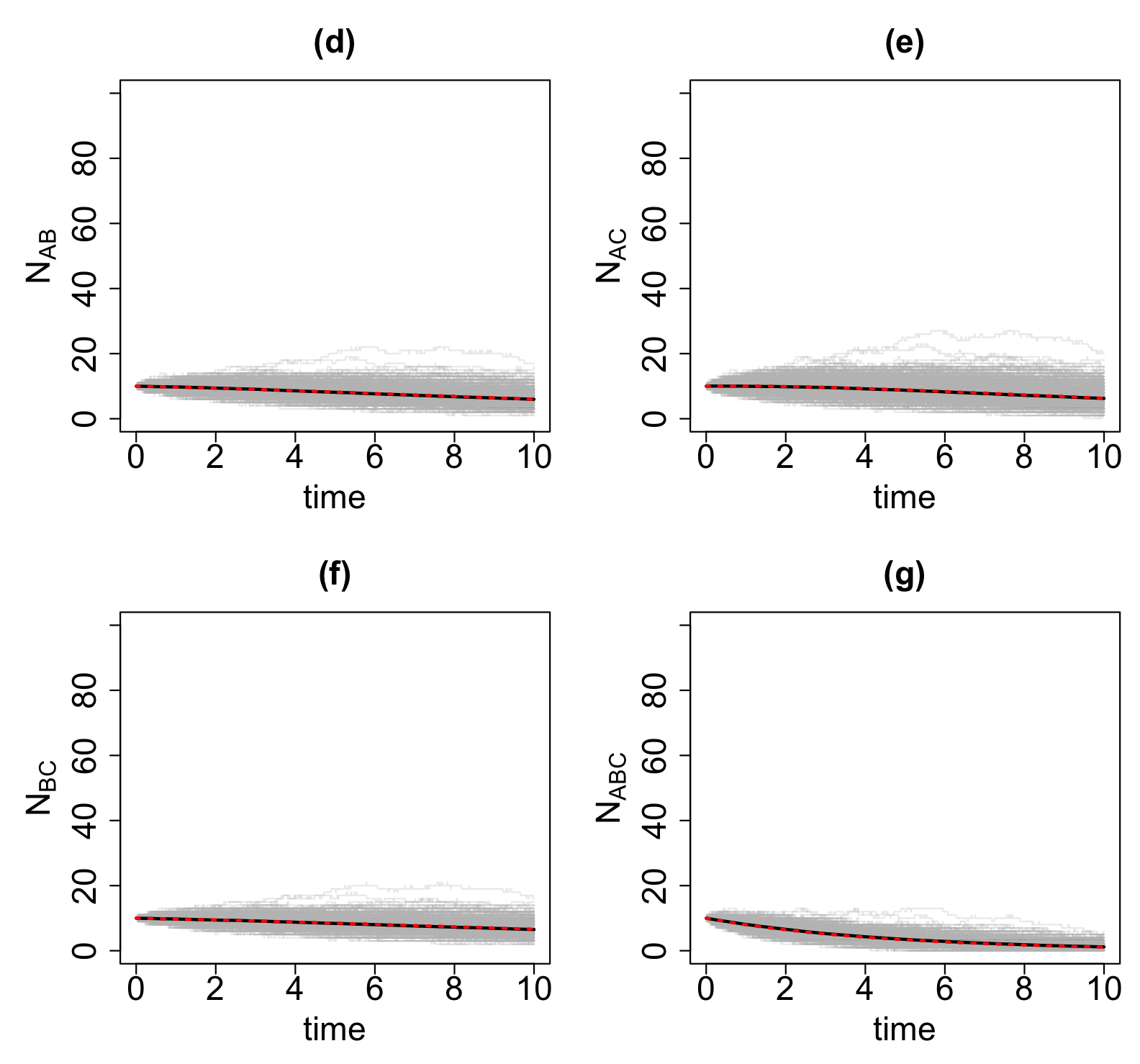}
    \end{subfigure}
    \begin{subfigure}{.6\textwidth}
        \includegraphics[scale=0.35]{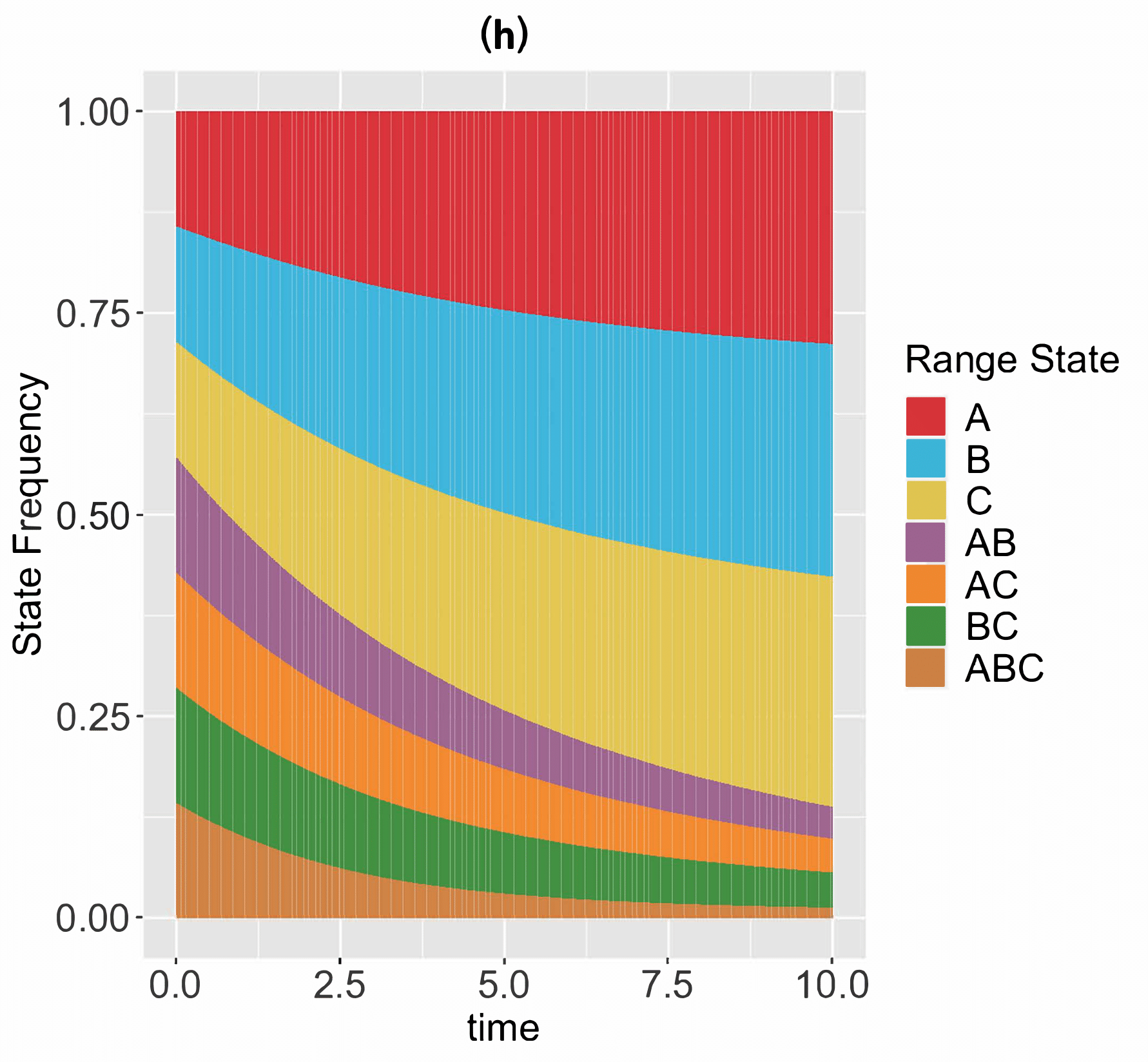}
    \end{subfigure}%
    \begin{subfigure}{.5\textwidth}
        \includegraphics[scale=0.35]{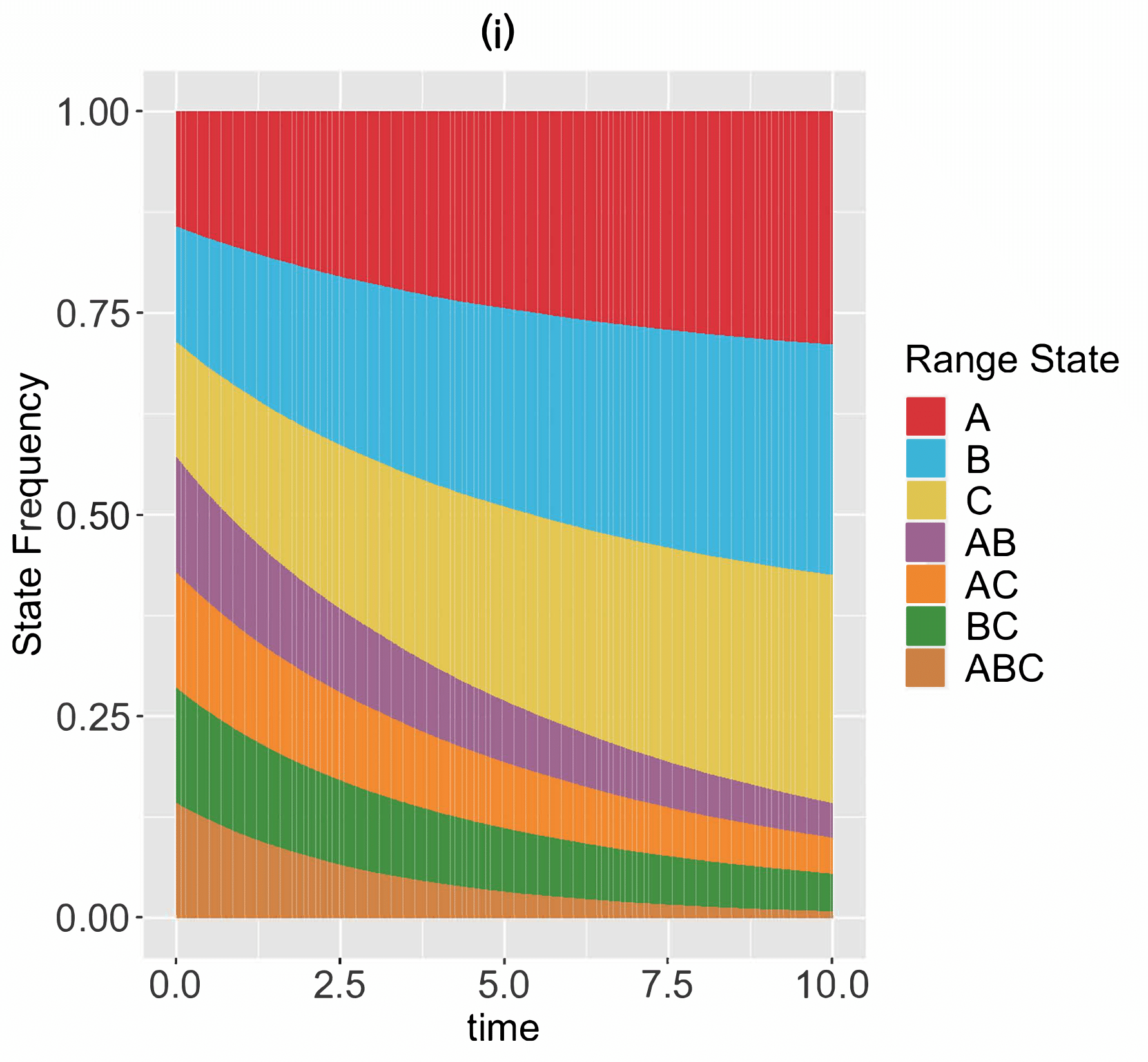}
    \end{subfigure}
	\nobreak
	\caption{Top \& middle panels: the trajectories of average count of range states for endemic species (Figs.~\ref{fig:geossewrbet}(a)-(c)) and widespread species (Figs.~\ref{fig:geossewrbet}(d)-(g)) over $[0,10]$ time interval and over 1000 simulations runs for the three-region GeoSSE model as described in Example~\ref{ex::geossewrbe} each simulated under both diffusion-based process (red line) and tree-based process (black line). The gray trajectories show the dynamics across 1000 replicates simulated under diffusion-based process. Bottom panel: stacked bar chart showing the state frequencies over time using diffusion-based approach (Fig.~\ref{fig:geossewrbet}(h)) and tree-based approach (Fig.~\ref{fig:geossewrbet}(i)). In both approaches, we start the process with $N(0) = 40$ and the following initial state frequencies: $\Pi_{\{A\}}(0) = \Pi_{\{B\}}(0) = \Pi_{\{C\}}(0) = \Pi_{\{A,B\}}(0) = \Pi_{\{A,C\}}(0) = \Pi_{\{B,C\}}(0) = \Pi_{\{A,B,C\}}(0) = \frac{1}{7}$. At $t=10$, the mean frequencies for each range state from both diffusion-based and tree-based simulations are as follows:\\ 
    $\bar{\Pi}^{diffusion}_{\{A\}}= 0.29$, $\bar{\Pi}^{tree}_{\{A\}}= 0.29$; $\bar{\Pi}^{diffusion}_{\{B\}}= 0.29$, $\bar{\Pi}^{tree}_{\{B\}}= 0.29$;\\ 
    $\bar{\Pi}^{diffusion}_{\{C\}}= 0.29$, $\bar{\Pi}^{tree}_{\{C\}}= 0.28$; $\bar{\Pi}^{diffusion}_{\{A,B\}}= 0.04$, $\bar{\Pi}^{tree}_{\{A,B\}}= 0.04$;\\ $\bar{\Pi}^{diffusion}_{\{A,C\}}= 0.04$, $\bar{\Pi}^{tree}_{\{A,C\}}= 0.04$; $\bar{\Pi}^{diffusion}_{\{B,C\}}= 0.04$, $\bar{\Pi}^{tree}_{\{B,C\}}= 0.05$;\\ $\bar{\Pi}^{diffusion}_{\{A,B,C\}}= 0.01$, $\bar{\Pi}^{tree}_{\{A,B,C\}}= 0.01$.\\Simulations are conducted using the following parameter values: $w_{A}=w_{B}=w_{C}=0.03,b^{A}_{B}=0.08,b^{A}_{C}=0.10,b^{B}_{C}=0.06,b^{A}_{BC}=0.04,b^{B}_{AC}=0.12,b^{C}_{AB}=0.06,e_{A}=e_{B}=e_{C}=0,d_{AB}=d_{BA}=d_{AC}=d_{CA}=d_{BC}=d_{CB}=0$}
	\label{fig:geossewrbet}
\end{figure}

\begin{figure}[htbp]
	\centering
    \begin{subfigure}{.6\textwidth}
        \includegraphics[scale=0.14]{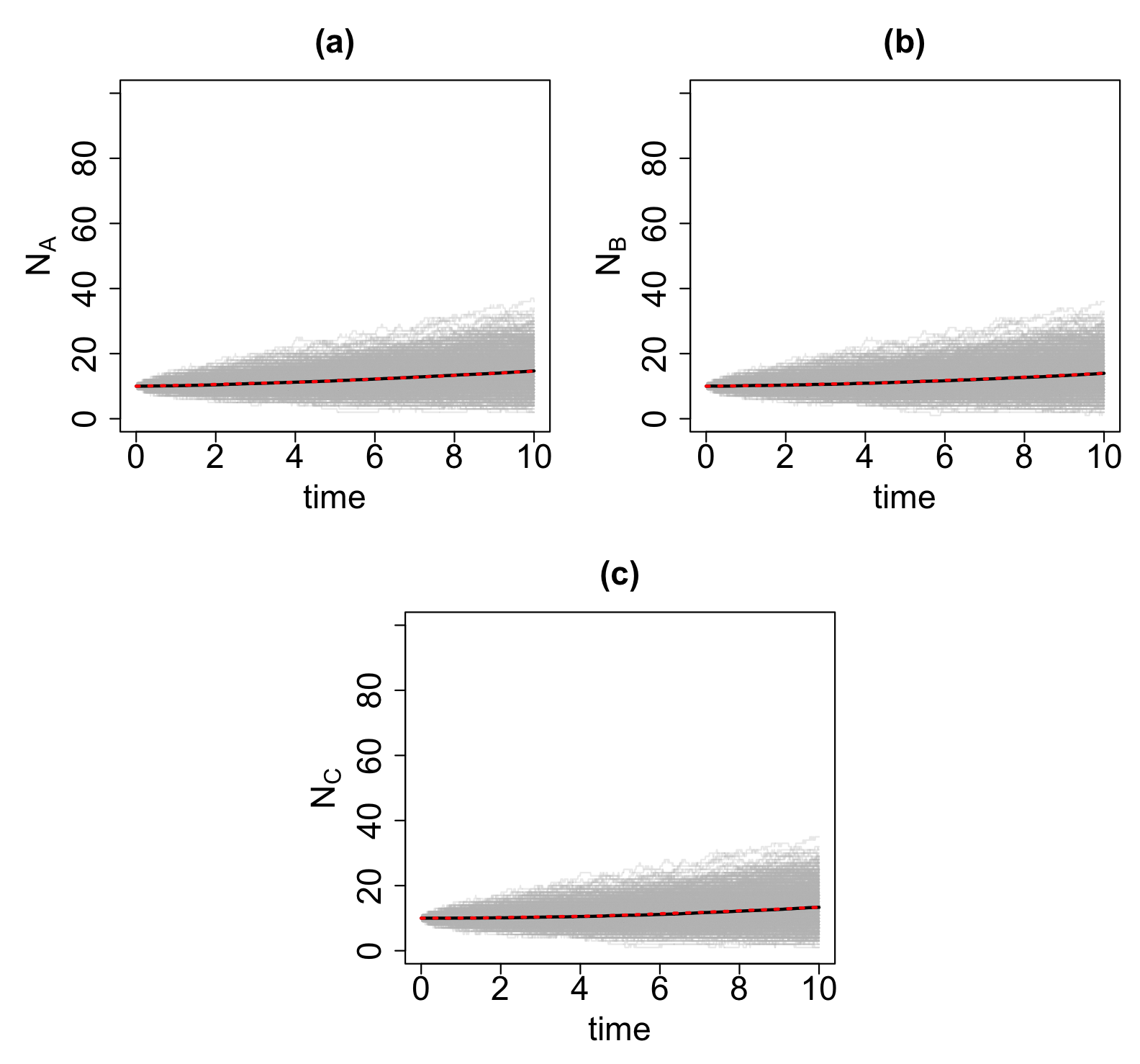}
    \end{subfigure}%
    \begin{subfigure}{.5\textwidth}
        \includegraphics[scale=0.14]{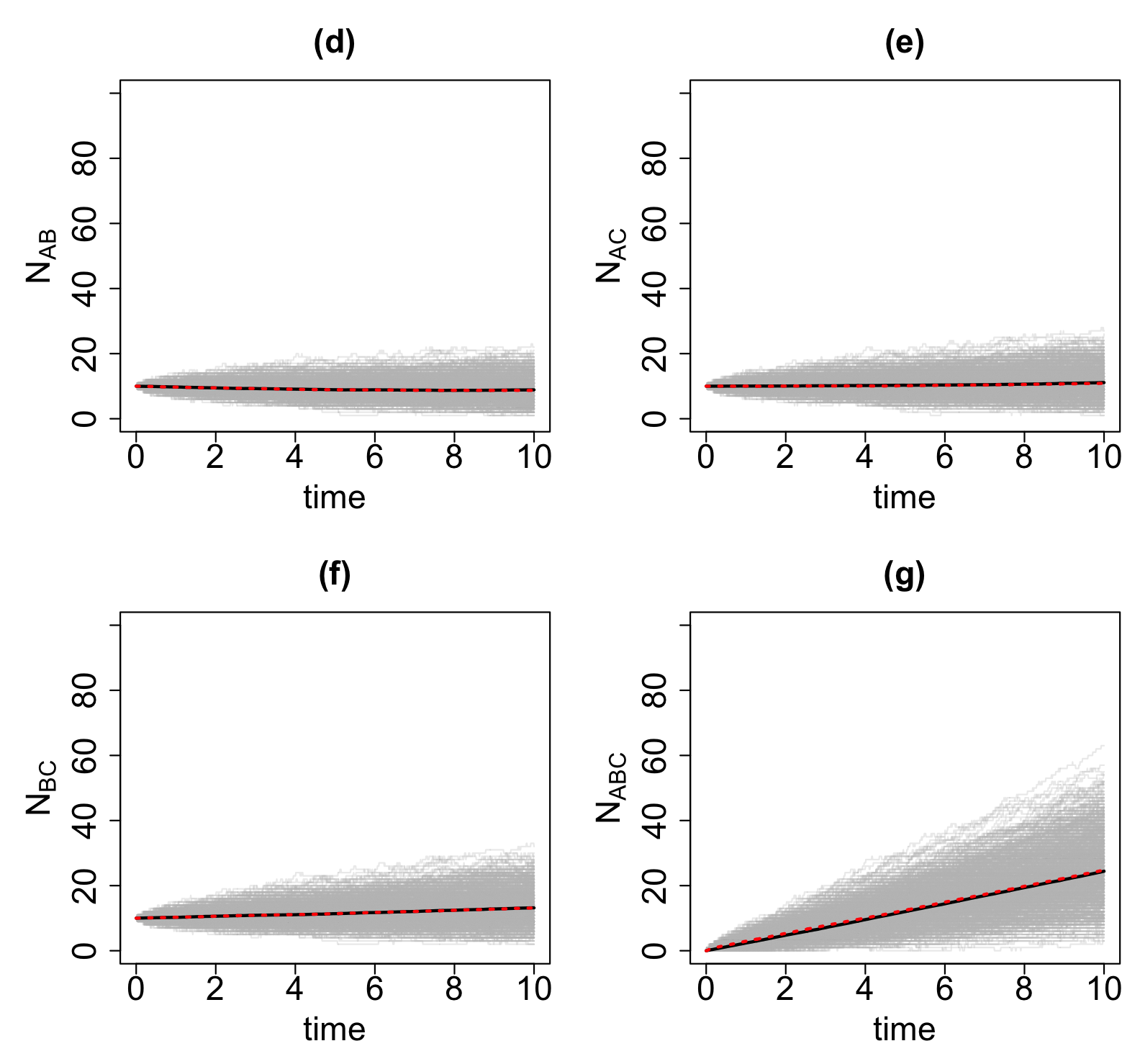}
    \end{subfigure}
    \begin{subfigure}{.6\textwidth}
        \includegraphics[scale=0.35]{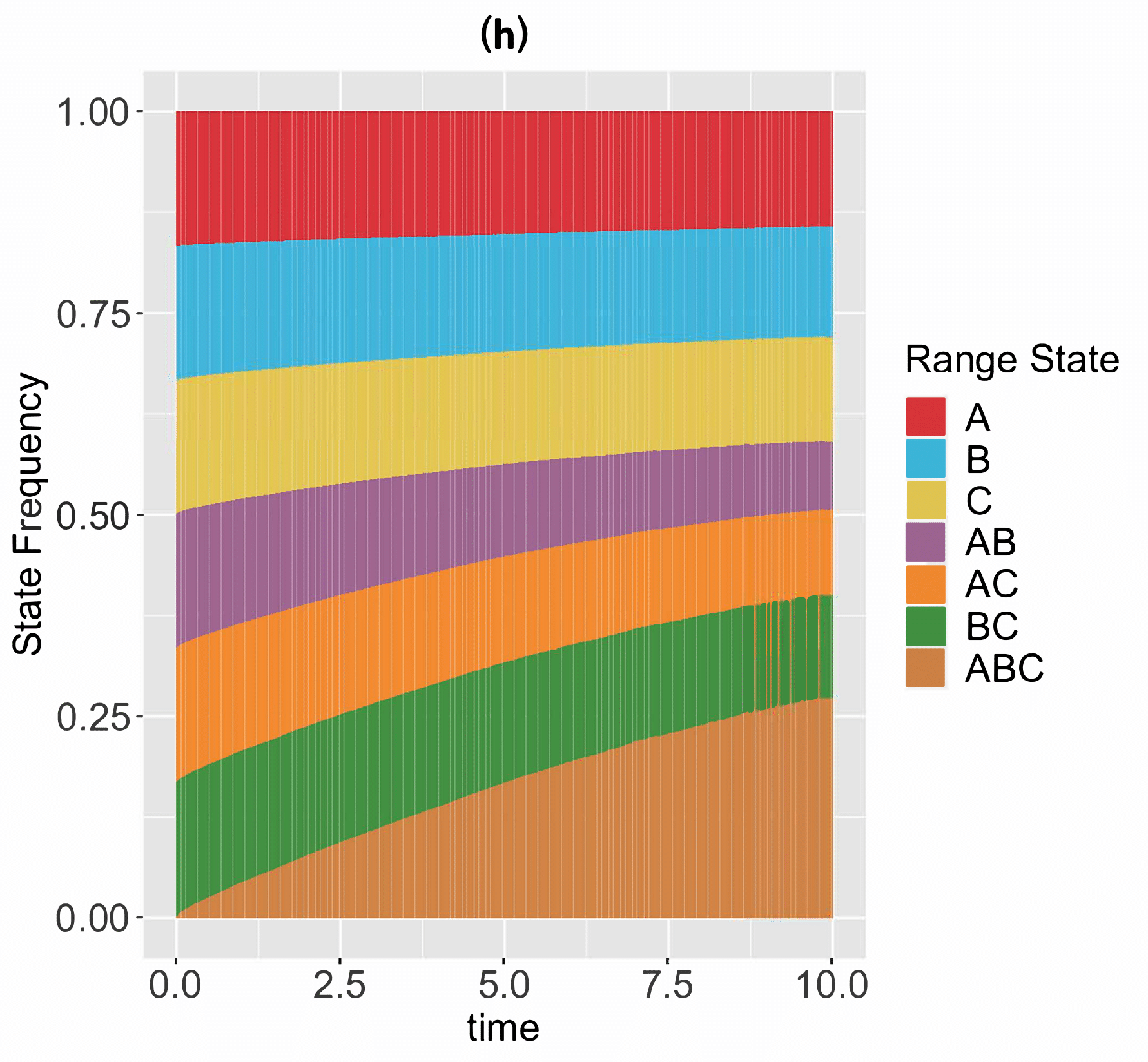}
    \end{subfigure}%
    \begin{subfigure}{.5\textwidth}
        \includegraphics[scale=0.35]{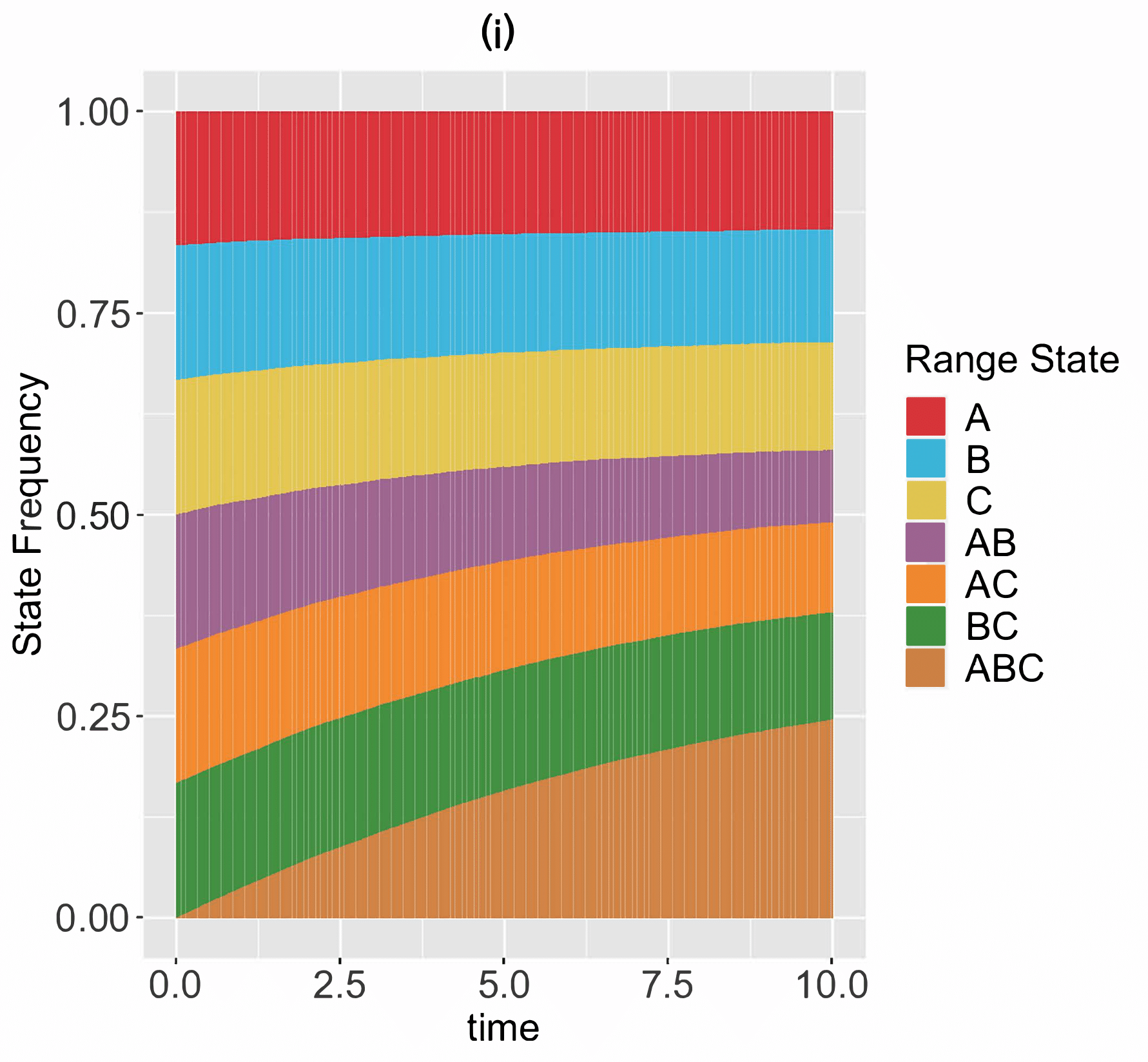}
    \end{subfigure}
	\nobreak
	\caption{Top \& middle panels: the trajectories of average count of range states for endemic species (Figs.~\ref{fig:geossewrdi}(a)-(c)) and widespread species (Figs.~\ref{fig:geossewrdi}(d)-(g)) over $[0,10]$ time interval and over 1000 simulations runs for the three-region GeoSSE model as described in Example~\ref{ex::geossewrdi} simulated under both diffusion-based process (red line) and tree-based process (black line). The gray trajectories show the dynamics across 1000 replicates simulated under diffusion-based process. Bottom panel: stacked bar chart showing the state frequencies over time using diffusion-based approach (Fig.~\ref{fig:geossewrdi}(h)) and tree-based approach (Fig.~\ref{fig:geossewrdi}(i)). In both approaches, we start the process with $N(0) = 40$ and the following initial state frequencies: $\Pi_{\{A\}}(0) = \Pi_{\{B\}}(0) = \Pi_{\{C\}}(0) = \Pi_{\{A,B\}}(0) = \Pi_{\{A,C\}}(0) = \Pi_{\{B,C\}}(0) = \frac{1}{6}$, $\Pi_{\{A,B,C\}}(0) =0$. At $t=10$, the mean frequencies for each range state from both diffusion-based and tree-based simulations are as follows:\\ 
    $\bar{\Pi}^{diffusion}_{\{A\}}= 0.14$, $\bar{\Pi}^{tree}_{\{A\}}= 0.15$; $\bar{\Pi}^{diffusion}_{\{B\}}= 0.14$, $\bar{\Pi}^{tree}_{\{B\}}= 0.14$;\\ 
    $\bar{\Pi}^{diffusion}_{\{C\}}= 0.13$, $\bar{\Pi}^{tree}_{\{C\}}= 0.13$; $\bar{\Pi}^{diffusion}_{\{A,B\}}= 0.08$, $\bar{\Pi}^{tree}_{\{A,B\}}= 0.09$;\\ $\bar{\Pi}^{diffusion}_{\{A,C\}}= 0.11$, $\bar{\Pi}^{tree}_{\{A,C\}}= 0.11$; $\bar{\Pi}^{diffusion}_{\{B,C\}}= 0.13$, $\bar{\Pi}^{tree}_{\{B,C\}}= 0.13$;\\ $\bar{\Pi}^{diffusion}_{\{A,B,C\}}= 0.27$, $\bar{\Pi}^{tree}_{\{A,B,C\}}= 0.25$.\\Simulations are conducted using the following parameter values: $w_{A}=w_{B}=w_{C}=0.03,b^{A}_{B}=b^{A}_{C}=b^{B}_{C}=b^{A}_{BC}=b^{B}_{AC}=b^{C}_{AB}=0,e_{A}=e_{B}=e_{C}=0,d_{AB}=d_{BA}=0.03,d_{AC}=d_{CA}=0.04,d_{BC}=d_{CB}=0.05$}
	\label{fig:geossewrdi}
\end{figure}

\begin{figure}[htbp]
	\centering
    \begin{subfigure}{.6\textwidth}
        \includegraphics[scale=0.14]{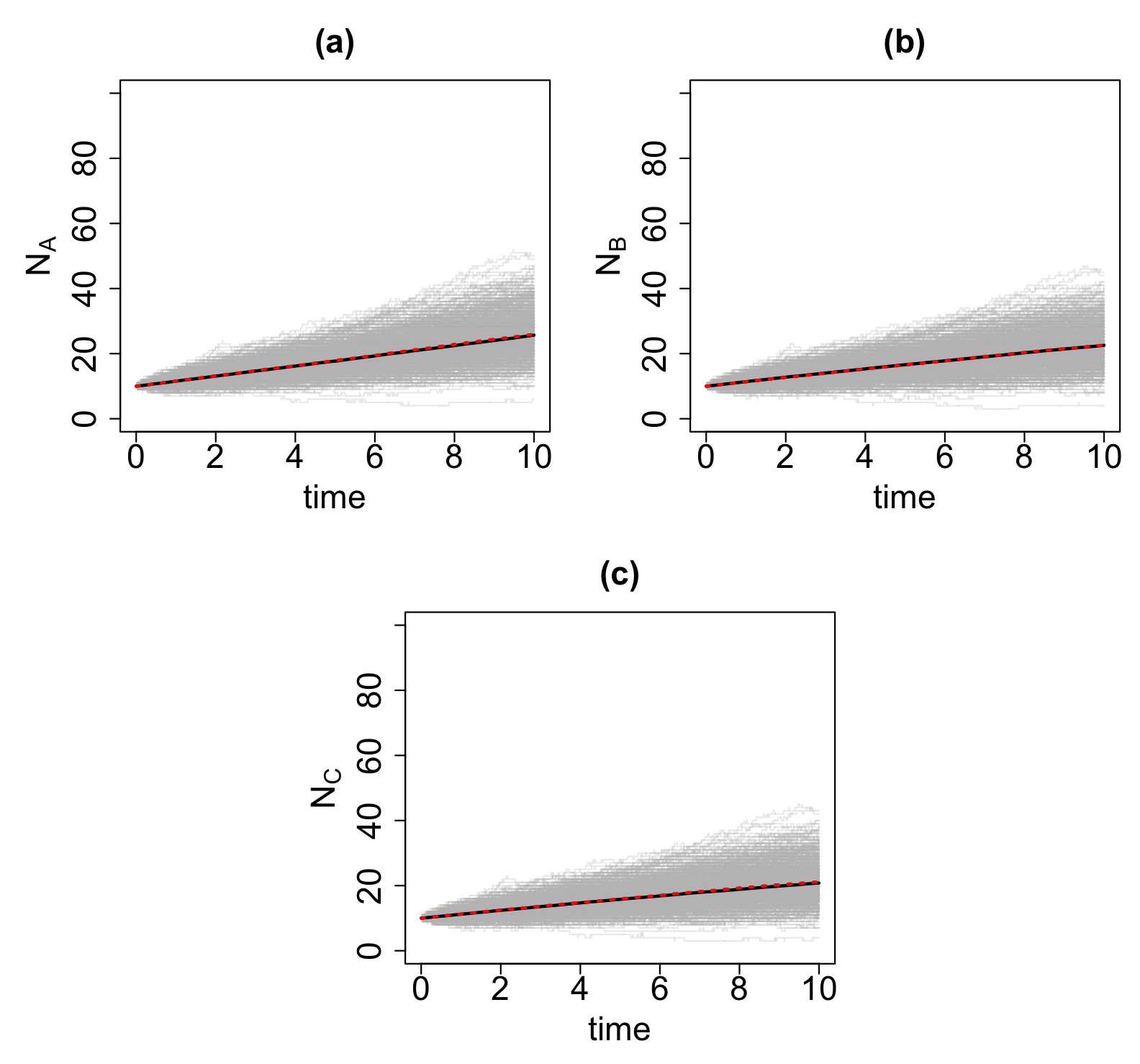}
    \end{subfigure}%
    \begin{subfigure}{.5\textwidth}
        \includegraphics[scale=0.14]{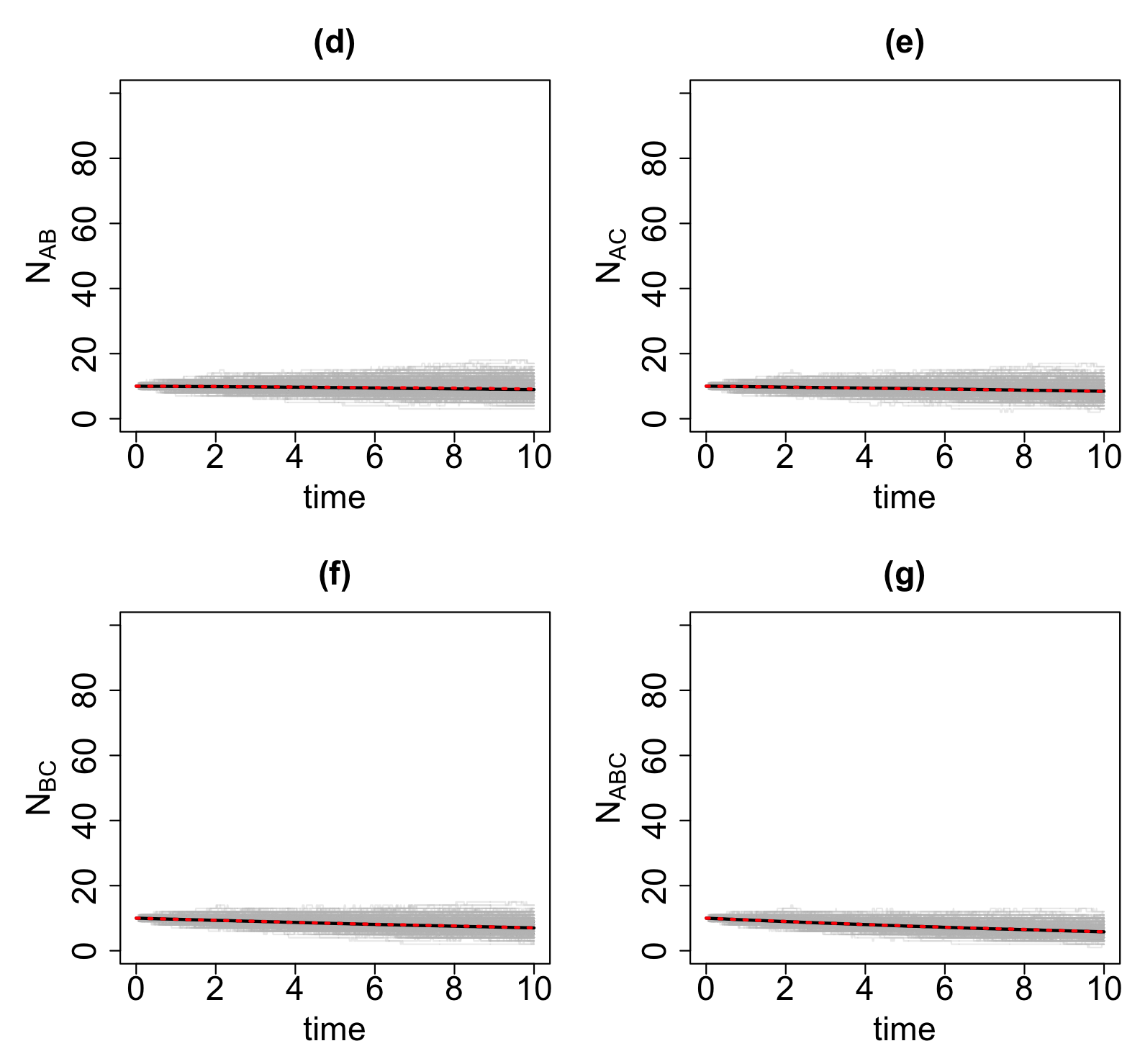}
    \end{subfigure}
    \begin{subfigure}{.6\textwidth}
        \includegraphics[scale=0.35]{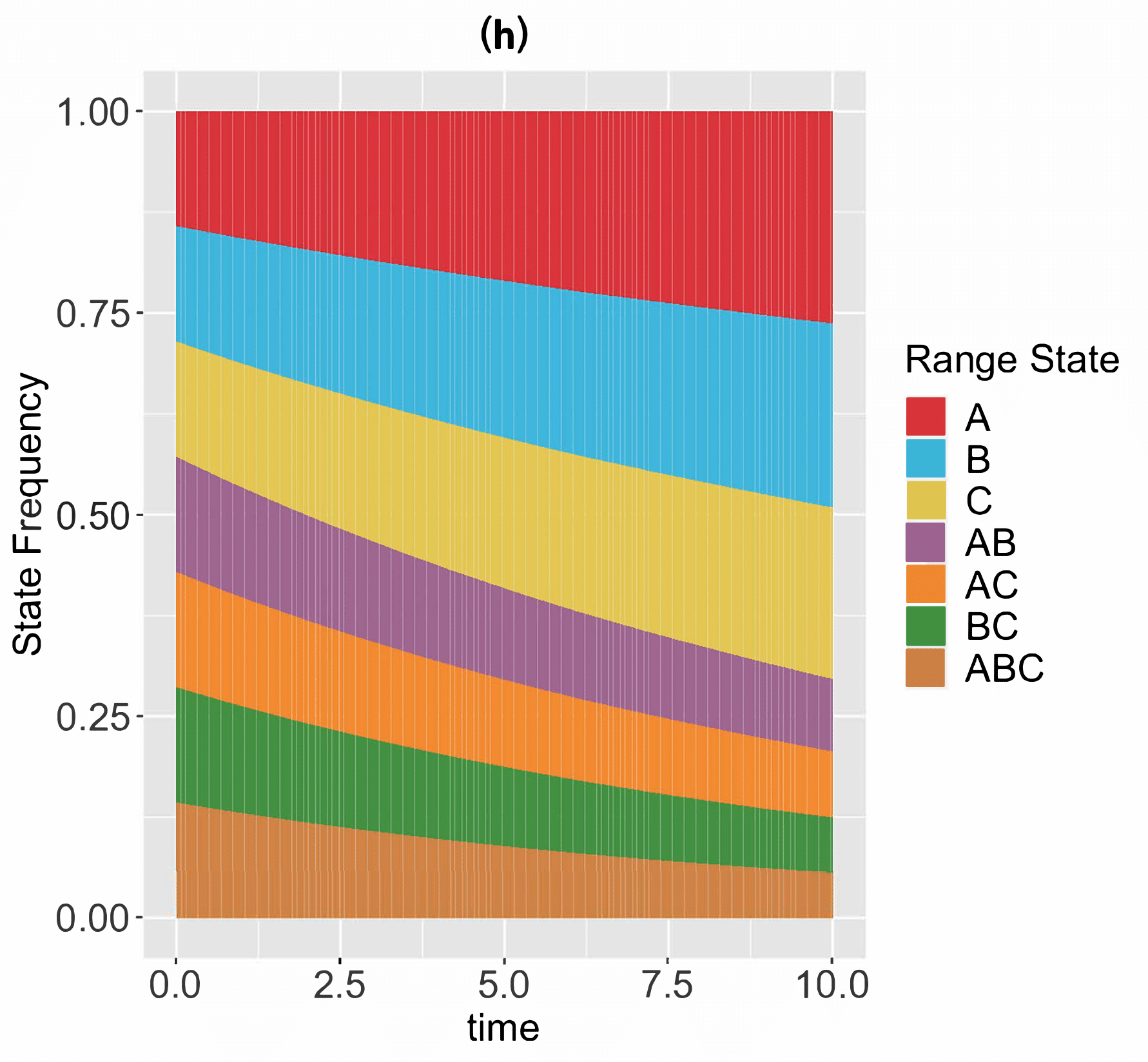}
    \end{subfigure}%
    \begin{subfigure}{.5\textwidth}
        \includegraphics[scale=0.35]{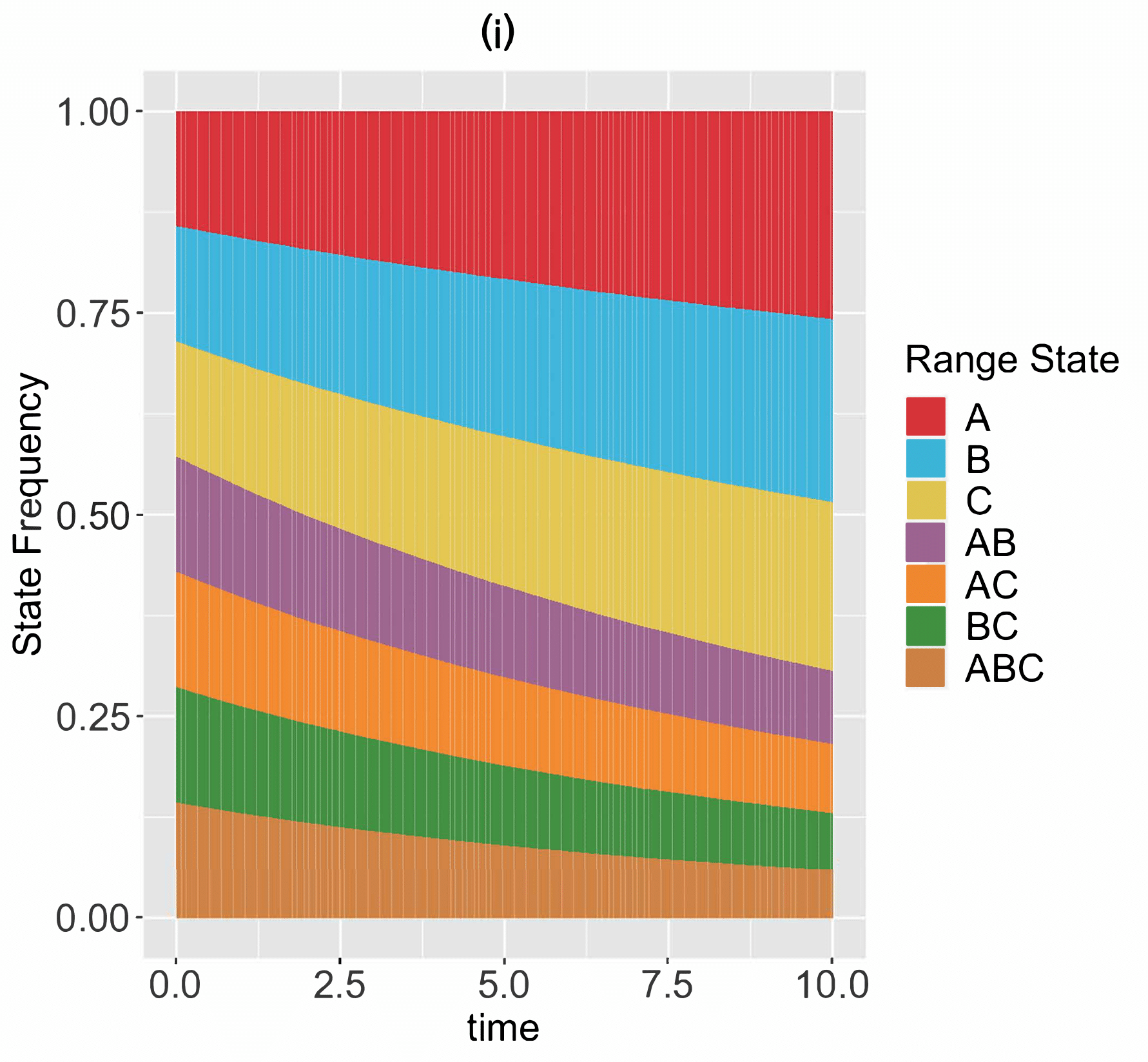}
    \end{subfigure}
	\nobreak
	\caption{Top \& middle panels: the trajectories of average count of range states for endemic species (Figs.~\ref{fig:geossewrext}(a)-(c)) and widespread species (Figs.~\ref{fig:geossewrext}(d)-(g)) over $[0,10]$ time interval and over 1000 simulations runs for the three-region GeoSSE model as described in Example~\ref{ex:geossewrext} simulated under both diffusion-based process (red line) and tree-based process (black line). The gray trajectories show the dynamics across 1000 replicates simulated under diffusion-based process. Bottom panel: stacked bar chart showing the state frequencies over time using diffusion-based approach (Fig.~\ref{fig:geossewrext}(h)) and tree-based approach (Fig.~\ref{fig:geossewrext}(i)). In both approaches, we start the process with $N(0) = 40$ and the following initial state frequencies: $\Pi_{\{A\}}(0) = \Pi_{\{B\}}(0) = \Pi_{\{C\}}(0) = \Pi_{\{A,B\}}(0) = \Pi_{\{A,C\}}(0) = \Pi_{\{B,C\}}(0) = \Pi_{\{A,B,C\}}(0) = \frac{1}{7}$, the mean frequencies for each range state from both diffusion-based and tree-based simulations are as follows:\\ 
    $\bar{\Pi}^{diffusion}_{\{A\}}= 0.26$, $\bar{\Pi}^{tree}_{\{A\}}= 0.26$; $\bar{\Pi}^{diffusion}_{\{B\}}= 0.23$, $\bar{\Pi}^{tree}_{\{B\}}= 0.23$;\\ 
    $\bar{\Pi}^{diffusion}_{\{C\}}= 0.21$, $\bar{\Pi}^{tree}_{\{C\}}= 0.21$; $\bar{\Pi}^{diffusion}_{\{A,B\}}= 0.09$, $\bar{\Pi}^{tree}_{\{A,B\}}= 0.09$;\\ $\bar{\Pi}^{diffusion}_{\{A,C\}}= 0.08$, $\bar{\Pi}^{tree}_{\{A,C\}}= 0.08$; $\bar{\Pi}^{diffusion}_{\{B,C\}}= 0.07$, $\bar{\Pi}^{tree}_{\{B,C\}}= 0.07$;\\ $\bar{\Pi}^{diffusion}_{\{A,B,C\}}= 0.06$, $\bar{\Pi}^{tree}_{\{A,B,C\}}= 0.06$.\\Simulations are conducted using the following parameter values: $w_{A}=w_{B}=w_{C}=0.03,b^{A}_{B}=b^{A}_{C}=b^{B}_{C}=b^{A}_{BC}=b^{B}_{AC}=b^{C}_{AB}=0,e_{A}=0.01,e_{B}=0.02,e_{C}=0.025,d_{AB}=d_{BA}=d_{AC}=d_{CA}=d_{BC}=d_{CB}=0$}
	\label{fig:geossewrext}
\end{figure}

\begin{figure}[htbp]
	\centering
    \begin{subfigure}{.6\textwidth}
        \includegraphics[scale=0.14]{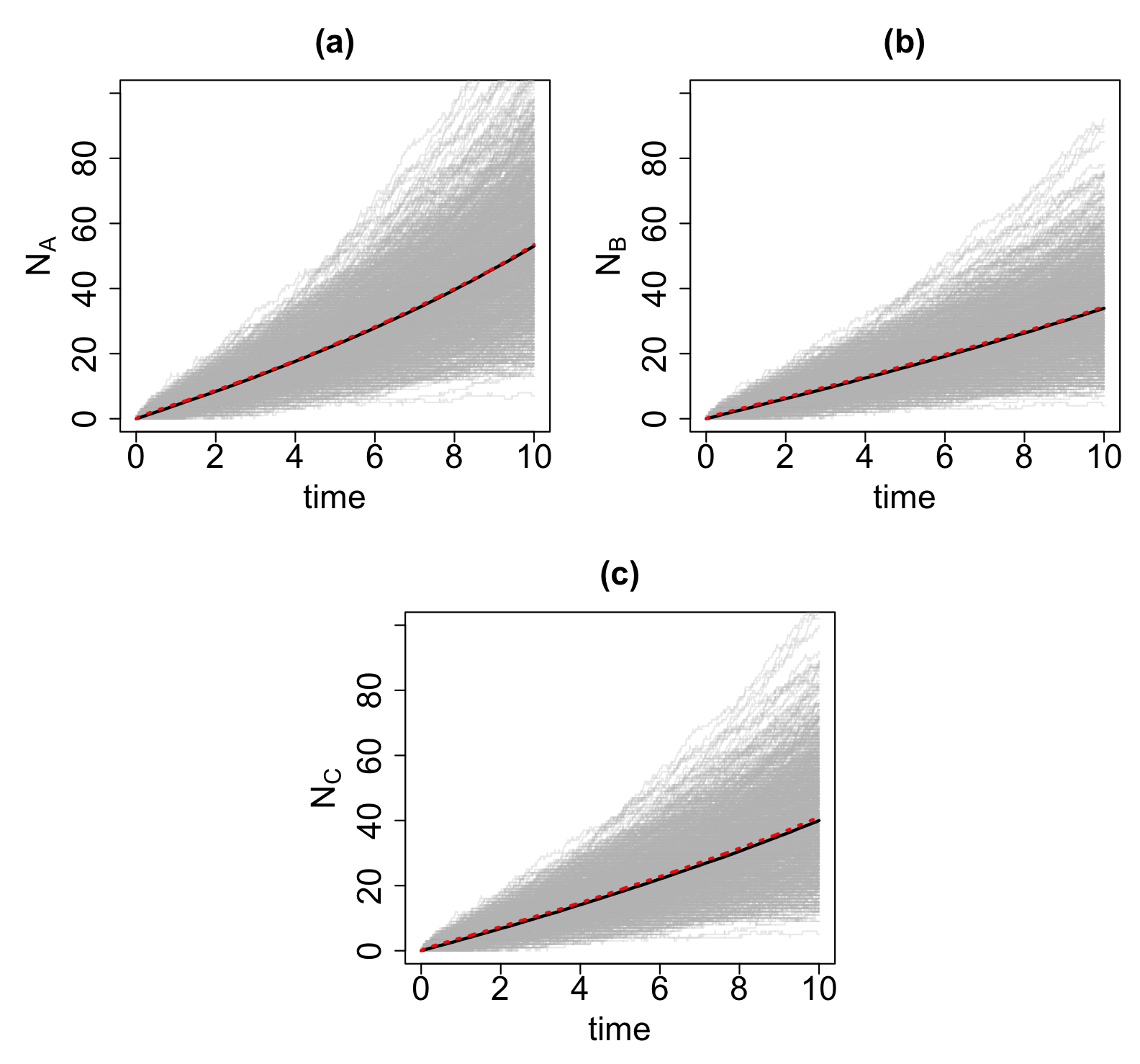}
    \end{subfigure}%
    \begin{subfigure}{.5\textwidth}
        \includegraphics[scale=0.14]{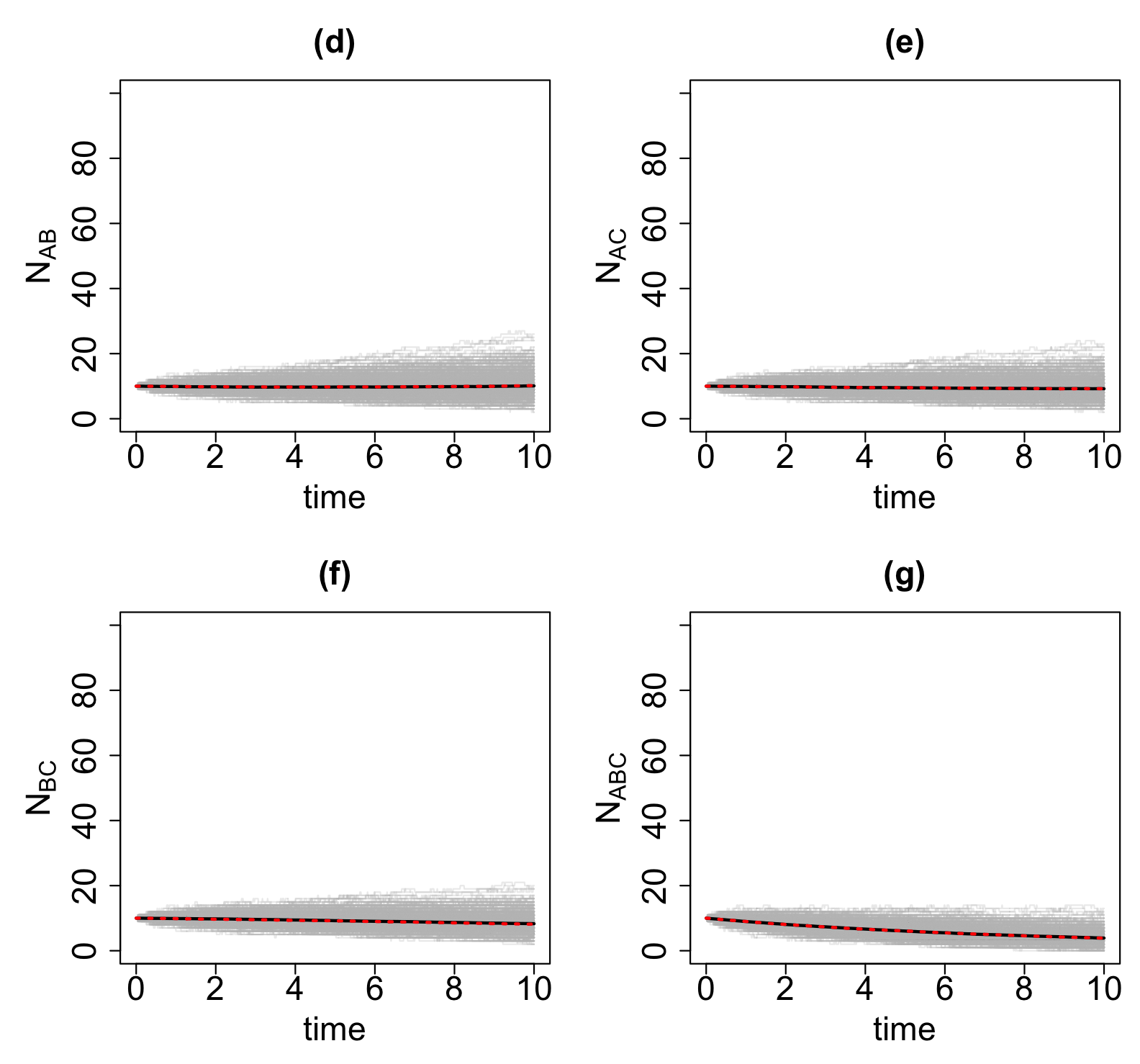}
    \end{subfigure}
    \begin{subfigure}{.6\textwidth}
        \includegraphics[scale=0.35]{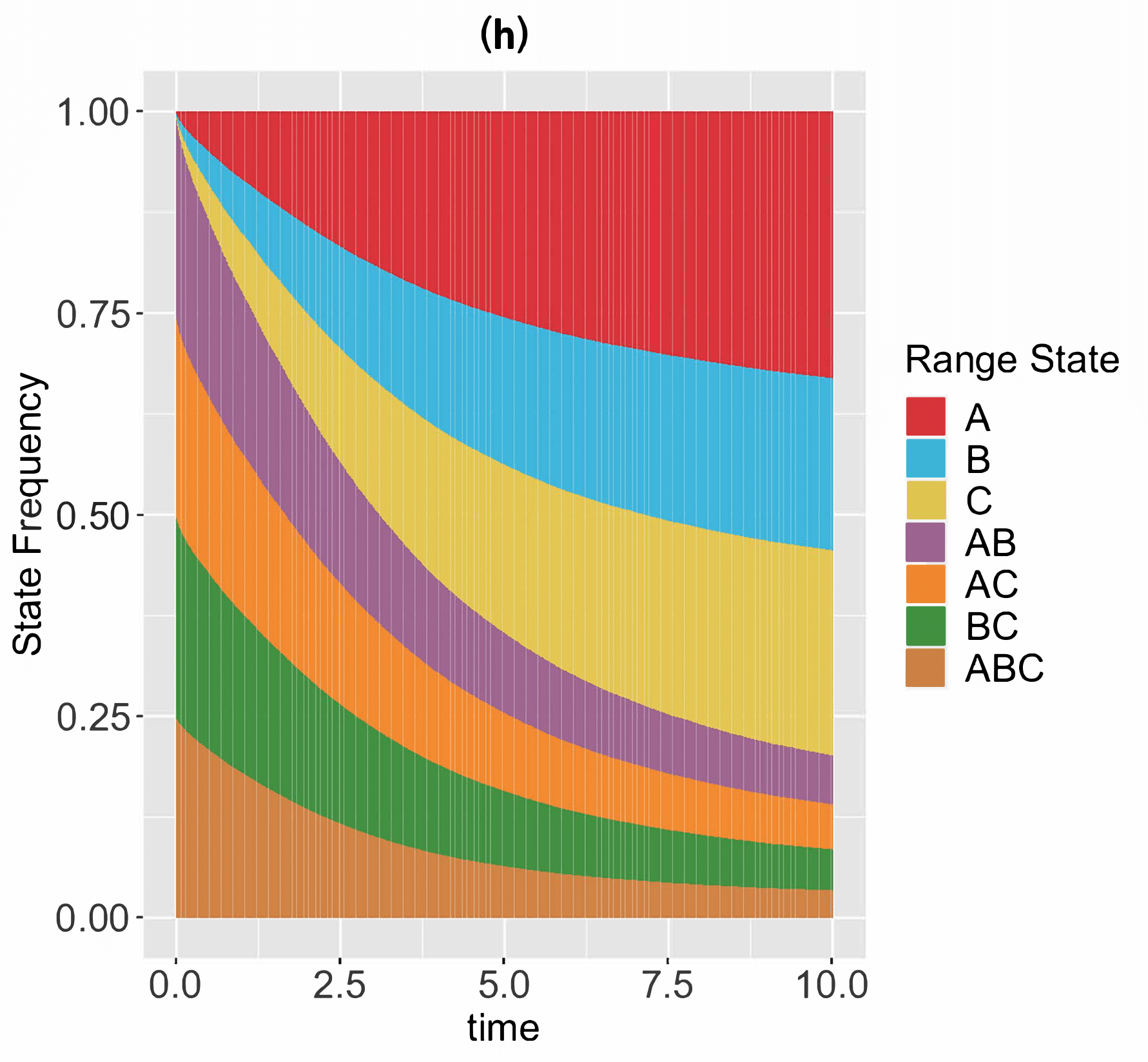}
    \end{subfigure}%
    \begin{subfigure}{.5\textwidth}
        \includegraphics[scale=0.35]{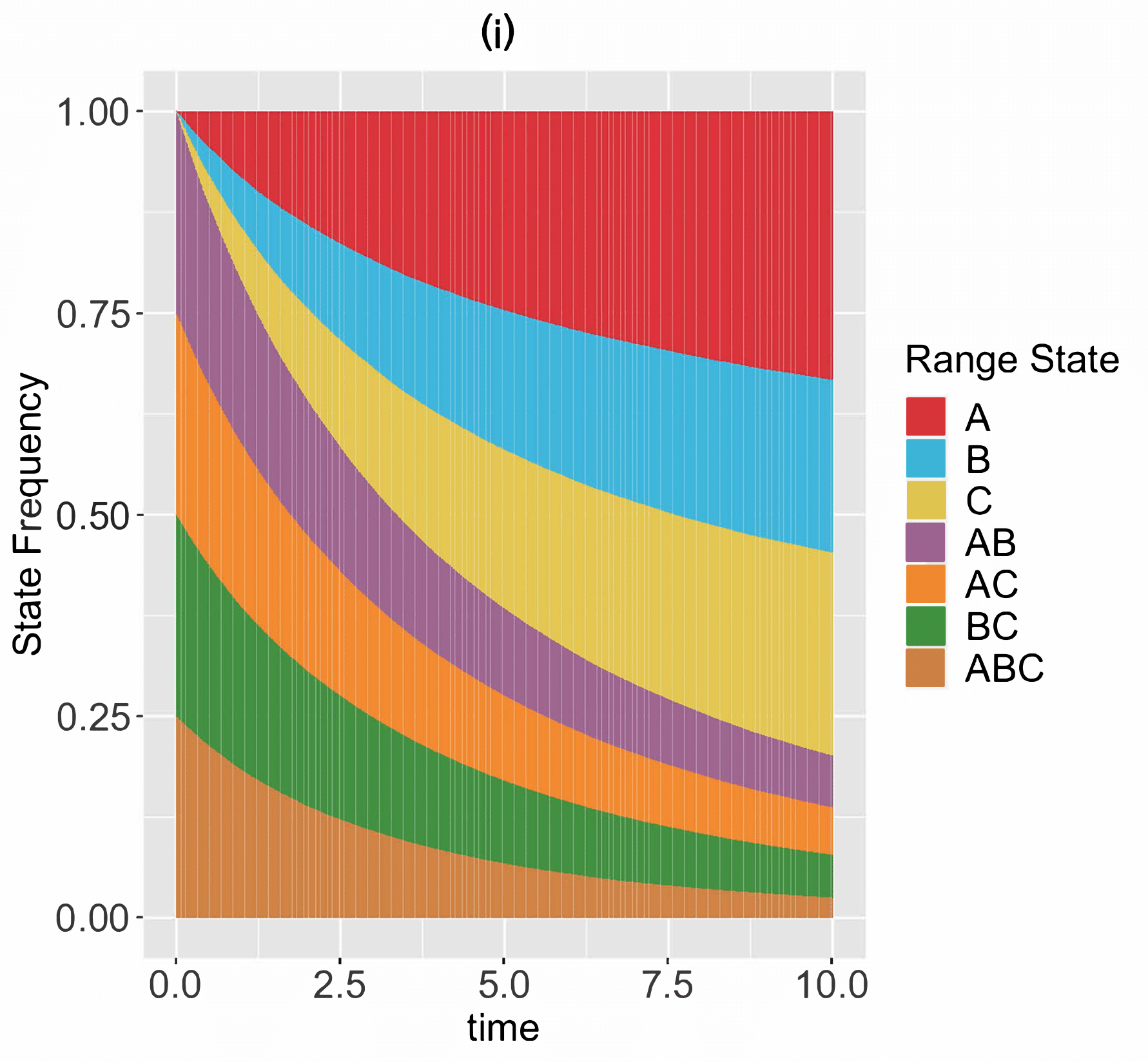}
    \end{subfigure}
	\nobreak
	\caption{Top \& middle panels: the trajectories of average count of range states for endemic species (Figs.~\ref{fig:geosseall}(a)-(c)) and widespread species (Figs.~\ref{fig:geosseall}(d)-(g)) over $[0,10]$ time interval and over 1000 simulations runs for the three-region GeoSSE model as described in Example~\ref{ex:geosseall} simulated under the diffusion-based process (red line) and tree-based process (black line). The gray trajectories show the dynamics across 1000 replicates simulated under diffusion-based process. Bottom panel: stacked bar chart showing the state frequencies over time using diffusion-based approach (Fig.~\ref{fig:geosseall}(h)) and tree-based approach (Fig.~\ref{fig:geosseall}(i)). In both approaches, we start the process with $N(0) = 40$ and the following initial state frequencies: $\Pi_{\{A,B\}}(0) = \Pi_{\{A,C\}}(0) = \Pi_{\{B,C\}}(0) = \Pi_{\{A,B,C\}}(0) = \frac{1}{4}$ and $\Pi_{\{A\}}(0) = \Pi_{\{B\}}(0) = \Pi_{\{C\}}(0) = 0$. At $t=10$, the mean frequencies for each range state from both diffusion-based and tree-based simulations are as follows:\\ 
    $\bar{\Pi}^{diffusion}_{\{A\}}= 0.33$, $\bar{\Pi}^{tree}_{\{A\}}= 0.33$; $\bar{\Pi}^{diffusion}_{\{B\}}= 0.21$, $\bar{\Pi}^{tree}_{\{B\}}= 0.21$;\\ 
    $\bar{\Pi}^{diffusion}_{\{C\}}= 0.25$, $\bar{\Pi}^{tree}_{\{C\}}= 0.25$; $\bar{\Pi}^{diffusion}_{\{A,B\}}= 0.06$, $\bar{\Pi}^{tree}_{\{A,B\}}= 0.06$;\\ $\bar{\Pi}^{diffusion}_{\{A,C\}}= 0.06$, $\bar{\Pi}^{tree}_{\{A,C\}}= 0.06$; $\bar{\Pi}^{diffusion}_{\{B,C\}}= 0.05$, $\bar{\Pi}^{tree}_{\{B,C\}}= 0.05$;\\ $\bar{\Pi}^{diffusion}_{\{A,B,C\}}= 0.03$, $\bar{\Pi}^{tree}_{\{A,B,C\}}= 0.03$.\\Simulations are conducted using the following parameter values: $w_{A}=0.09,w_{B}=0.06,w_{C}=0.07,b^{A}_{B}=b^{A}_{C}=b^{B}_{C}=b^{A}_{BC}=b^{B}_{AC}=b^{C}_{AB}=0.04,e_{A}=0.002,e_{B}=0.003,e_{C}=0.001,d_{AB}=d_{BA}=0.006,d_{AC}=d_{CA}=0.003,d_{BC}=d_{CB}=0.001$}
	\label{fig:geosseall}
\end{figure}

In this section, we visualize the range state dynamics using tree-based and diffusion-based approaches under several GeoSSE scenarios described in Section~\ref{sec::Pro1} (Figs.~\ref{fig:geossewrbet}-\ref{fig:geosseall}). In all these scenarios, we show that the null hypothesis that the average counts of the ranges states at the end of the simulation time between these approaches are equal cannot be rejected (Table~\ref{tab:welch}). This shows that the diffusion-based approach is a good approximation for means to the tree-based approach.\\ 

In most cases, we observe that data (state counts and frequencies) simulated under diffusion-based approach relatively have higher variances compared to data simulated under tree-based approach (Table~\ref{tab:welch}). The $95\%$ confidence interval for the ratio of two variances, shown in Table~\ref{tab:welch}, gives an interval estimate on how much variation one would expect to get for generating state patterns under the diffusion process. Moreover, assuming that data simulated using the MASTER package~\citep{vaughan2013stochastic} represent the true distribution of range state counts, this observation implies that diffusion process is not a good approximation for the second moment of the sampled state state frequencies. While this is not ideal, this is to be expected since diffusion is an approximation method to a generative model. Therefore, we should not expect state counts from both approaches to be drawn from the same distribution.\\
		\begin{table}[htbp]
			\centering
			\begin{tabular}{l l l l l l l l l l}
                    \hline
                \multicolumn{9}{l}{\textbf{Example~\ref{ex::geossewrbe}: GeoSSE with within-region and between-region speciation events}}\\
				\multirow{2}{*}{Range state} & \multicolumn{2}{c}{$\bar{N}_{i,end}$} & \multicolumn{2}{c}{Lower bound} & \multicolumn{2}{c}{Upper bound} & \multirow{2}{*}{$p_{mean}$} & \multirow{2}{*}{$p_{var}$} &
                \multirow{2}{*}{$95\%$ CI $r_{i,var}$}
                \\
                    \cline{2-7}
                    & tree & diffusion & tree & diffusion & tree & diffusion & & \\
				\hline
				$\{A\}$ & $40.835$ & $40.897$ & $40.517$ & $40.156$ & $41.153$ & $41.638$ & $0.880$ & $\ll 0.001$ & $[4.794,6.144]$ \\
				  $\{B\}$ & $40.240$ & $40.544$ & $39.906$ & $39.842$ & $40.574$ & $41.246$ & $0.444$ & $\ll 0.001$ & $[3.908,5.008]$ \\
				  $\{C\}$ & $39.875$ & $40.234$ & $39.555$ & $39.518$ & $40.195$ & $40.950$ & $0.370$ & $\ll 0.001$ & $[4.409,5.651]$\\
				$\{A,B\}$ & $5.980$ & $5.981$ & $5.858$ & $5.837$ & $6.102$ & $6.125$ & $0.992$ & $\ll 0.001$ & $[1.219,1.562]$\\
                    $\{A,C\}$ & $6.239$ & $6.305$ & $6.107$ & $6.128$ & $6.371$ & $6.481$ & $0.558$ & $\ll 0.001$ & $[1.587,2.033]$\\
                    $\{B,C\}$ & $6.506$ & $6.625$ & $6.391$ & $6.494$ & $6.621$ & $6.756$ & $0.182$ &$\ll 0.001$ & $[1.139,1.459]$\\
                    $\{A,B,C\}$ & $1.185$ & $1.112$ & $1.121$ & $1.048$ & $1.25$ & $1.176$ & $0.115$ & $0.782$ & $[0.899,1.152]$\\
				\hline
			\end{tabular}
                \begin{tabular}{l l l l l l l l l l}
                \multicolumn{9}{l}{\textbf{Example~\ref{ex::geossewrdi}: GeoSSE with within-region speciation and dispersal events}}\\
				\multirow{2}{*}{Range state} & \multicolumn{2}{c}{$\bar{N}_{i,end}$} & \multicolumn{2}{c}{Lower bound} & \multicolumn{2}{c}{Upper bound} & \multirow{2}{*}{$p_{mean}$} & \multirow{2}{*}{$p_{var}$} &
                \multirow{2}{*}{$95\%$ CI $r_{i,var}$}\\
                    \cline{2-7}
                    & tree & diffusion & tree & diffusion & tree & diffusion & & \\
				\hline
				$\{A\}$ & $14.689$ & $14.642$ & $14.450$ & $14.282$ & $14.928$ & $15.002$ & $0.831$ & $\ll 0.001$ & $[2.002,2.566]$ \\
				  $\{B\}$ & $13.960$ & $13.993$ & $13.728$ & $13.637$ & $14.192$ & $14.349$ & $0.879$ & $\ll 0.001$ & $[2.076,2.661]$\\
				  $\{C\}$ & $13.339$ & $13.387$ & $13.109$ & $13.035$ & $13.568$ & $13.739$ & $0.823$ & $\ll 0.001$ & $[2.082,2.669]$\\
				$\{A,B\}$ & $8.870$ & $8.740$ & $8.707$ & $8.512$ & $9.033$ & $8.968$ & $0.363$ & $\ll 0.001$ & $[1.732,2.220]$\\
                    $\{A,C\}$ & $11.107$ & $10.870$ & $10.930$ & $10.596$ & $11.284$ & $11.144$ & $0.155$ &$\ll 0.001$ & $[2.113,2.709]$\\
                    $\{B,C\}$ & $13.175$ & $13.172$ & $12.985$ & $12.853$ & $13.365$ & $13.491$ & $0.987$ & $\ll 0.001$ & $[2.482,3.182]$\\
                    $\{A,B,C\}$ & $24.427$ & $24.790$ & $24.189$ & $24.172$ & $24.665$ & $25.408$ & $0.283$ & $\ll 0.001$ & $[5.968,7.649]$\\
				\hline
                \end{tabular}
                \begin{tabular}{l l l l l l l l l l}
                \multicolumn{9}{l}{\textbf{Example~\ref{ex:geossewrext}: GeoSSE with within-region speciation and local extinction events}}\\
				\multirow{2}{*}{Range state} & \multicolumn{2}{c}{$\bar{N}_{i,end}$} & \multicolumn{2}{c}{Lower bound} & \multicolumn{2}{c}{Upper bound} & \multirow{2}{*}{$p_{mean}$} & \multirow{2}{*}{$p_{var}$} &\multirow{2}{*}{$95\%$ CI $r_{i,var}$} \\
                    \cline{2-7}
                    & tree & diffusion & tree & diffusion & tree & diffusion & & \\
				\hline
				$\{A\}$ & $25.672$ & $25.950$ & $25.385$ & $25.550$ & $25.959$ & $26.350$ & $0.269$ & $\ll 0.001$ & $[1.714,2.196]$\\
				  $\{B\}$ & $22.540$ & $22.630$ & $22.266$ & $22.262$ & $22.814$ & $22.998$ & $0.701$ & $\ll 0.001$ & $[1.592,2.040]$\\
				  $\{C\}$ & $20.804$ & $21.179$ & $20.536$ & $20.825$ & $21.072$ & $21.533$ & $0.098$ & $\ll 0.001$ & $[1.547,1.983]$\\
				$\{A,B\}$ & $8.960$ & $9.105$ & $8.843$ & $8.973$ & $9.077$ & $9.237$ & $0.108$ & $< 0.001$ & $[1.111,1.425]$\\
                    $\{A,C\}$ & $8.467$ & $8.370$ & $8.355$ & $8.244$ & $8.579$ & $8.496$ & $0.260$ & $<0.001$ & $[1.109,1.421]$\\
                    $\{B,C\}$ & $7.007$ & $7.024$ & $6.902$ & $6.911$ & $7.112$ & $7.137$ & $0.829$ & $0.027$ & $[1.016,1.302]$\\
                    $\{A,B,C\}$ & $5.805$ & $5.811$ & $5.703$ & $5.711$ & $5.907$ & $5.911$ & $0.934$ & $0.469$ & $[0.844,1.081]$\\
				\hline
			\end{tabular}
                \begin{tabular}{l l l l l l l l l l}
                \multicolumn{9}{l}{\textbf{Example~\ref{ex:geosseall}: GeoSSE with full events}}\\
				\multirow{2}{*}{Range state} & \multicolumn{2}{c}{$\bar{N}_{i,end}$} & \multicolumn{2}{c}{Lower bound} & \multicolumn{2}{c}{Upper bound} & \multirow{2}{*}{$p_{mean}$}  & \multirow{2}{*}{$p_{var}$} & $\multirow{2}{*}{$95\%$ CI $r_{i,var}$}$ \\
                    \cline{2-7}
                    & tree & diffusion & tree & diffusion & tree & diffusion & & \\
				\hline
				$\{A\}$ & $53.067$ & $53.494 $ & $52.420$ & $52.235$ & $53.714$ & $54.753$ & $0.555$ & $\ll 0.001$ & $[3.347,4.290]$\\
				  $\{B\}$ & $33.919$ & $34.425$ & $33.472$ & $33.575$ & $34.366$ & $35.275$ & $0.302$ & $\ll 0.001$ & $[3.193,4.092]$\\
				  $\{C\}$ & $39.981$ & $41.044$ & $39.476$ & $40.060$ & $40.486$ & $42.028$ & $0.060$ & $\ll 0.001$ & $[3.353,4.297]$\\
				$\{A,B\}$ & $10.096$ & $10.193$ & $9.942$ & $9.968$ & $10.250$ & $10.418$ & $0.486$ & $\ll 0.001$ & $[1.880,2.409]$\\
                    $\{A,C\}$ & $9.229$ & $9.224$ & $9.091$ & $9.028$ & $9.367$ & $9.420$ & $0.967$ & $\ll 0.001$ & $[1.772,2.271]$\\
                    $\{B,C\}$ & $8.309$ & $8.138$ & $8.181$ & $7.969$ & $8.437$ & $8.307$ & $0.115$ & $\ll 0.001$ & $[1.526,1.956]$\\
                    $\{A,B,C\}$ & $3.897$ & $3.890$ & $3.789$ & $3.767$ & $4.005$ & $4.013$ & $0.933$ & $\ll 0.001$ & $[1.149,1.472]$\\
				\hline
			\end{tabular}
   
			\caption{The sample mean count for each range state at the end of simulation time, $\Bar{N}_{i,end}$, computed under tree-based and diffusion-based simulations across different GeoSSE scenarios described in Section~\ref{sec::Pro1}. The ``Lower bound" and ``Upper bound" represent the $95\%$ confidence interval of the average count for each range state using diffusion and tree based approaches. The ``$95\%$ CI $r_{i,var}$" correspond to the $95\%$ confidence interval of the ratio of two sample variances from diffusion and tree based approaches for range state $i$. $p_{var}$ and $p_{mean}$ correspond to $p$ value from the $F$ test and the Welch's unequal variances t-test, respectively}
			\label{tab:welch}
		\end{table}
\clearpage
\subsection{Multiple rate scenarios lead to the same stationary state frequencies}
\label{subsec::resultstationary}

We apply the theoretical results from Sections~\ref{subsec::derivestationary}-\ref{subsec::derivestationary-time} for a 2-region GeoSSE model. The different sets of relationships between rate parameters given stationary frequencies in Example~\ref{ex::3} are derived using Mathematica~\citep{Mathematica}. In this example, we show that there exist alternative rate scenarios leading to the same stationary frequencies. Furthermore, using Lemma~\ref{lemma::solve}, we confirm that the stationary frequencies observed from simulations converge to  the theoretical frequencies given the rate parameters, which are derived using Lemma~\ref{lemma::system}. Using the procedure described in Section~\ref{subsec::derivestationary-time}, we compute time to stationary frequencies in Example~\ref{ex::3} for each rate scenario and different sets of initial frequencies.\\
\begin{Example}
\label{ex::3}
We consider a 2-region GeoSSE model with range state space $S = \{\{A\},\{B\},\{A,B\}\}$. We find a set of rate parameters and initial state frequencies that give the following stationary range state frequencies,
\begin{eqnarray*}
    \hat{\Pi}_{\{A\}} = \frac{1}{3}, \hat{\Pi}_{\{B\}} = \frac{1}{3},\hat{\Pi}_{\{A,B\}} = \frac{1}{3}.
\end{eqnarray*}
That is, by Eq.~\eqref{eq::statio_set}, we have,
\begin{eqnarray}
\label{eq::equal_stat}
    \frac{2}{3}w_A +\frac{1}{3}b^{A}_{B} + \frac{1}{3}e_B &=& \frac{1}{3}(d_{AB} + e_A) \nonumber \\
    \frac{2}{3}w_B + \frac{1}{3}b^{A}_{B} + \frac{1}{3}e_A &=& \frac{1}{3}(d_{BA} + e_B) \nonumber \\
    \frac{1}{3}(d_{AB}+d_{BA}) &=&\frac{1}{3}(b^{A}_{B}+e_{A}+e_{B}) \nonumber \\
    2w_{A}+b^{A}_{B}+e_{B}-e_{A}-d_{AB} &=& 2w_{B}+b^{A}_{B} + e_{A}-e_{B}-d_{BA} \nonumber \\
    2w_{B}+b^{A}_{B} + e_{A}-e_{B}-d_{BA} &=& d_{AB}+d_{BA}-b^{A}_{B}-(e_{A}+e_{B}) \nonumber\\
    \sum_{i \in S}\Pi_{i}(0) &=&1, \: \Pi_{\{A\}}(0),\Pi_{\{B\}}(0),\Pi_{\{A,B\}}(0) \geq 0 \nonumber \\
    w_A,w_B,e_A,e_B,d_{AB},d_{BA},b^{A}_{B} &>& 0.
\end{eqnarray}
We found a set of solutions to Eq.~\eqref{eq::equal_stat}. That is, 
\begin{eqnarray}
    && w_{A} = \frac{1}{2}\left(-2b^{A}_{B} + 2 d_{AB} + d_{BA} -2 e_B\right)\nonumber\\
    && w_{B} = \frac{1}{2}\left(-d_{AB}+2e_B\right)\nonumber\\
    && e_A = -b^{A}_{B} + d_{AB} + d_{BA} - e_{B}\nonumber
\end{eqnarray}
\begin{eqnarray}
    && 0 < b^{A}_{B} \leq d_{AB}-e_B, \: e_B < d_{AB} < 2e_B\nonumber\\
    &&d_{BA}>0, \: e_{B}>0.
    \label{eq::2geosse_param_ex3}
\end{eqnarray}
Another set of solutions is given by,
\begin{eqnarray}
    && w_{A} = \frac{1}{2}\left(-2b^{A}_{B} + 2 d_{AB} + d_{BA} -2 e_B\right)\nonumber\\
    && w_{B} = \frac{1}{2}\left(-d_{AB}+2e_B\right)\nonumber\\
     && e_A = -b^{A}_{B} + d_{AB} + d_{BA} - e_{B}\nonumber
\end{eqnarray}
\begin{eqnarray}
     && b^{A}_{B} > 0, \: 0 < d_{AB} \leq e_B\nonumber\\
     && d_{BA} > 2\left(b^A_B - d_{AB} + e_B\right), \: e_{B} > 0.
    \label{eq::2geosse_param_ex3_alt}
\end{eqnarray}

Next, we simulate the range state dynamics, shown in Fig.~\ref{fig:2geosse_ex3}, using the method described in Section.~\ref{sec::Pro1} and rate parameters chosen according to Eq.~\eqref{eq::2geosse_param_ex3}.\\
\begin{figure}[h!]
	\centering
    \begin{subfigure}{0.55\textwidth}
        \includegraphics[width=1\linewidth]{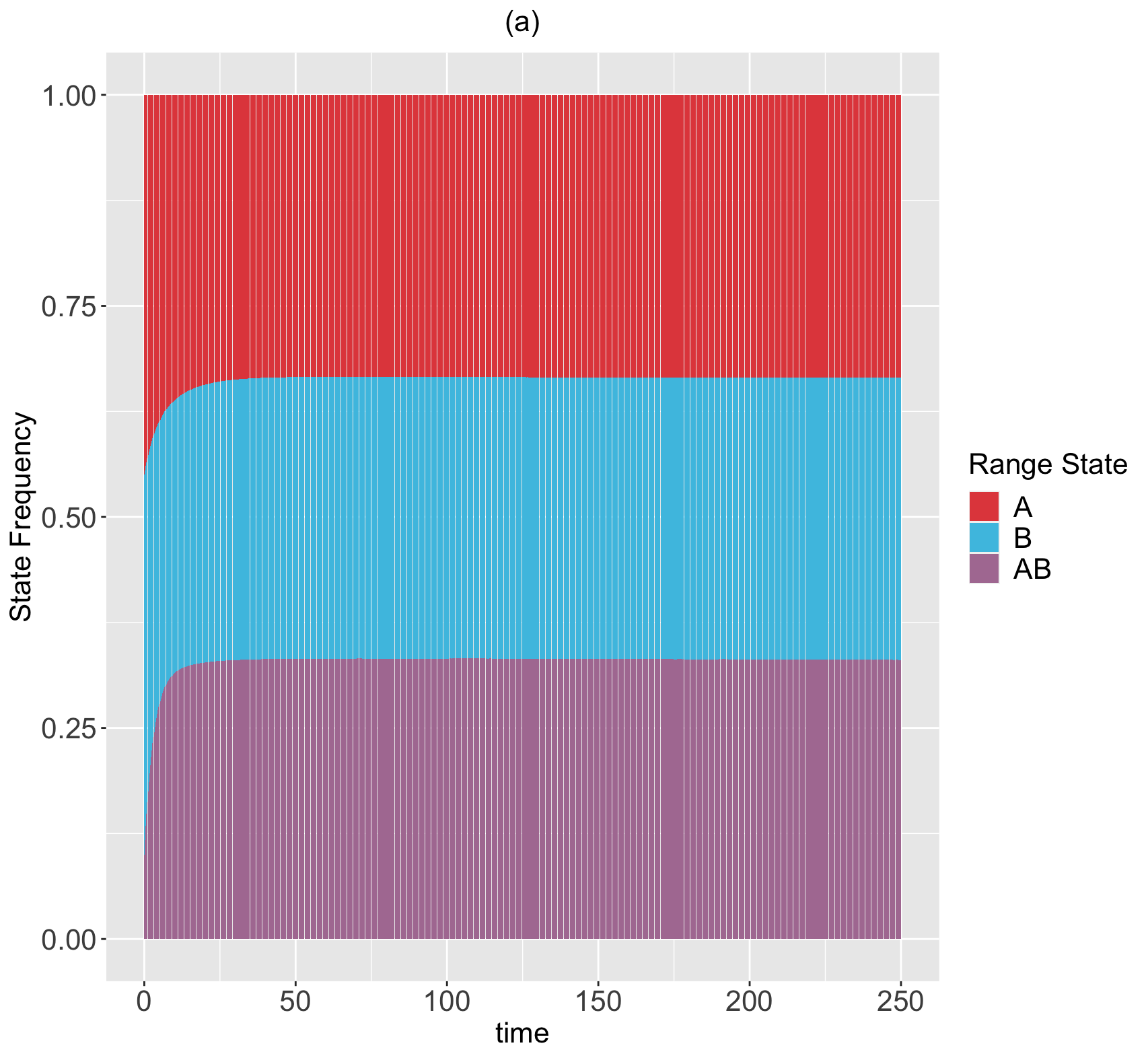}
    \end{subfigure}%
    ~
    \begin{subfigure}{0.55\textwidth}
        \includegraphics[width=1\linewidth]{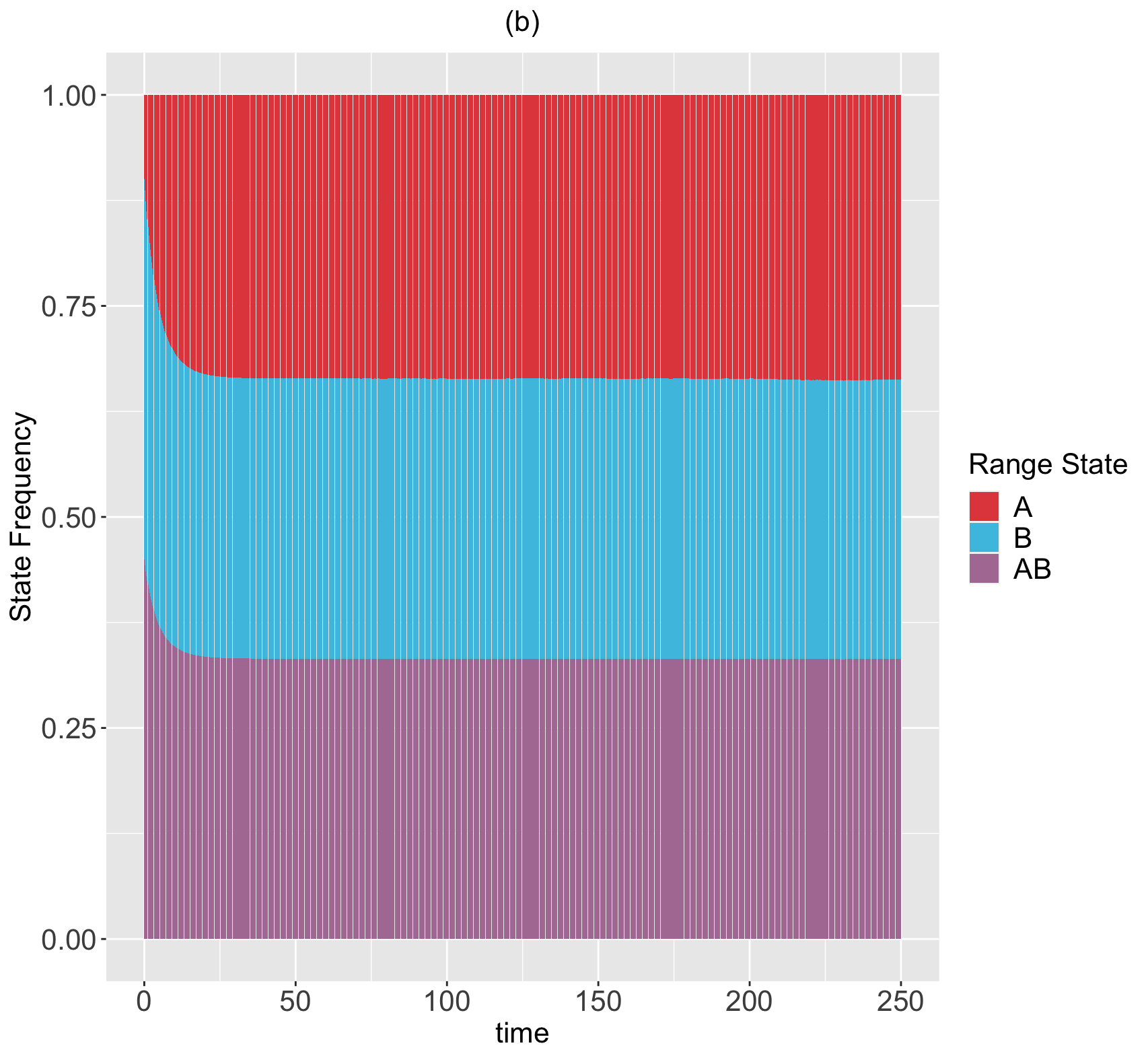}
    \end{subfigure}
	\nobreak
	\caption{The expected range state dynamics over [0,250] time interval and over 100 trajectories for the two-region GeoSSE model as described in Example.~\ref{ex::3}. Each process is simulated under the following initial state frequencies and rate parameters according to Eq.~\eqref{eq::2geosse_param_ex3}: (Left panel) $\Pi_{\{A\}}(0)=\Pi_{\{B\}}(0)=0.45, \Pi_{\{A,B\}}(0)=0.1,w_{A}=0.090,e_{A}=0.176,w_{B}\approx 0,e_{B}=0.008,d_{AB}=0.015,d_{BA}=0.173, b^{A}_{B}= 0.004$ ; (Right panel) $\Pi_{\{A\}}(0)=0.1,\Pi_{\{B\}}(0)=\Pi_{\{A,B\}}(0)=0.45,w_{A}=0.160,e_{A}=0.315,w_{B}= 0.002,e_{B}=0.009,d_{AB}=0.014, d_{BA}=0.310,b^{A}_{B}=0.001$. In both panels, $\mathbb{E}(\hat{\Pi}_{\{A\}})\rightarrow \frac{1}{3}, \mathbb{E}(\hat{\Pi}_{\{B\}})\rightarrow \frac{1}{3}, \mathbb{E}(\hat{\Pi}_{\{A,B\}})\rightarrow \frac{1}{3}$. Using Lemma~\ref{lemma::solve}, we confirm that these expected stationary frequencies from simulations converge to the theoretical, and true stationary frequencies given these sets of rates. Furthermore, using the procedure described in Section~\ref{subsec::derivestationary-time} with $\epsilon = 10^{-9}$, we found that the stationary frequencies are reached at: $t^{*}_{A}=114.114, \: t^{*}_{B}=111.862,t^{*}_{AB}=102.603$ (Left panel); $t^{*}_{A}=76.827, \: t^{*}_{B}=75.576,t^{*}_{AB}=70.320$ (Right panel)}
	\label{fig:2geosse_ex3}
\end{figure}

To show that there are multiple rate scenarios that lead to the same stationary distribution, we simulate the range state dynamics, shown in Fig.~\ref{fig:2geosse_ex3_alt}, using rate parameters that satisfy the alternative set of solutions described in Eq.~\eqref{eq::2geosse_param_ex3_alt}, but do not satisfy Eq.~\eqref{eq::2geosse_param_ex3}. 
\begin{figure}[h!]
	\centering
    \begin{subfigure}{0.55\textwidth}
        \includegraphics[width=1\linewidth]{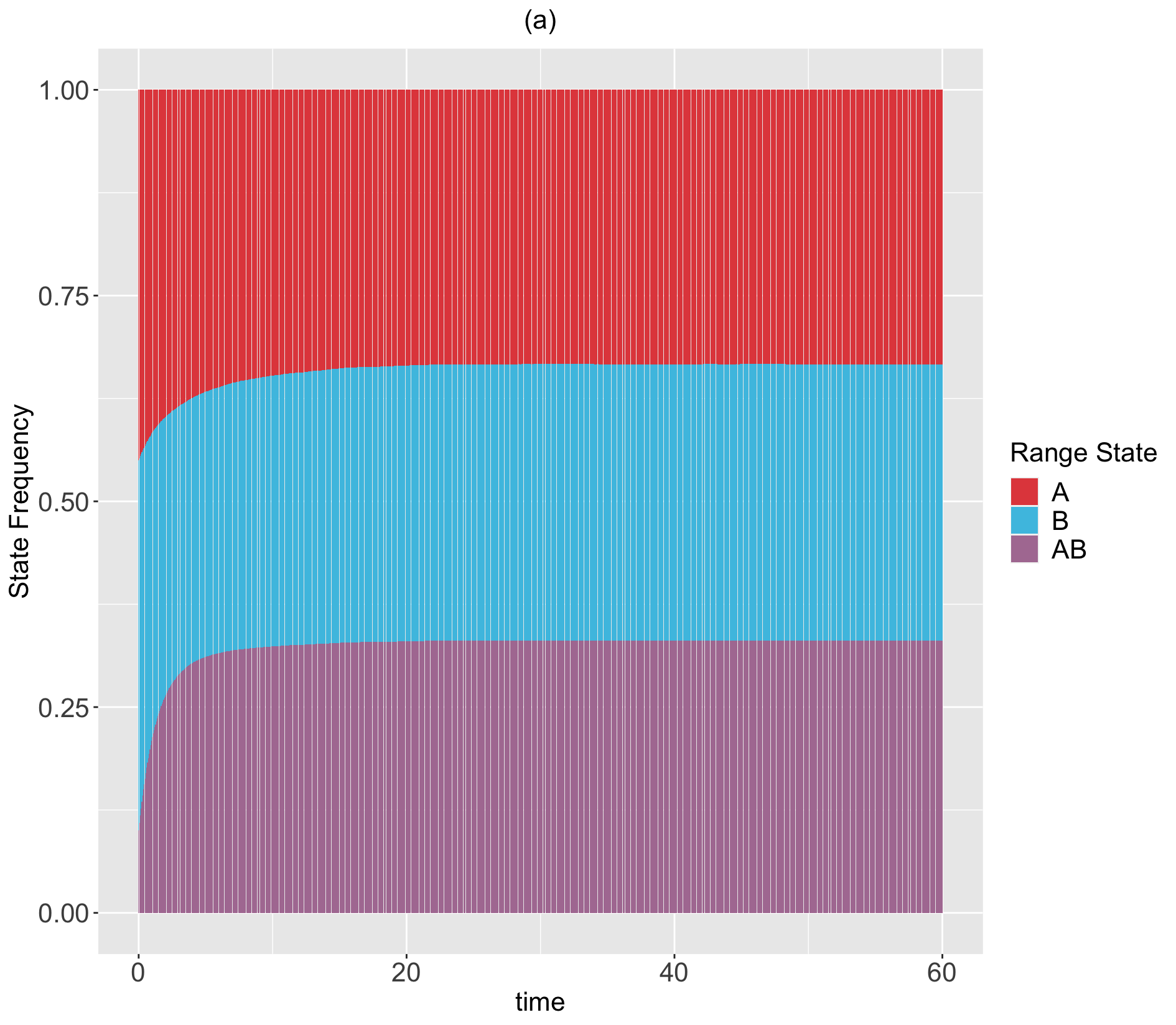}
    \end{subfigure}%
    ~
    \begin{subfigure}{0.55\textwidth}
        \includegraphics[width=1\linewidth]{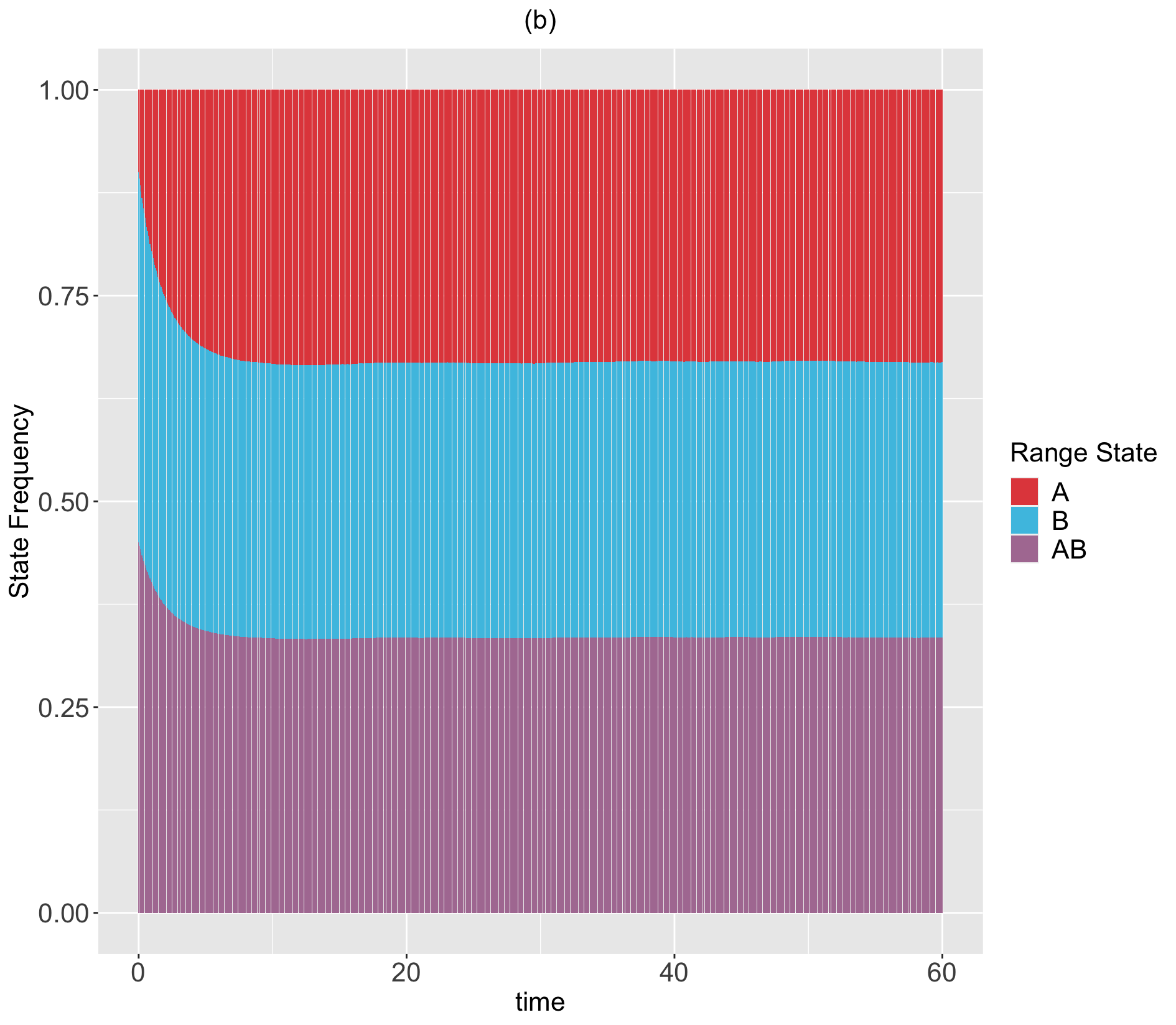}
    \end{subfigure}
	\nobreak
	\caption{The expected range state dynamics over [0,60] time interval and over 100 trajectories for the two-region GeoSSE model as described in Example.~\ref{ex::3}. Each process is simulated under the following initial state frequencies and rate parameters according to Eq.~\eqref{eq::2geosse_param_ex3_alt}: (Left panel) $\Pi_{\{A\}}(0)=\Pi_{\{B\}}(0)=0.45, \Pi_{\{A,B\}}(0)=0.1,w_{A}=0.107,e_{A}=0.309,w_{B}= 0.008,e_{B}=0.008,d_{AB}=0.001,d_{BA}=0.405, b^{A}_{B}= 0.089$ ; (Right panel) $\Pi_{\{A\}}(0)=0.1, \Pi_{\{B\}}(0)=\Pi_{\{A,B\}}(0)=0.45,w_{A}=0.049,e_{A}=0.470,w_{B}=0.005,e_{B}=0.008,d_{AB}=0.006, d_{BA}=0.843,b^{A}_{B}=0.371$. In both panels, $\mathbb{E}(\hat{\Pi}_{\{A\}})\rightarrow \frac{1}{3}, \mathbb{E}(\hat{\Pi}_{\{B\}})\rightarrow \frac{1}{3}, \mathbb{E}(\hat{\Pi}_{\{A,B\}})\rightarrow \frac{1}{3}$. Using Lemma~\ref{lemma::solve}, we confirm that these expected stationary frequencies from simulations converge to the theoretical, and true stationary frequencies given these sets of rates. Furthermore, using the procedure described in Section~\ref{subsec::derivestationary-time} with $\epsilon = 10^{-9}$, we found that the stationary frequencies are reached at: $t^{*}_{A}=53.153, \: t^{*}_{B}=51.952,t^{*}_{AB}=48.048$ (Left panel); $t^{*}_{A}=30.781, \: t^{*}_{B}=30.330,t^{*}_{AB}=28.378$ (Right panel)}
	\label{fig:2geosse_ex3_alt}
\end{figure}
\end{Example}

\subsection{Comparing our method of computing stationary state frequencies with existing literature}
\label{subsec:comparing}

In this section, we compare our method for computing stationary state frequencies from rate parameters introduced in Section~\ref{subsec::derivestationary-inverse} with another method used in diversitree package~\citep{fitzjohn2012diversitree} for the ClaSSE~\citep{goldberg2012tempo} and GeoSSE~\citep{goldberg2011phylogenetic} models. Although the technique used in diversitree has not been discussed in any SSE papers, such as the papers introducing the MuSSE~\citep{fitzjohn2012diversitree}, ClaSSE~\citep{goldberg2012tempo}, and GeoSSE~\citep{goldberg2011phylogenetic} models, the technique applies projection matrix models that are widely used in the context of population biology to obtain ClaSSE and GeoSSE stationary frequencies (pers. comm. E. E. Goldberg and R. FitzJohn). Originally developed for applications in discrete-time models with either size-structured or age-structured population~\citep{van1988projection}, this approach has also been adapted for continuous-time models with the latter structured population~\citep{kapur1979continuous}. Under this approach, one would create a square matrix with entries that map the state of a structured population from one time to the next. Then, the dominant eigenvalue of such matrix represents the overall population growth rate with its eigenvector represents the stable stage distribution~\citep{van1988projection}.  \\

Through examples below we find that our method returns similar state frequencies to those computed under the projection matrix model in diversitree package~\citep{fitzjohn2012diversitree}. For example, under the following rate parameters in a two-region GeoSSE model,
\begin{eqnarray*}
    w_A = 0.01, \: w_B = 0.02, \: b^{A}_{B} = 0.003, \: e_A = 0.169, \: e_B = 0.008, \: d_{AB} = 0.002, \: d_{BA} = 0.178,
\end{eqnarray*}
our method gives $\hat{\Pi}_{\{A\}} \approx 0.057, \: \hat{\Pi}_{\{B\}} \approx 0.506, \: \hat{\Pi}_{\{A,B\}} \approx 0.437 $ while the projection matrix approach implemented in diversitree returns $\hat{\Pi}_{\{A\}} \approx 0.055, \: \hat{\Pi}_{\{B\}} \approx 0.490, \: \hat{\Pi}_{\{A,B\}} \approx 0.455$. Another example using the following rate parameters,
\begin{eqnarray*}
    w_A \approx 0.0006, \: w_B \approx 0.0003, \: b^{A}_{B} \approx 0, \: e_A \approx 0.0048, \: e_B \approx 0.0045, \: d_{AB} \approx  0.0370, \: d_{BA} \approx 0.03703,
\end{eqnarray*}
we have $\hat{\Pi}_{\{A\}} \approx 0.0996, \: \hat{\Pi}_{\{B\}} \approx 0.0996, \: \hat{\Pi}_{\{A,B\}} \approx 0.8008 $ while the other method produces $\hat{\Pi}_{\{A\}} \approx 0.0997, \:  \hat{\Pi}_{\{B\}} \approx 0.0997, \: \hat{\Pi}_{\{A,B\}} \approx 0.8006$.

\clearpage

\section{Discussion and Conclusion}\label{sec:conclusion}

In our paper, we have constructed a general framework using diffusion processes to study state dynamics over time from a general state-dependent speciation and extinction model with both anagenetic and cladogenetic state transitions, making it suitable for studying members of the ClaSSE model family~\citep{goldberg2012tempo, magnuson2012, goldberg2011phylogenetic, freyman2018}. We have applied this framework under various diversification scenarios for the GeoSSE model~\citep{goldberg2011phylogenetic}, a special case of the ClaSSE model, as described in Sections~\ref{subsec::diffusiongeosse}-\ref{sec::Pro1}. Our framework can easily be applied to other discrete state-dependent diversification models, such as simpler BiSSE and MuSSE models~\citep{maddison2007estimating,fitzjohn2012diversitree} and Markovian Binary Tree (MBT) models~\citep{Kontoleon,2009HLR,soewongsono2023matrix}. Through simulations and statistical analyses, we have shown that state dynamics simulated under diffusion-based approach and tree-based approach are comparable, given that we start the simulations with relative large clade size (Figs.~\ref{fig:geossewrbet}-\ref{fig:geosseall}, Table~\ref{tab:welch}). We also obtain good agreement between diffusion-based and tree-based simulations when beginning the process with a single species in random state, after applying a model-based correction procedure (Appendix~\ref{subsec::singlespec}, Fig.~\ref{fig:compare_correct}). We also show, using a statistical test, that our diffusion framework offers a good approximation for the mean of state counts. This result allows one to understand how data generating process i.e. rate parameters from a diversification model can explain observed state patterns without using phylogenetic information. For inferring rates using empirical state data at present, this diffusion-based approach to simulate state dynamics could be treated as a way to validate whether rates estimated from biological datasets using phylogenetic methods are sensible. \\

Moreover, in Sections~\ref{subsec::derivestationary}-\ref{subsec::derivestationary-inverse}, we have derived theoretical results to deduce the expected state frequencies generated by a set of rates, and what possible rates will generate a given set of expected state frequencies. These results are generalizable to accommodate a system having more states, and provide an alternative way to validate the correctness of SSE simulation and inference methods. Additionally, in Section~\ref{subsec::derivestationary-time}, we described a procedure to compute the minimum time for an SSE process to reach stationarity in its state frequencies. We have applied these results for a 2-region GeoSSE model. As seen in Figs.~\ref{fig:2geosse_ex3}-\ref{fig:2geosse_ex3_alt}, we showed that there exist multiple different rate scenarios that can lead to the same stationary behaviour of state pattern. Our framework also creates an alternative mathematical approach to tree-based models that could help establish conditions for which SSE model parameters are and are not identifiable.\\

We next plan to study the time for perturbed SSE models to reach stationarity. This would help biologists understand how evolutionary systems re-equilibrate and how long that re-equilibration takes following perturbation. In particular, we plan to apply this framework to study scenarios where SSE rates shift across time \citep{condamine2013,quintero2023}. Scenarios with time-heterogeneous rates are particularly interesting for GeoSSE model variants, mainly because regions experience changes in their features (e.g., region size, distance with nearby regions, separation types) over time. As studied in~\cite{landis2022phylogenetic, swiston2023testing}, paleogeographically-changing regional features should influence rates of speciation, extinction, and dispersal over time. Mathematical knowledge of expected state (range) frequencies for arbitrary biogeographical systems could help biodiversity researchers assess whether certain clades of regions are within or between states of equilibrium.

\section{Funding}
This research was funded by the National Science Foundation (NSF Award DEB-2040347), the Fogarty International Center at the National Institutes of Health (Award Number R01 TW012704) as part of the joint NIH-NSF-NIFA Ecology and Evolution of Infectious Disease program, and the Washington University Incubator for Transdisciplinary Research.

\section{Acknowledgements}
We thank members of the Landis Lab for valuable feedback that improved the quality of this study. We also thank Emma Goldberg, Richard FitzJohn, and Mike May for directing us to literature used for computing stationary frequencies in the diversitree package.

\section*{Data availability}
\label{sec::data}

The datasets and all the relevant code are publicly available on \url{https://github.com/alberts2/Diffusion_GeoSSE.git}.

\section*{Conflict of interest}
\label{sec::conflic}

The authors declare that they have no conflict of interest.

\section{Appendix}
\label{sec:appendix}

\subsection{Proof of Lemma~\ref{lemma::stochtrans}}
\label{subsec::stochtrans}

From Eq.~\eqref{eq1}, we compute $g'(N)$ and $g''(N)$,
\begin{eqnarray}
    g'(N) = \frac{\partial g}{\partial N} &=& \frac{\partial \left(\sum_{i}h(n_{i})\right)}{\partial N}\nonumber\\
    &=& \frac{\partial \left(\sum_{i}h(n_{i})\right)}{\partial n_{i}} \, \text{(For each i, the other partial derivatives w.r.t $j\neq i$ equals 0)} \nonumber\\
    &=& \frac{\partial g}{\partial n_{i}}\nonumber \\
    &=& \frac{\partial X}{\partial n_{i}}.
\end{eqnarray}
Similarly, we have,
\begin{eqnarray}
    g''(N) = \frac{\partial^{2}X}{\partial n_{i}^{2}}.
\end{eqnarray}
Now applying Theorem~\eqref{thm:1} to Eq.~\eqref{eq1}, we have,
\begin{eqnarray}
    \mu_{X} &=& \sum_{i}\frac{\partial X}{\partial n_{i}}\mu_{i} + \frac{1}{2}\sum_{i}\frac{\partial^{2}X}{\partial n_{i}^{2}}\sigma^{2}_{i}, \\
    \sigma^{2}_{X} &=& \sum_{i}\left(\frac{\partial X}{\partial n_{i}}\right)^{2}\sigma^{2}_{i}.
\end{eqnarray}\qed

\subsection{Proof of Lemma~\ref{lemma::geosseNpar}}
\label{subsec::geosseNpar}

By definition, following Eqs.~1.2--1.3 of Chapter~5 in~\cite{karlin1981second}, we define the infinitesimal mean $\mu_i$ and infinitesimal variance $\sigma_{i}^{2}$ of the stochastic process $\{N_i(t) : t > 0\}$ as follows,
\begin{eqnarray*}
    \mu_{i} &=& \lim_{\Delta t \rightarrow 0}\frac{1}{\Delta t} \mathbb{E}\left(N_{i}(t+\Delta t)-N_{i}(t) \vert N_{i}(t) = N_i\right) \\
    &=& \lim_{\Delta t \rightarrow 0}\frac{1}{\Delta t}\{\mathbb{E}(N_{i}(t+\Delta t)) - \mathbb{E}(N_{i}(t) \vert N_{i}(t) = N_i)\} \\
    &=& \lim_{\Delta t \rightarrow 0}\frac{1}{\Delta t}\{\mathbb{E}(N_{i}(t+\Delta t))-N_{i}\}, \\
    \sigma_{i}^{2}  &=& \lim_{\Delta t \rightarrow 0}\frac{1}{\Delta t}\mathbb{E}\left((N_{i}(t+\Delta t)-N_{i}(t))^{2} \vert N_{i}(t) = N_i\right) \\
    &=& \lim_{\Delta t \rightarrow 0}\frac{1}{\Delta t}\mathbb{E}\left(N_{i}^{2}(t+\Delta t) - 2 N_{i}(t)N_{i}(t+\Delta t) + N_{i}^{2}(t) \vert N_{i}(t)\right) \\ 
    &=& \lim_{\Delta t \rightarrow 0}\frac{1}{\Delta t}\{\mathbb{E}\left(N_{i}^{2}(t+\Delta t)\right)-2 N_{i}\mathbb{E}\left(N_{i}(t+\Delta t)\right)+ N_{i}^{2}\}.
\end{eqnarray*}

By definition of the first-order and second-order moments we have,
\begin{eqnarray*}
\mathbb{E}(N_{i}(t+\Delta t))   &=& (N_{i}+1)\mathbb{P}_{i}^{+}\Delta t +                                        (N_{i}-1)\mathbb{P}_{i}^{-}\Delta t + (N_{i})\mathbb{P}_{i}\Delta t \\
                                &=& N_{i}\left(\mathbb{P}_{i}^{+} + \mathbb{P}_{i}^{-} + \mathbb{P}_{i}\right)\Delta t + \left(\mathbb{P}_{i}^{+}-\mathbb{P}_{i}^{-}\right)\Delta t\\
                                &=& N_{i} + \left(\mathbb{P}_{i}^{+} - \mathbb{P}_{i}^{-}\right)\Delta t, \\
\mathbb{E}(N_{i}^{2}(t+\Delta t) &=& (N_{i}+1)^{2}\mathbb{P}_{i}^{+}\Delta t +                                        (N_{i}-1)^{2}\mathbb{P}_{i}^{-}\Delta t + (N_{i})^{2}\mathbb{P}_{i}\Delta t.\\
\end{eqnarray*}

Thus, 
\begin{eqnarray}
    \mu_{i}         &=& \mathbb{P}_{i}^{+} - \mathbb{P}_{i}^{-}, \\
    \sigma_{i}^2    &=& \mathbb{P}_{i}^{+} + \mathbb{P}_{i}^{-}.
\end{eqnarray}\qed

\subsection{Proof of Lemma~\ref{Lemma::geossepipars}}
\label{subsec::geossepipars}

We compute the following and substitute to Eqs.~\eqref{eq::itomean}--\eqref{eq::itovariance}.

\begin{eqnarray*}
\frac{\partial \Pi_{i}}{\partial N_{i}}     &=& \frac{\partial}{\partial N_{i}}\left(\frac{N_{i}}{\sum_{\substack{k \in s \\ k \neq i}}N_{k}+N_{i}}\right)\\
                                            &=& \frac{\sum_{\substack{k \in S \\ k \neq i}}N_{k}}{\left(\sum_{k \in S}N_{k}\right)^{2}}\\
                                            &=& \frac{1-\Pi_{i}}{N},\\
\frac{\partial \Pi_{i}}{\partial N_{j}}     &=& \frac{\partial}{\partial N_{j}}\left(\frac{N_{i}}{\sum_{\substack{k \in s \\ k \neq i}}N_{k}+N_{i}}\right), \: j\neq i\\
                                            &=& -\frac{N_{i}}{N^{2}}\\
                                            &=& -\frac{\Pi_{i}}{N},\\
\frac{\partial^{2}\Pi_{i}}{\partial N_{i}^{2}}  &=& \frac{\partial}{\partial N_{i}}\left(\frac{1-\Pi_{i}}{N}\right)\\
                                                &=& \frac{-\frac{\partial \Pi_{i}}{\partial N_{i}}N - (1-\Pi_{i})\frac{\partial N}{\partial N_{i}})}{N^{2}}\\
                                                &=& \frac{-(1-\Pi_{i})-(1-\Pi_{i})}{N^{2}}\\
                                                &=& -\frac{2(1-\Pi_{i})}{N^{2}},\\
\frac{\partial^{2}\Pi_{i}}{\partial N_{j}^{2}}  &=& \frac{\partial}{\partial N_{j}}\left(-\frac{N_{i}}{N^2}\right), \: j\neq i\\
                                                &=& \frac{N_{i}(2N)}{N^4}\\
                                                &=& \frac{2\Pi_{i}N}{N^3}\\
                                                &=& \frac{2\Pi_i}{N^2}.
\end{eqnarray*}
Thus, 
\begin{eqnarray}
\mu_{\Pi_{i}}       &=& \sum_{j \in S}\frac{\partial \Pi_i}{\partial N_j}\mu_{j} + \frac{1}{2}\sum_{j \in S}\frac{\partial^{2}\Pi_i}{\partial N_{j}^2}\sigma_{j}^{2} \nonumber \\
                    &=& \frac{\partial \Pi_{i}}{\partial N_{i}}\mu_{i} + \sum_{\substack{j \in S \\ j \neq i}}\frac{\partial \Pi_i}{\partial N_j}\mu_{j} + \frac{1}{2}\frac{\partial^{2}\Pi_i}{\partial N_{i}^2}\sigma_{i}^{2} + \frac{1}{2}\sum_{\substack{j \in S \\ j \neq i}}\frac{\partial^{2}\Pi_i}{\partial N_{j}^2}\sigma_{j}^{2} \nonumber \\
                    &=& \frac{1-\Pi_i}{N}\mu_{i} + \sum_{\substack{j \in S \\ j \neq i}}\frac{-\Pi_i}{N}\mu_{j} - \frac{1-\Pi_i}{N^2}\sigma_{i}^{2} +  \sum_{\substack{j \in S \\ j \neq i}}\frac{\Pi_i}{N^2}\sigma_{j}^{2} \nonumber \\
                    &=& \left(\frac{1-\Pi_i}{N}\right)\left(\mu_{i}-\frac{\sigma_{i}^2}{N}\right) + \sum_{\substack{j \in S \\ j \neq i}}\frac{\Pi_i}{N}\left(-\mu_j + \frac{\sigma_{j}^2}{N}\right) \nonumber \\
                    &=& \frac{1}{N}\left(\mu_i - \frac{\sigma^{2}_{i}}{N}\right) + \frac{\Pi_i}{N}\sum_{j \in S}\left(-\mu_j + \frac{\sigma_{j}^2}{N}\right),\\
\sigma_{\Pi_{i}}^{2} &=& \sum_{j\in S}\left(\frac{\partial \Pi_i}{\partial N_{j}}\right)^{2}\sigma_{j}^{2} \nonumber \\
                     &=& \left(\frac{\partial \Pi_i}{\partial N_{i}}\right)^{2}\sigma_{i}^{2} + \sum_{\substack{j\in S \\ j\neq i}}\left(\frac{\partial \Pi_i}{\partial N_{j}}\right)^{2}\sigma_{j}^{2} \nonumber \\
                     &=& \left(\frac{1-\Pi_i}{N}\right)^{2}\sigma_{i}^{2} + \sum_{\substack{j\in S \\ j\neq i}}\left(\frac{\Pi_i}{N}\right)^{2}\sigma_{j}^2 \nonumber \\
                     &=& \left(\frac{\sigma_i}{N}\right)^{2}(1-2\Pi_i) + \left(\frac{\Pi_i}{N}\right)^{2}\sum_{j\in S}\sigma_{j}^2,
\end{eqnarray}

where  $\mu_{i}$ and $\sigma_{i}^{2}$ follow Eqs.~\eqref{eq::mu_geosse}--\eqref{eq::var_geosse}, respectively. \qed

\subsection{Proof of Lemma~\ref{lemma::solve}}
\label{subsec::prooflemma5}

We find $\hat{\Pi}_{i}$ such that $\lim_{t \rightarrow \infty} \Pi_{i}(t) = \hat{\Pi}_{i}$ for all $i \in S, \: S = \{\{A\},\{B\},\{A,B\}\}$. \\

For $i = \{A\}$ we have,
\begin{eqnarray}
    \hat{\mu}_{\Pi_A} &=&  \hat{\mathbb{P}}_{A}^{+} - \hat{\mathbb{P}}_{A}^{-} \nonumber \\
    &=& \left[w_{A}\left(\Pi_{\{A\}}+\Pi_{\{A,B\}}\right)+ \Pi_{\{A,B\}}b^{A}_{B} + e_{B}\Pi_{\{A,B\}}\right] - \left[d_{AB}\Pi_{\{A\}}+ e_A \Pi_{\{A\}}\right]\nonumber \\
    &=& \Pi_{\{A\}}\left(w_A - e_A - d_{AB}\right) + \Pi_{\{A,B\}}\left(w_A + b^{A}_{B} + e_B\right).
    \label{eq::pia_1}
\end{eqnarray}

For $i = \{B\}$ we have,
\begin{eqnarray}
    \hat{\mu}_{\Pi_B} &=& \hat{\mathbb{P}}_{B}^{+} - \hat{\mathbb{P}}_{B}^{-} \nonumber \\
&=&\left[w_{B}\left(\Pi_{\{B\}}+\Pi_{\{A,B\}}\right)+ \Pi_{\{A,B\}}b^{A}_{B} + e_{A}\Pi_{\{A,B\}}\right] - \left[d_{BA}\Pi_{\{B\}}+ e_B \Pi_{\{B\}}\right] \nonumber \\
    &=&\Pi_{\{B\}}\left(w_B - e_B - d_{BA}\right) + \Pi_{\{A,B\}}\left(w_B + b^{A}_{B} + e_A\right).
    \label{eq::pib_2}
\end{eqnarray}

For $i = \{A,B\}$ we have,
\begin{eqnarray}
    \hat{\mu}_{\Pi_{AB}} &=& \hat{\mathbb{P}}_{AB}^{+} - \hat{\mathbb{P}}_{AB}^{-}  \nonumber \\
    &=& \left[d_{AB}\Pi_{\{A\}} + d_{BA}\Pi_{\{B\}}\right] - \left[b^{A}_{B}\Pi_{\{A,B\}} + (e_A + e_B)\Pi_{\{A,B\}}\right] \nonumber \\
    &=& \left(d_{AB}\Pi_{\{A\}} + d_{BA}\Pi_{\{B\}}\right) - \Pi_{\{A,B\}}\left(b^{A}_{B} + e_A + e_B\right).
    \label{eq::piAb_2}
\end{eqnarray}

Thus, we want to find the general solution for the following system of differential equations 
\begin{eqnarray}
    \frac{d\Pi_{\{A\}}}{d t} &=& \Pi_{\{A\}}\left(w_A - e_A - d_{AB}\right) + \Pi_{\{A,B\}}\left(w_A + b^{A}_{B} + e_B\right), \\
    \frac{d\Pi_{\{B\}}}{d t} &=& \Pi_{\{B\}}\left(w_B - e_B - d_{BA}\right) + \Pi_{\{A,B\}}\left(w_B + b^{A}_{B} + e_A\right), \\
    \frac{d\Pi_{\{A,B\}}}{d t} &=& \left(d_{AB}\Pi_{\{A\}} + d_{BA}\Pi_{\{B\}}\right) - \Pi_{\{A,B\}}\left(b^{A}_{B} + e_A + e_B\right).
\end{eqnarray}
given initial state frequencies $\Pi_{\{A\}}(0) = \Pi_{A}^{0},\Pi_{\{B\}}(0) = \Pi_{B}^{0},\Pi_{\{A,B\}}(0) = \Pi_{AB}^{0}$.\\

However, since $\hat{\Pi}_{\{A\}} + \hat{\Pi}_{\{B\}} + \hat{\Pi}_{\{A,B\}} = 1$, we can always derive $\hat{\Pi}_{\{A,B\}}$ using $\hat{\Pi}_{\{A\}}$ and $\hat{\Pi}_{\{B\}}$. Therefore, we want to solve the following system instead. 
\begin{eqnarray}
    \frac{d\Pi_{\{A\}}}{d t} &=& \Pi_{\{A\}}\left(w_A - e_A - d_{AB}\right) + \Pi_{\{A,B\}}\left(w_A + b^{A}_{B} + e_B\right), \\
    \frac{d\Pi_{\{B\}}}{d t} &=& \Pi_{\{B\}}\left(w_B - e_B - d_{BA}\right) + \Pi_{\{A,B\}}\left(w_B + b^{A}_{B} + e_A\right).
\end{eqnarray}
Since $\Pi_{\{A\}}(t) + \Pi_{\{B\}}(t) + \Pi_{\{A,B\}}(t) = 1$, we have,
\begin{eqnarray}
    \frac{d\Pi_{\{A\}}}{d t} &=& \Pi_{\{A\}}\left(-e_A -d_{AB}-b^{A}_{B} - e_B\right) - \Pi_{\{B\}}\left(w_A + b^{A}_{B} + e_B\right) + \left(w_A + b^{A}_{B} + e_B\right), \nonumber \\
    \\
    \frac{d\Pi_{\{B\}}}{d t} &=& \Pi_{\{B\}}\left(-e_B - d_{BA} - b^{A}_{B} - e_A\right) - \Pi_{\{A\}}\left(w_B + b^{A}_{B} + e_A\right) + \left(w_B + b^{A}_{B} + e_A\right). \nonumber \\
    \label{system_ode}
\end{eqnarray}

We write the above system in a matrix form as follows, 
\begin{eqnarray}
    \bm{\Pi^{'}} &=& M\bm{\Pi} + \bm{r},
    \label{eq::matrixform}
\end{eqnarray}
where
\begin{eqnarray}
    &&\bm{\Pi^{'}} = \begin{bmatrix}
         \Pi_{\{A\}}^{'}\\
         \Pi_{\{B\}}^{'}
     \end{bmatrix}, \: 
         \bm{\Pi} = \begin{bmatrix}
         \Pi_{\{A\}}\\
         \Pi_{\{B\}}
     \end{bmatrix}, \: 
     M = \begin{bmatrix}
         -\left(e_A + d_{AB} + b^{A}_{B} + e_B\right) & -\left(w_A + b^{A}_{B} + e_B\right)\\
         -\left(w_B + b^{A}_{B} + e_A\right) & -\left(e_B + d_{BA} + b^{A}_{B} + e_A\right)
     \end{bmatrix}, \: \nonumber\\[10pt]
     &&\bm{r} = \begin{bmatrix}
         w_A + b^{A}_{B} + e_B \\
         w_B + b^{A}_{B} + e_A
     \end{bmatrix}.
\end{eqnarray}

First, we find the complimentary solution to the following equation using eigenvalues and eigenvectors of matrix $M$,
\begin{eqnarray}
    \bm{\Pi^{'}} &=& M\bm{\Pi}. 
    \label{eq::complimentary}
\end{eqnarray}
Using Mathematica, the eigenvalues ($\lambda_1,\lambda_2$) and eigenvectors ($\bm{\nu}_1,\bm{\nu}_2$) of $M$ are given by, 
\begin{eqnarray}
    \bm{\lambda} &=& \begin{bmatrix}
        \lambda_1 \\
        \lambda_2
    \end{bmatrix}\\
    &=& \begin{bmatrix}
        \frac{1}{2}\left(-2 b^{A}_{B} - d_{AB} -d_{BA} - 2e_{A} - 2e_{B} - R\right)\\[10pt]
        \frac{1}{2}\left(-2 b^{A}_{B} - d_{AB} -d_{BA} - 2e_{A} - 2e_{B} + R\right)
    \end{bmatrix}, \\[10pt]
    \bm{\nu}_1 &=& \begin{bmatrix}
        -\frac{1}{2\left(b^{A}_{B} + e_A + w_B\right)}\left(-d_{AB} + d_{BA} -R\right)\\[10pt]
        1
    \end{bmatrix}, \\[10pt]
    \bm{\nu}_2 &=& \begin{bmatrix}
        -\frac{1}{2\left(b^{A}_{B} + e_A + w_B\right)}\left(-d_{AB} + d_{BA} + R\right)\\[10pt]
        1
    \end{bmatrix}, 
\end{eqnarray}
where
\begin{eqnarray}
    &&R = \sqrt{R_{1}+R_{2}},\nonumber \\[10pt]
    &&R_{1} = 4 \left(b^{A}_{B}\right)^2 + 4\left(b^{A}_{B}e_{A} + b^{A}_{B}e_{B} + b^{A}_{B}w_A + b^{A}_{B}w_B\right) + 4\left(e_{A}e_{B} + e_{A}w_{A} + e_{B}w_{B} + w_{A}w_{B} \right),\nonumber\\[10pt]
    &&R_{2} = - 2 d_{AB}d_{BA} + \left(d_{AB}^{2} + d_{BA}^{2}\right).\nonumber
\end{eqnarray}
The complimentary solution for Eq.~\eqref{eq::complimentary} is given by,
\begin{eqnarray}
    \bm{\Pi}_{C} &=& C_{1}\bm{\nu}_{1}e^{\lambda_{1}t} + C_{2}\bm{\nu}_{2}e^{\lambda_{2}t},
    \label{eq::complimentary_sol}
\end{eqnarray}
where $C_1$ and $C_2$ are arbitrary constants.\\

Next, we find the particular solution $\bm{\Pi}_{P}$ for Eq.~\eqref{eq::matrixform} using the method of undetermined coefficients. Suppose the solution $\bm{\Pi}_{P}$ is of the form 
\begin{eqnarray}
    \bm{\Pi}_{P} = \begin{bmatrix}
        K_{1} \\
        K_{2}
    \end{bmatrix}, \:
    \bm{\Pi}_{P}^{'} = \begin{bmatrix}
        0 \\
        0
    \end{bmatrix}.
    \label{eq::particular_subs}
\end{eqnarray}

Substitute Eq.~\eqref{eq::particular_subs} to Eq.~\eqref{eq::matrixform} we have,
\begin{eqnarray}
      \bm{\Pi^{'}}_{P} &=& M\bm{\Pi}_{P} + \bm{r}.
\end{eqnarray}
That is, we want to solve the following system of linear equations, 
\begin{eqnarray}
    \left(e_A + d_{AB} + b^{A}_{B} + e_B\right)K_{1} + \left(w_A + b^{A}_{B} + e_B\right)K_{2} &=& w_A + b^{A}_{B} + e_B    \label{linear_ode_1} \\
    \left(w_B + b^{A}_{B} + e_A\right)K_{1} + \left(e_B + d_{BA} + b^{A}_{B} + e_A\right)K_{2} &=& w_B + b^{A}_{B} + e_A.
    \label{linear_ode_2}
\end{eqnarray}
Thus, 
\begin{eqnarray}
    K_1 &=& \frac{\left(w_A + b^{A}_{B} + e_B\right)\left(e_B + d_{BA} - w_B\right)}{\left(e_A + d_{AB} + b^{A}_{B} + e_B\right)\left(e_B + d_{BA} + b^{A}_{B} + e_A\right)-\left(w_B + b^{A}_{B} + e_A\right)\left(w_A + b^{A}_{B} + e_B\right)} \nonumber\\ [10pt]
    &=& \frac{num_A}{denom_A}.
    \label{sol_K1}
\end{eqnarray}
Substitute Eq.~\eqref{sol_K1} to Eq.~\eqref{linear_ode_1} to get, 
\begin{eqnarray}
    K_2 = 1 - \left(\frac{e_A + d_{AB} + b^{A}_{B} + e_B}{w_A + b^{A}_{B} + e_B}\right) \left(\frac{num_A}{denom_A}\right).
\end{eqnarray}
Therefore, the general solution is given by,
\begin{eqnarray}
    \bm{\Pi} &=& \bm{\Pi}_{C} + \bm{\Pi}_{P} \nonumber\\
    &=& C_{1}\bm{\nu}_{1}e^{\lambda_{1}t} + C_{2}\bm{\nu}_{2}e^{\lambda_{2}t} + \bm{K},
    \label{eq::sol_general_ode}
\end{eqnarray}
where 
\begin{eqnarray}
    \bm{K} = \begin{bmatrix}
        K_1 \\
        K_2
    \end{bmatrix}.
\end{eqnarray}
By taking a limit of Eq.~\eqref{eq::sol_general_ode} as $t \rightarrow \infty$, the exponential terms in Eq.~\eqref{eq::sol_general_ode} will approach $0$. Therefore, we have,
\begin{eqnarray}
    \bm{\hat{\Pi}} &=& \lim_{t \rightarrow \infty} \bm{\Pi} \nonumber \\
    \begin{bmatrix}
        \hat{\Pi}_{\{A\}}\\
        \hat{\Pi}_{\{B\}}
    \end{bmatrix} 
    & = & 
    \begin{bmatrix}
        K_1 \\
        K_2
    \end{bmatrix}.
    \label{eq::stationary_proof}
\end{eqnarray}
$\hat{\Pi}_{\{A\}}$ and $\hat{\Pi}_{\{B\}}$ from Eq.~\eqref{eq::stationary_proof} are the stationary frequencies for state $\{A\}$ and $\{B\}$, respectively, as shown in Lemma~\ref{lemma::solve}. \\

Next, to get the general solution to the system in Eq.~\eqref{eq::matrixform}, we find the constants, $C_1$ and $C_2$ by substituting the initial value condition to Eq.~\eqref{eq::sol_general_ode} for $t=0$. We have,
\begin{eqnarray}
    C_{1}\begin{bmatrix}
        -\frac{1}{2\left(b^{A}_{B} + e_A + w_B\right)}\left(-d_{AB} + d_{BA} -R\right)\\[10pt]
        1
    \end{bmatrix}
    +
    C_{2}\begin{bmatrix}
        -\frac{1}{2\left(b^{A}_{B} + e_A + w_B\right)}\left(-d_{AB} + d_{BA} + R\right)\\[10pt]
        1
    \end{bmatrix}
    +
    \begin{bmatrix}
        K_1 \\[10pt]
        K_2
    \end{bmatrix}
    = 
    \begin{bmatrix}
        \Pi_{A}^{0} \\[10pt]
        \Pi_{B}^{0}
    \end{bmatrix}.\nonumber\\
\end{eqnarray}
That is, we solve the following system of linear equations, 
\begin{eqnarray}
    C_{1}\left(\frac{R + d_{AB} - d_{BA}}{2\left(b^{A}_{B} + e_A + w_B\right)}\right) + C_{2}\left(\frac{d_{AB} - d_{BA} - R}{2\left(b^{A}_{B} + e_{A} + w_{B}\right)}\right) &=& \Pi_{A}^{0} - K_{1}, \\
    \label{eq::constant_1}
    C_{1} + C_{2} &=& \Pi_{B}^{0} - K_{2}.
    \label{eq::constant_2}
\end{eqnarray}
Thus, 
\begin{eqnarray}
    C_{1} = \frac{\left(\Pi_{A}^{0} - K_1\right)\left(b^{A}_{B} + e_{A} + w_{B}\right)}{R} - \frac{\left(\Pi_{B}^{0}-K_{2}\right)\left(d_{AB} - d_{BA} - R\right)}{2 R}.
    \label{eq::constant_1_sol}
\end{eqnarray}
Then, substitute Eq.~\eqref{eq::constant_1_sol} to Eq.~\eqref{eq::constant_2} we get,
\begin{eqnarray}
    C_{2} = \frac{\left(K_{1}-\Pi_{A}^{0}\right)\left(b^{A}_{B} + e_{A} + w_{B}\right)}{R} + \left(\Pi_{B}^{0} - K_{2}\right)\left(1 + \frac{d_{AB}-d_{BA}-R}{2 R}\right).
\end{eqnarray}
\qed

\subsection{Simulating range state dynamics using MASTER}
\label{subsec::master}

We can express events in MASTER for GeoSSE using the following reaction equations.
\begin{eqnarray*}
    \hat{S}[\{i\}] &\xrightarrow{e_{i}}& \hat{R}[i] + L[i], \: (\textit{extinction of endemic species in region i}) \\
    \hat{S}[\{i\}] &\xrightarrow{d_{ij}}& \hat{S}[\{i\} \cup \{j\}] + G[j], \: (\textit{dispersal from region i to region j})\\
    \hat{S}\left[\bigcup_{i \in R}\{i\}\right] &\xrightarrow{e_{j}}& \hat{S}\left[\bigcup_{i \in R;i \neq j}\{i\}\right] + L[j], \: (\textit{local extinction in region j})\\
    \hat{S}\left[\bigcup_{i \in R}\{i\}\right] &\xrightarrow{b^{i}_{j}}& \hat{S}\left[\bigcup_{i \in R;i \neq j}\{i\}\right]  + \hat{S}[\{j\}], \: (\textit{between-region speciation into ranges $\{i\}$ and $\{j\}$})\\
    \hat{S}\left[\bigcup_{i \in R}\{i\}\right] &\xrightarrow{w_j}& \hat{S}\left[\bigcup_{i \in R}\{i\}\right] + S[{j}] + G[j], \: (\textit{within region speciation in region $j$ for a widespread species})\\
    \hat{S}[\{i\}] &\xrightarrow{w_i}& \hat{S}[\{i\}] + \hat{S}[\{i\}] + G[i], \: (\textit{within region speciation in region $i$ for an endemic species}),
\end{eqnarray*}
where $\hat{S}[\{i\}]$ indicates the number of endemic species with range $\{i\}$, $\hat{S}\left[\bigcup_{i \in R}\{i\}\right]$ indicates the number of widespread species with range $\bigcup_{i \in R}\{i\}$, $\hat{R}[i]$ indicates the number of species in region $i$, $L[i]$ indicates a species lost in region $i$, and $G[i]$ indicates a species gain in region $i$.\\

\subsection{Simulation under the diffusion model starting with a single species}
\label{subsec::singlespec}

\begin{figure}[h!]
	\centering
    \begin{subfigure}{0.55\textwidth}
        \includegraphics[width=1\linewidth]{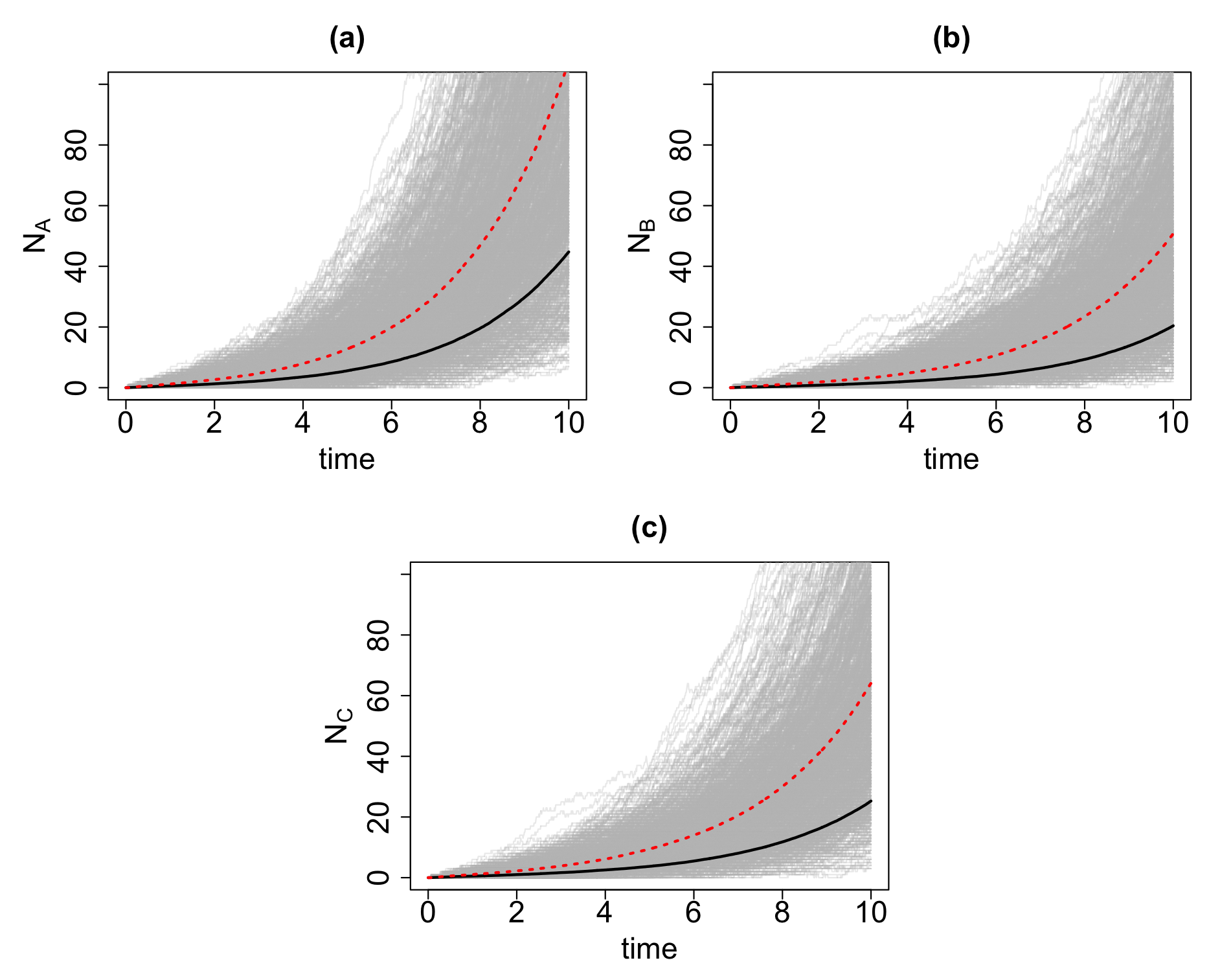}
    \end{subfigure}%
    ~
    \begin{subfigure}{0.55\textwidth}
        \includegraphics[width=1\linewidth]{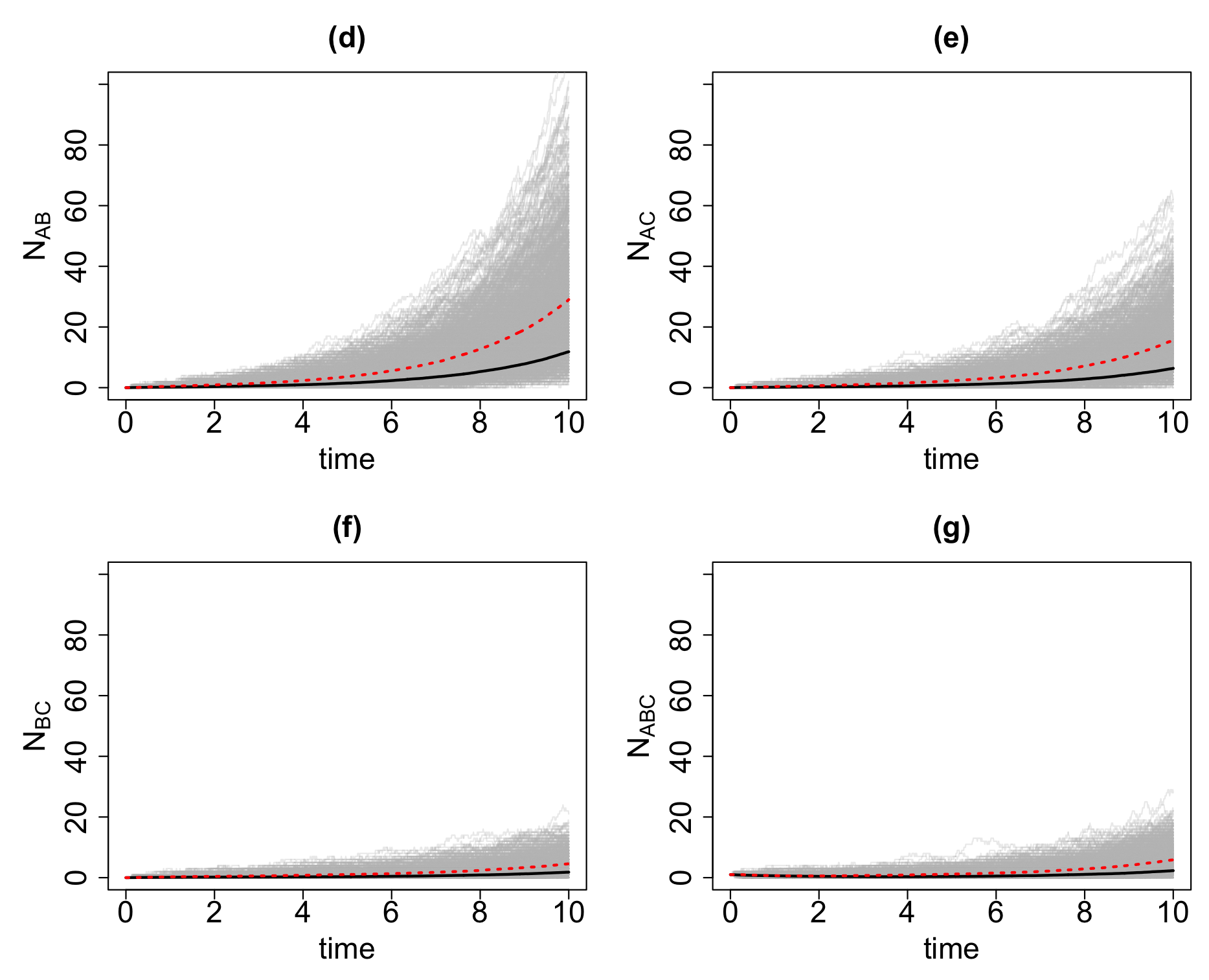}
    \end{subfigure}
	\nobreak
	\caption{The trajectories of mean counts of range states for endemic species (a-c) and widespread species (d-g) over the $[0,10]$ time interval. Trajectories were simulated under the diffusion-based process (red line) and tree-based process (black line), starting with 1 species in a random state. For each starting state, we simulate 150 trajectories (1050 trajectories in total) for each approach. The gray trajectories show the dynamics across 1050 replicates simulated under diffusion-based process. Simulations are conducted using the following parameter values: $w_{A}=0.36,w_{B}=0.24,w_{C}=0.28,b^{A}_{B}=b^{A}_{C}=b^{B}_{C}=b^{A}_{BC}=b^{B}_{AC}=b^{C}_{AB}=0.16,e_{A}=0.02,e_{B}=0.03,e_{C}=0.01,d_{AB}=d_{BA}=0.12,d_{AC}=d_{CA}=0.06,d_{BC}=d_{CB}=0.02$ }
	\label{fig:compare_nocorrect}
\end{figure}

 We propose a procedure to correct the diffusion-based simulation for state dynamics when starting with a single species in random states under an SSE model, as demonstrated for a GeoSSE model. Note that without this correction procedure, the bias between the tree-based and diffusion-based approaches is apparent as described in Example~\ref{ex::bias_small}.\\
         \begin{Example}
            \label{ex::bias_small}
            Consider a 3-region GeoSSE system with state space \\
            $S = \{\{A\},\{B\},\{C\},\{A,B\},\{A,C\},\{B,C\},\{A,B,C\}\}$. Suppose we start with a single species in state $\{A,B\}$. \\
            
            \noindent That is, at $t=0$ we have
            \begin{eqnarray*}
                N_{\{A\}} = 0, \: N_{\{B\}} = 0, N_{\{C\}} = 0, \: N_{\{A,B\}} = 1, \: N_{\{A,C\}} = 0, \: N_{\{B,C\}} = 0, \:  N_{\{A,B,C\}} = 0.
            \end{eqnarray*}
            It is very likely, given a small total species number, for diffusion-based simulation to give a pattern at the next time step such as the one below, 
            \begin{eqnarray}
                N_{\{A\}} = 0, \: N_{\{B\}} = 0, N_{\{C\}} = 1, \: N_{\{A,B\}} = 1, \: N_{\{A,C\}} = 0, \: N_{\{B,C\}} = 0, \:  N_{\{A,B,C\}} = 0.\nonumber \\
                \label{eq::unlikely}
            \end{eqnarray}
            The transition described in Eq.~\eqref{eq::unlikely} cannot be explained using just one GeoSSE event. Furthermore, since the diffusion parameters for simulating species count across states, described in Eqs~\eqref{eq::mu_geosse}-\eqref{eq::var_geosse}, dynamically depend on the number of species in each state, the difference between diffusion-based and tree-based simulations becomes more apparent over time (Fig.~\ref{fig:compare_nocorrect}).\\
        \end{Example}
In general, the correction procedure works by preventing a diffusion-based path to enter a disallowed regime when there are few species (Table~\ref{table:path_difference}). This procedure only applies for simulating state counts. However, we can get the state frequencies from species counts in each state using the following formula, 
    \begin{eqnarray*}
        \Pi_{i}(t) = \frac{N_{i}(t)}{\sum_{i\in S}N_{i}(t)}.
    \end{eqnarray*}
\begin{enumerate}
    \item For each simulated trajectory for state counts under the diffusion-based approach where we start with a single species in a random state, we simulate the next path according to the procedure described in Section~\ref{sec::Pro1}.\\
    \item Next, we compute the difference in paths at $t$ and $t + \Delta t$. That is, we find
    \begin{eqnarray}
        \bm{N}(t+\Delta t) - \bm{N}(t) = \{N_{i}(t+\Delta t) - N_{i}(t)\}_{\forall i \in S},
        \label{eq:diff_small_path}
    \end{eqnarray}
    where $S$ is the range state space of the GeoSSE model. Then, round each element of the vector in Eq.~\eqref{eq:diff_small_path} to the nearest integer value.\\
    \item We check whether this difference in paths is in the list described in Table~\ref{table:path_difference}. If it is not in the list, we reject this future path and re-sample until it satisfies the table. In effect, this correction approximates model-based conditional sampling for diffusion-based paths. \\
    \item Repeat steps~$1-2$ until we reach a path where every state has at least 1 representative species, $\forall i : N_i(t) > 0$. Then, we generate the future paths as usual, according to the procedure described in Section~\ref{sec::Pro1}.
\end{enumerate}
\begin{figure}[h!]
	\centering
    \begin{subfigure}{0.55\textwidth}
        \includegraphics[width=1\linewidth]{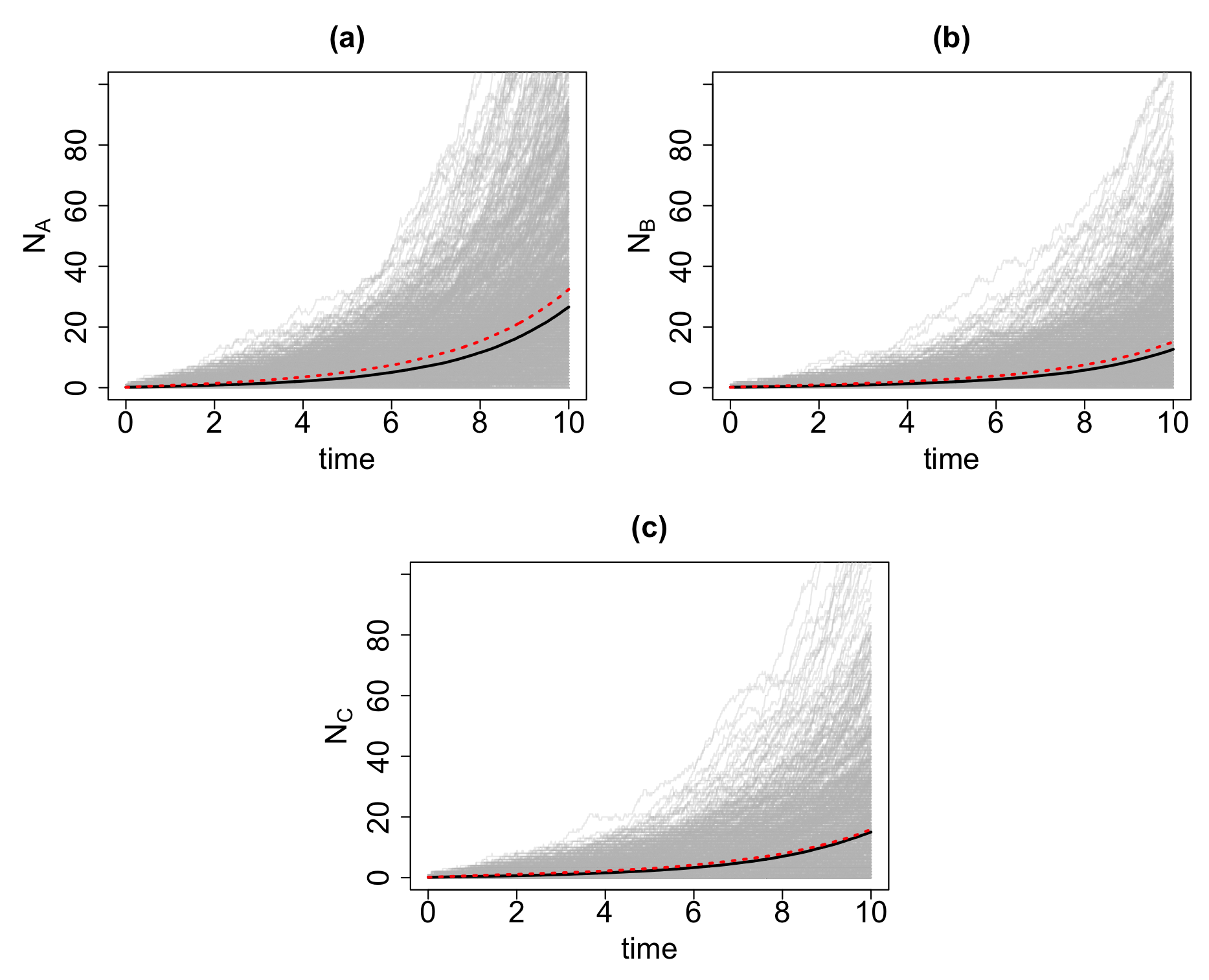}
    \end{subfigure}%
    ~
    \begin{subfigure}{0.55\textwidth}
        \includegraphics[width=1\linewidth]{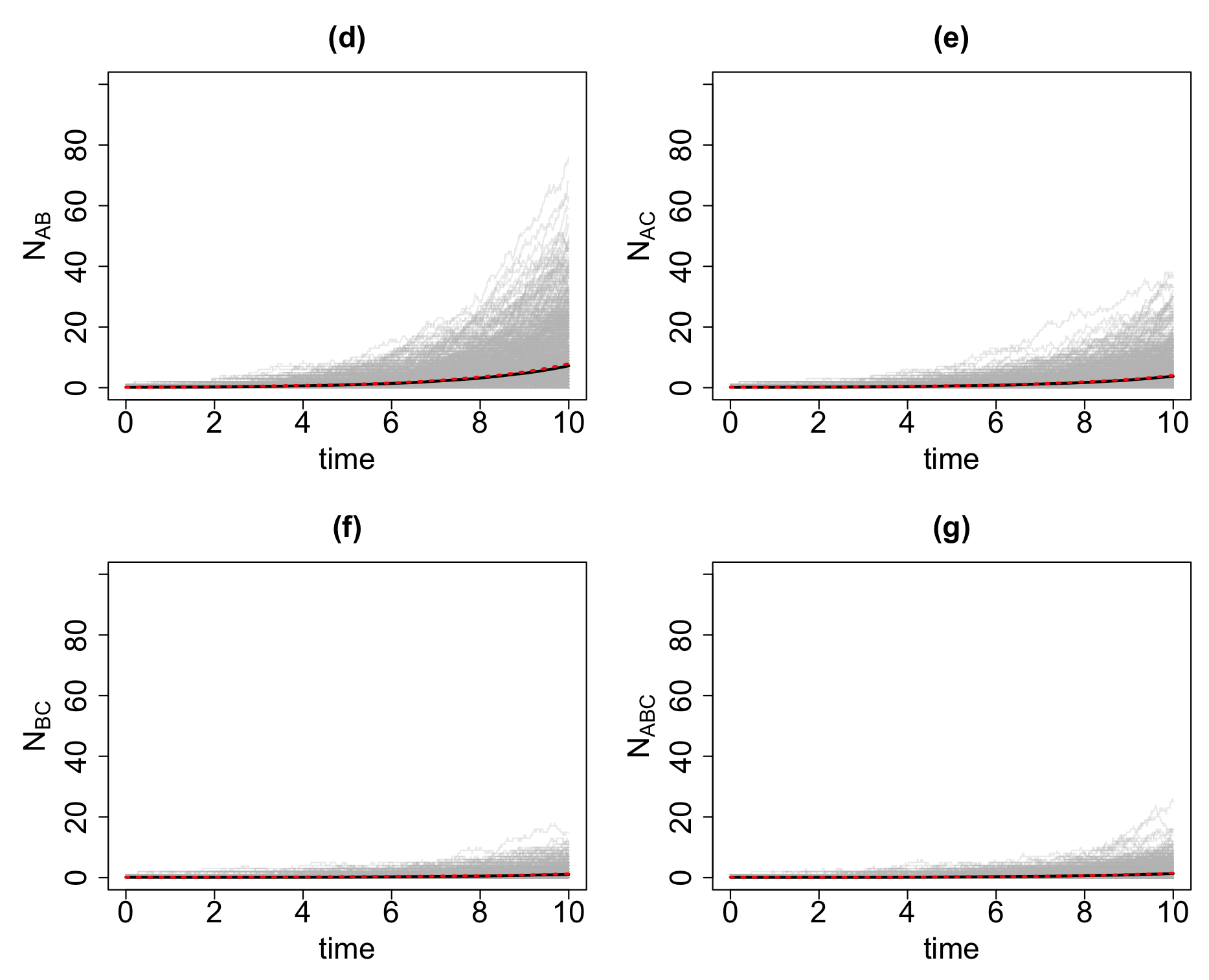}
    \end{subfigure}
	\nobreak
	\caption{The trajectories of average count of range states for endemic species (a-c) and widespread species (d-g) over the $[0,10]$ time interval simulated under the diffusion-based process (red line) after applying the correction procedure and tree-based process (black line), starting with 1 species in a random state. For each starting state, we simulate 150 trajectories (1050 trajectories in total) for each approach. The gray trajectories show the dynamics across 1050 replicates simulated under diffusion-based process. Simulations are conducted using the following parameter values: $w_{A}=0.36,w_{B}=0.24,w_{C}=0.28,b^{A}_{B}=b^{A}_{C}=b^{B}_{C}=b^{A}_{BC}=b^{B}_{AC}=b^{C}_{AB}=0.16,e_{A}=0.02,e_{B}=0.03,e_{C}=0.01,d_{AB}=d_{BA}=0.12,d_{AC}=d_{CA}=0.06,d_{BC}=d_{CB}=0.02$ }
	\label{fig:compare_correct}
\end{figure}
As seen in Fig.~\ref{fig:compare_correct}, the difference between diffusion-based and tree-based simulation is reduced  after applying the correction procedure. Note that it is possible to apply this correction procedure to any discrete-state SSE models by modifying the table (Table~\ref{table:path_difference}) according to events described by the model. Also, this procedure only minimizes the difference between diffusion-based and tree-based trajectories as it is still possible for a diffusion-based simulation to construct disallowed paths at a later time, despite the starting condition. However, these differences become negligible for as $N(t) >> 0$ for our example.
\begin{table}[!htbp]
	\caption{List of allowed differences between two consecutive paths in a 3-region GeoSSE model}
	\label{table:path_difference}
		\centering
    		\begin{tabular}{lccccccc}
    			\hline
    			\multirow{2}{*}{Condition} & \multicolumn{7}{c}{$\bm{N}_{i}(t+\Delta t) - \bm{N}_{i}(t)$}\\
                    \cline{2-8}
    			& $\{A\}$ & $\{B\}$ & $\{C\}$ & $\{A,B\}$ & $\{A,C\}$ & $\{B,C\}$ & $\{A,B,C\}$ \\
       			\hline
                \multicolumn{8}{c}{\textbf{Within region speciation}}\\
                    \hline
                If $N_{\{A\}}(t) \neq 0, \: N_{j}(t) = 0, \forall j \in S \setminus {\{A\}} $ & 1 & 0 & 0 & 0 & 0 & 0 & 0\\
                    \hline
                If $N_{\{B\}}(t) \neq 0, \: N_{j}(t) = 0, \forall j \in S \setminus {\{B\}} $ & 0 & 1 & 0 & 0 & 0 & 0 & 0\\
                    \hline
                If $N_{\{C\}}(t) \neq 0, \: N_{j}(t) = 0, \forall j \in S \setminus {\{C\}} $ & 0 & 0 & 1 & 0 & 0 & 0 & 0\\
                    \hline
                \multirow{2}{*}{\raggedleft If $N_{\{A,B\}} \neq 0, \: N_{\{A,B,C\}} = 0 $} & 1 & 0 & 0 & 0 & 0 & 0 & 0\\
                & 0 & 1 & 0 & 0 & 0 & 0 & 0\\
                    \cline{1-8}
                \multirow{2}{*}{\raggedleft If $N_{\{A,C\}} \neq 0, \: N_{\{A,B,C\}} = 0 $} & 1 & 0 & 0 & 0 & 0 & 0 & 0\\
                & 0 & 0 & 1 & 0 & 0 & 0 & 0\\
                    \cline{1-8}
                \multirow{2}{*}{\raggedleft If $N_{\{B,C\}} \neq 0, \: N_{\{A,B,C\}} = 0 $} & 0 & 1 & 0 & 0 & 0 & 0 & 0\\
                & 0 & 0 & 1 & 0 & 0 & 0 & 0\\
                    \cline{1-8}
                \multirow{3}{*}{\raggedleft If $N_{\{A,B,C\}} \neq 0 $} & 1 & 0 & 0 & 0 & 0 & 0 & 0\\
                & 0 & 1 & 0 & 0 & 0 & 0 & 0\\
                & 0 & 0 & 1 & 0 & 0 & 0 & 0\\
                    \cline{1-8}
                \multicolumn{8}{c}{\textbf{Extinction in species with range size 1}}\\
                \hline
                \multirow{3}{*}{} & -1 & 0 & 0 & 0 & 0 & 0 & 0\\
                & 0 & -1 & 0 & 0 & 0 & 0 & 0\\
                & 0 & 0 & -1 & 0 & 0 & 0 & 0\\
                    \cline{1-8}
                \multicolumn{8}{c}{\textbf{Extinction in species with range size 2}}\\
                \hline
                \multirow{6}{*}{} & 1 & 0 & 0 & -1 & 0 & 0 & 0\\
                & 1 & 0 & 0 & 0 & -1 & 0 & 0\\
                & 0 & 1 & 0 & -1 & 0 & 0 & 0\\
                & 0 & 1 & 0 & 0 & 0 & -1 & 0\\
                & 0 & 0 & 1 & 0 & -1 & 0 & 0\\
                & 0 & 0 & 1 & 0 & 0 & -1 & 0\\
                    \cline{1-8}
                \multicolumn{8}{c}{\textbf{Extinction in species with range size 3}}\\
                \hline
                \multirow{3}{*}{} & 0 & 0 & 0 & 1 & 0 & 0 & -1\\
                & 0 & 0 & 0 & 0 & 1 & 0 & -1\\
                & 0 & 0 & 0 & 0 & 0 & 1 & -1\\
                    \cline{1-8}
                \multicolumn{8}{c}{\textbf{Between-region speciation}}\\
                \hline
                \multirow{6}{*}{} & 1 & 1 & 0 & -1 & 0 & 0 & 0\\
                & 1 & 0 & 1 & 0 & -1 & 0 & 0\\
                & 0 & 1 & 1 & 0 & 0 & -1 & 0\\
                & 1 & 0 & 0 & 0 & 0 & 1 & -1\\
                & 0 & 1 & 0 & 0 & 1 & 0 & -1\\
                & 0 & 0 & 1 & 1 & 0 & 0 & -1\\
                    \cline{1-8}
                \multicolumn{8}{c}{\textbf{Dispersal in species with range size 1}}\\
                \hline
                \multirow{6}{*}{} & -1 & 0 & 0 & 1 & 0 & 0 & 0\\
                & -1 & 0 & 0 & 0 & 1 & 0 & 0\\
                & 0 & -1 & 0 & 1 & 0 & 0 & 0\\
                & 0 & -1 & 0 & 0 & 0 & 1 & 0\\
                & 0 & 0 & -1 & 0 & 1 & 0 & 0\\
                & 0 & 0 & -1 & 0 & 0 & 1 & 0\\
                    \cline{1-8}
                \multicolumn{8}{c}{\textbf{Dispersal in species with range size 2}}\\
                \hline
                \multirow{3}{*}{} & 0 & 0 & 0 & -1 & 0 & 0 & 1\\
                & 0 & 0 & 0 & 0 & -1 & 0 & 1\\
                & 0 & 0 & 0 & 0 & 0 & -1 & 1\\
    		\end{tabular}
\end{table}

 \clearpage

	\bibliographystyle{abbrv}
	\bibliography{main}
\end{document}